\documentclass[twocolumn]{aastex63}

\usepackage{savesym}
\savesymbol{tablenum}
\usepackage{siunitx}
\usepackage{nicefrac}
\usepackage{bm}
\usepackage{mathtools}
\usepackage{multirow}
\usepackage{CJK}         

\usepackage{lineno}

\begin{document}
\begin{CJK*}{UTF8}{gbsn}

\title{Negative Lags on the Viscous Timescale in Quasar Photometry and Prospects for Detecting More with LSST}

\author[0000-0002-1174-2873]{Amy Secunda}
\affil{Department of Astrophysical Sciences, Princeton University, Peyton Hall, Princeton, NJ 08544, USA}

\author[0000-0002-5612-3427]{Jenny E. Greene}
\affil{Department of Astrophysical Sciences, Princeton University, Peyton Hall, Princeton, NJ 08544, USA}

\author[0000-0002-2624-3399]{Yan-Fei Jiang (姜燕飞)}
\affil{Center for Computational Astrophysics, Flatiron Institute, New York, NY 10010, USA}

\author[0000-0003-3024-7218]{Philippe Z. Yao}
\affil{Department of Astrophysical Sciences, Princeton University, Peyton Hall, Princeton, NJ 08544, USA}

\author[0000-0002-0572-9613]{Abderahmen Zoghbi}
\affil{Department of Astronomy, University of Maryland, College Park, MD 20742}
\affil{HEASARC, Code 6601, NASA/GSFC, Greenbelt, MD 20771}
\affil{CRESST II, NASA Goddard Space Flight Center, Greenbelt, MD 20771}

\begin{abstract}
    The variability of quasar light curves can be used to study the structure of quasar accretion disks. For example, continuum reverberation mapping uses delays between variability in short and long wavelength bands (``short'' lags) to measure the radial extent and temperature profile of the disk. Recently, a potential reverse lag, where variations in shorter wavelength bands lag the longer wavelength bands at the much longer viscous timescale, was detected for Fairall 9. Inspired by this detection, we derive a timescale for these ``long'' negative lags from fluctuation propagation models and recent simulations. We use this timescale to forecast our ability to detect long lags using the Vera Rubin Legacy Survey of Space and Time (LSST). After exploring several methods, including the interpolated cross-correlation function, a Von-Neumann estimator, {\sc javelin}, and a maximum-likelihood Fourier method, we find that our two main methods, {\sc javelin} and the maximum-likelihood method, can together detect long lags of up to several hundred days in mock LSST light curves. Our methods work best on proposed LSST cadences with long season lengths, but can also work for the current baseline LSST cadence, especially if we add observations from other optical telescopes during seasonal gaps. We find that LSST has the potential to detect dozens to hundreds of additional long lags. Detecting these long lags can teach us about the vertical structure of quasar disks and how it scales with different quasar properties.
\end{abstract}

\keywords{Accretion (14), Active galactic nuclei (16), Black hole physics (159), Black holes (162), Seyfert galaxies (1447), Supermassive black holes (1663)}

\section{Introduction}

Quasars are among the most luminous objects in the Universe. As a result, they are important to a variety of fields within astrophysics, from the study of accretion disks to galaxy evolution \citep{Fabian2012, Kormendy2013}. With the exception of the extraordinary efforts by the \cite{EHT2019a} and \cite{Gravity:2018}, it is generally not possible to resolve the microarcsecond scales of the accretion disk, broad line region (BLR), and event horizon. Since spatial resolution is so difficult to achieve and quasars vary over a range of frequencies, temporal resolution is instead often used to study the accretion disks of quasars.

In particular, \cite{Blandford1982} proposed a technique known as reverberation mapping to measure distances to the BLR, by measuring time lags (of order 100 days) between the continuum emission and broad emission lines \citep[see also][]{Kaspi_2000, Peterson2004, Bentz_2015, Grier:2017}. A second type of reverberation mapping, first introduced by \cite{Collier:1999} and known as disk continuum reverberation mapping, uses time lags between variability in different wavebands to measure the radial extent of the disk, using the lamppost model \citep[e.g.,][]{Sergeev:2005, Cackett:2007, Cackett:2018, derosa2015, Edelson:2015, Edelson2017, Edelson:2019, Jiang:2017,  Fausnaugh2016, Starkey2017,Homayouni:2022}. The lamp-post model proposes that high-frequency (X-ray or hard-UV) emission from the corona will be reprocessed by the disk and re-emitted as UV/optical light. Because it takes light time to travel from the inner hotter regions of the disk, where it is re-emitted in shorter wavebands, to the outer cooler regions of the disk, where it is re-emitted in longer wavebands, there will be a time lag of order days between variability in short wavelength bands and long wavelength bands. Because this lag is on the order of days, we refer to it here as the ``short'' lag.

Using the \cite{ShakuraSunyaev1973} accretion disk model of an optically thick, geometrically thin disk to determine the temperature dependence of the disk, this time lag should scale as $\tau_{\rm c} \propto \lambda^{4/3}$, and also depend on the mass of the supermassive black hole (SMBH) and accretion rate \citep[see also,][]{LyndenBell1969,PringleRees1972,NovikovThorne1973,Friedjung1985}. However, recent observations have suggested that the radial extent of the disk is larger than predicted by the standard thin disk model \citep{Jiang:2017,Cackett2018,Mudd:2018,Guo:2022}, although this discrepancy may be due to nuclear extinction \citep{Gaskell2017,Gaskell2022} or an underestimation of SMBH masses due to unknown BLR geometry \citep{Pozo2019}. Other models suggest that AGN disks may appear more extended due to large azimuthal temperature fluctuations \citep{Dexter2012}, disk winds \citep{LaorDavis2014}, or magnetically elevated accretion \citep{DexterBegelman2019}, none of which are included in the standard thin disk model.

There is also growing evidence that the lamp-post model may be insufficient to fully explain the optical/UV variability in quasar light curves \cite[e.g.,][]{Arevalo:2009,Shappee:2014,Edelson2019,Yu:2022} Alternative sources of optical/UV light curve variability, which may occur on the thermal timescale, are magneto-rotational instability turbulence \citep{BalbusHawley1991,Burkeetal2021} and convection due to enhanced opacity in the optical/UV region of the disk \citep{JiangBlaes2020}. As these turbulent perturbations are accreted inwards on the viscous timescale, they can also modify the variability of light curves in inner regions of the disk \citep{Lyubarskii1997}. Because these perturbations are accreted inwards the lag will be ``negative'' compared to the short lag. In other words, light curve variability from disk turbulence in short wavelength bands will lag variability in long wavelength bands. The viscous timescale is \citep{Davis_2020},
\begin{equation}
\label{eq:ttherm}
\tau_{\rm{visc}} = \frac{1}{\alpha \Omega}\left(\frac{h}{r}\right)^{-2},
\end{equation}
where $\alpha$ is the \cite{ShakuraSunyaev1973} stress parameter, $\Omega=(GM/r^3)^{1/2}$ is the orbital frequency, $M$ is the mass of the SMBH, $G$ is the gravitational constant, $h$ is the scale height of the disk, and $r$ is the radial position in the disk. 

In the standard thin disk model $\tau_{\rm visc}$ can be of order hundreds of years, because in the radiation-dominated region of the disk $h$ is constant as a function of $r$, and $h/r \lesssim 0.01$. However, a recent simulation by \cite{JiangBlaes2020} suggests that due to variations in the scale height of the disk, the viscous timescale could be on the order of 100~days. Furthermore, two recent independent works have suggested the detection of negative lags in the nearby quasar Fairall 9 \citep{F92020,Yao:2022}. In particular, using the \emph{SWIFT} and Las Cumbres Observatory (LCO) light curves with a baseline of $\sim 270$~days and sub-daily cadence from \cite{F92020}, \cite{Yao:2022} found a long negative lag of $\sim -70$~days between the \emph{W2}- and \emph{z}-band. This sub-100 day negative lag supports the conclusions in the simulation in \cite{JiangBlaes2020} and suggests that it may be possible to detect long lags in many more quasars.

In this paper, we derive a new empirical model for the ``long'' negative lag based on the model fit to the long lags detected for Fairall 9 by \cite{Yao:2022}. We then use this model to examine our ability to detect long lags in the future. Detecting more long lags will help to constrain our model and provide information on the vertical structure of disks. Additional lag measurements would also allow us to model how the vertical structure of disks change as a function of quasar properties such as SMBH mass and accretion rate. 

Because long lags can last tens to hundreds of days, a long baseline of high cadence observations is necessary to detect them. The Vera Rubin Legacy Survey of Space and Time \citep[LSST,][]{Ivezi2019} will carry out a comprehensive survey of the southern sky and will contain over 100 million quasars. LSST will provide photometric light curves in \emph{ugrizy} bands, allowing us to study the wavelength dependence of quasar lags. \cite{Kovacevic:2022}, \cite{PozoNunez:2023}, and others found that LSST should be able to detect short lags in quasar light curves. Here, we generate mock LSST quasar light curves reprocessed with several different short and long lags. We then test our ability to detect long lags using several lag detection methods, primarily focusing on two: {\sc javelin} \citep{Zu2011} and a maximum-likelihood Fourier method developed by \cite{Miller2010} and adapted for use on quasars by \cite{Zoghbi2013} and \cite{Cackett:2022}.

We derive our model in Section \ref{sec:derive}, describe in detail how we generate our mock LSST light curves in Section \ref{sec:mocks}, and describe our lag detection methods in Section \ref{sec:detection}. In Section \ref{sec:detect} we report the accuracy of our lag detection methods for light curves reprocessed with only long lags (Section \ref{sec:long}), light curves reprocessed with long and short lags (Section \ref{sec:short_long}), light curves observed with different proposed LSST cadences (Section \ref{sec:lco}), light curves from the simulation in \cite{JiangBlaes2020} (Section \ref{sec:light_curves}), and light curves for dimmer magnitude quasars (Section \ref{sec:i21}) We provide a brief summary of the accuracy of our lag detection methods for all light curves in Section \ref{sec:result summary}. In Section \ref{sec:number} we make a prediction for the number of long lags detectable by LSST over its ten year baseline. We summarize our results in Section \ref{sec:discuss} and discuss the importance of improved light curve models (Section \ref{sec:discuss:lcs}) and detection methods (Section \ref{sec:discuss:detect}).

\section{What is the Long Lag Timescale?}
\label{sec:derive}

\begin{figure*}
    \centering
    \includegraphics[width=\textwidth]{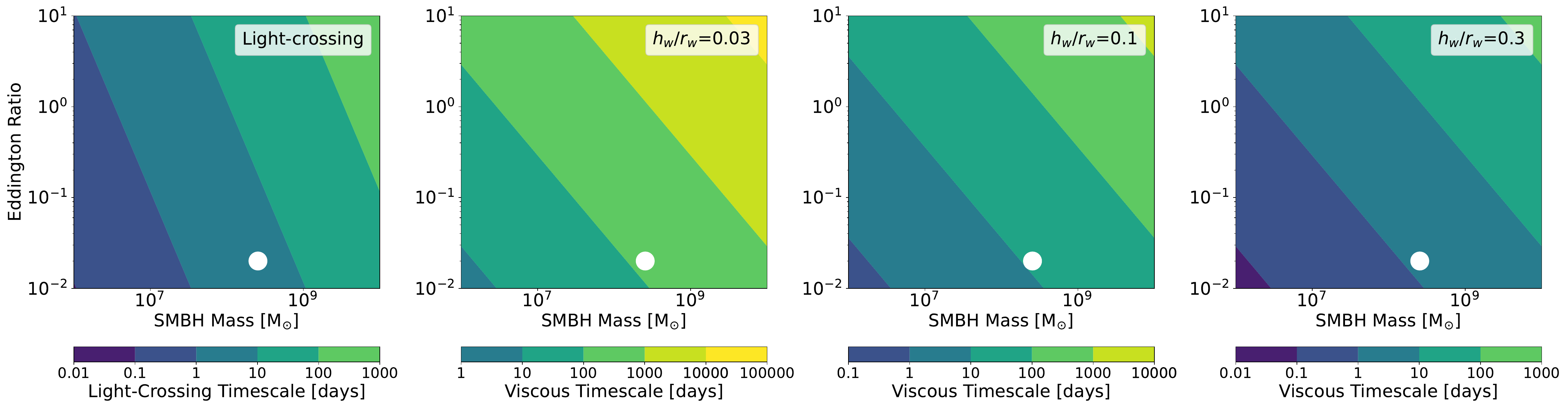}
    \caption{The leftmost panel shows the rest-frame short lag between the \emph{u}- and \emph{y}-band in days calculated using the standard thin disk model and the remaining panels show the long lag timescale between the \emph{u}- and \emph{y}-band in days for disks with aspect ratios from left to right of 0.03, 0.1, and 0.3 as a function of Eddington ratio and SMBH mass in M$_{\rm \sun}$. The white point shows the approximate SMBH mass and Eddington ratio of Fairall 9 \citep{Peterson2004,F92020}.}
    \label{fig:tinf}
\end{figure*}

\begin{figure*}
    \centering
    \includegraphics[width=\textwidth]{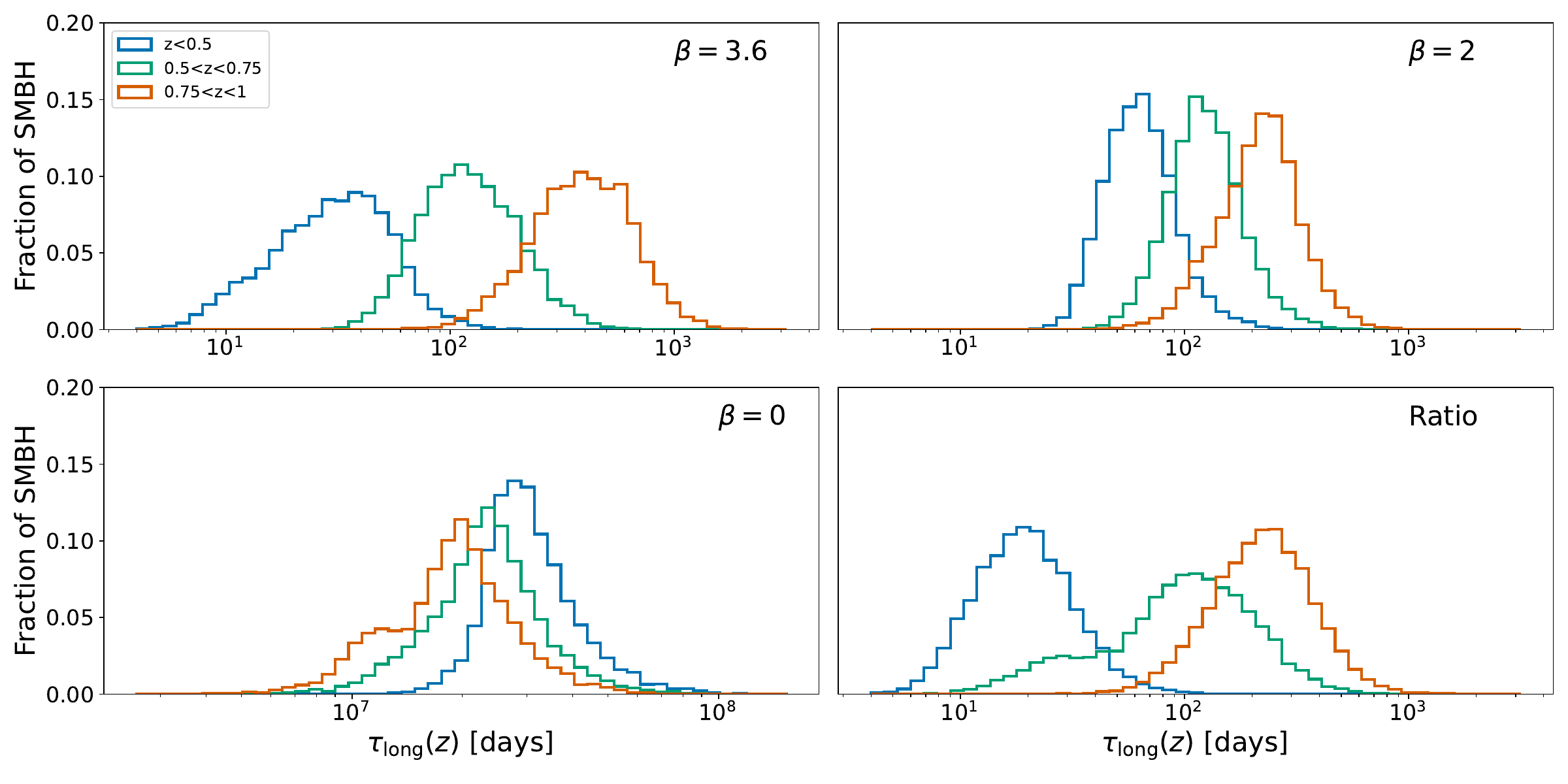}
    \caption{The normalized distribution of $\tau_{\rm long}(z)$ for quasars in SDSS DR7 \citep{Shen_2011} that fall within different redshift ranges. We assume $\alpha=0.1$, and take the masses, Eddington ratios, and redshifts from \cite{Shen_2011}. In the top left panel we use $\beta =3.6$ as in \cite{Yao:2022}; in the top right panel we use $\beta=2$, which is a more conservative estimate that the aspect ratio increases linearly with radius; in the bottom left panel we use $\beta=0$ and $h/r=0.01$, which are typical values in the standard thin disk model and lead to much long lag timescales (see different x-axis range)}; and in the bottom right panel we use the ratios between the long and short lag for different wavebands in \cite{Yao:2022}.
    \label{fig:t0z}
\end{figure*}

The long lag timescale recovered by \cite{Yao:2022} is much shorter (70~days versus 100~years) than the viscous timescale predicted by the standard \cite{ShakuraSunyaev1973} thin-disk model. However, this timescale is in good agreement with the recent simulation by \cite{JiangBlaes2020} that shows $h/r > 0.01$ in the radiation pressure dominated region. Disks with larger aspect ratios are also often invoked to explain line driving broad absorption line quasars \cite[e.g.][]{Murray1995,Leighly2004}. If this shorter timescale for the long lag is accurate, it should be possible to detect long lags with long baseline campaigns, like LSST, depending on the properties of the quasar and waveband coverage. We therefore use the results from \cite{Yao:2022} to derive a new empirical model for the long lag timescale.

As in \cite{Yao:2022}, we start by converting the viscous timescale to a velocity, $v_{\rm visc}=r/\tau_{\rm visc}$, and integrate from  $r_0$, the radius corresponding to the reference band, to $r_x$, the radius corresponding to band $x$,
\begin{equation}
\begin{split}
    \label{eq:dr/v}
    \tau_{\rm long} = \int_{r_0}^{r_x} \frac{dr}{v_{\rm visc}} = \frac{r_0^{2\beta}}{h_0^2 \alpha \sqrt{GM}} \int_{r_0}^{r_x} r^{5/2 -2\beta} dr \\ =\frac{r_0^{2\beta}}{(7/2-2\beta)h_0^2 \alpha \sqrt{GM}} \left[r_x^{7/2-2\beta}-r_0^{7/2-2\beta}\right].
    \end{split}
\end{equation}
Here we have written $h \equiv h_0(r/r_0)^{\beta}$ in order to capture the $r$-dependence of $h$. In the standard thin disk model, $\beta=0$, as $h$ is independent of radius.

Following \cite{Fausnaugh:2016} and \cite{Jiang:2017} the radius at which the disk radiates at a wavelength $\lambda$ is,
\begin{equation}
\label{eq:radius}
    r(\lambda) = \left(\frac{Xk_B \lambda}{\hat{h}c}\right)^{4/3} \left(\frac{f_iGM\Dot{M}}{8\pi\sigma}\right)^{1/3},
\end{equation}
where $k_B$ is Boltzmann's constant, $\hat{h}$ is Planck's constant, $c$ is the speed of light, $\sigma$ is the Stefan-Boltzmann constant, and $\Dot{M}$ is the accretion rate of the SMBH. $X=2.49$ is a factor to account for emission at this wavelength from a range of temperatures. $f_i$ is a factor to account for the irradiation from an X-ray/UV source near the SMBH by changing the normalization of the temperature profile. We set $f_i=1$, because three dimensional radiation magneto-hydrodynamic (MHD) simulations show that the UV/Optical region of the disk is very optically thick, making it difficult for X-ray photons to convert to UV photons, because  the free-free absorption opacity for X-rays is very small and the Compton process is not important \citep{Jiang2016,jiang2019,JiangBlaes2020}. As a result, the UV/Optical emission should be dominated by photons generated by the local disk ($f_i=1$). $f_i=1$ implies that X-ray light will not be directly reprocessed by absorption and re-emission in the disk. Instead in this model the X-ray short timescale variability is reprocessed by perturbing the local photosphere via scattering on the X-ray variability time scale.

While Equation (\ref{eq:radius}) is derived for the standard thin disk model, $r \propto \lambda^{4/3}$ does not rely on the assumption that $h/r \ll 1$, only the assumption that the emission is local. Since we care most about the wavelength dependence of the long lag timescale we use Equation (\ref{eq:radius}) here. However, the mass and accretion rate dependence of the long lag may need to be corrected when a more accurate model for the long lag can be derived either through simulations or additional observations. In particular, we note that $h/r$ and thereby $\beta$ may depend on the accretion rate and mass of the quasar. For simplicity, here we will define $h/r$ and $\beta$ independently of any quasar property.

Substituting Equation (\ref{eq:radius}) into Equation (\ref{eq:dr/v}) gives a viscous lag timescale of the form,
\begin{equation}
\label{eq:func_form}
    \tau_{\rm long} = \tau_0\left[\left(\frac{\lambda}{\lambda_0}\right)^{4/3(7/2-2\beta)}-1\right],
 \end{equation}
where,
\begin{equation}
\begin{split}
\label{eq:tau0w}
    \tau_0=93 \mathrm{~days} \left(\frac{-4}{7/2-2\beta}\right) \left(\frac{\lambda_0}{\lambda_{\rm W}}\right)^{4/3(7/2-2\beta)} \left(\frac{h_{\rm W}/r_{\rm W}}{0.2}\right)^{-2} \\
    \frac{0.1}{\alpha} \left(\frac{l/l_{\rm edd}}{0.1}\right)^{1/2} \left(\frac{M}{10^8 \rm{~M}_{\sun}}\right)^{1/2},
    \end{split}
\end{equation}
$\lambda_{\rm W}=1928$~\AA, and $h_{\rm W}/r_{\rm W}$ is the aspect ratio at the disk radius corresponding to the \emph{W2}-band. The parameterizations and bands are based on the recent fitting that \cite{Yao:2022} performed on the wavelength-dependent long lag they detected for Fairall 9 to this function for a reference band $\lambda_{\rm W}$. They found $\tau_0=66.8 \pm 2.3$ and $\beta = 3.59 \pm 1.22 $. For Fairall 9, $M = 2.55 \pm 0.56 \times 10^8$~M$_{\sun}$ \citep{Peterson2004} and the Eddington ratio $l/l_{\rm edd} = 0.02 \pm 0.004$ \citep{F92020}, which gives $\tau_0 = 66.8\mathrm{~days~}$ for $\alpha=0.1$ and $h_{\rm W}/r_{\rm W}=0.2$.

The wavebands for LSST range from the \emph{u}-band to the \emph{y}-band. For $\lambda_{\rm u}=3600$~\AA\ we get,
\begin{equation}
\begin{split}
\label{eq:taug}
    \tau_0=3.6 \mathrm{~days} \left(\frac{-4}{7/2-2\beta}\right) \left(\frac{h_{\rm W}/r_{\rm W}}{0.2}\right)^{-2} \\
    \frac{0.1}{\alpha} \left(\frac{l/l_{\rm edd}}{0.1}\right)^{1/2} \left(\frac{M}{10^8 \rm{~M}_{\sun}}\right)^{1/2},
    \end{split}
\end{equation}
for our fiducial values.

We show the lag between the \emph{u}- and \emph{y}-band at zero redshift for a range of quasar masses and Eddington ratios in Figure \ref{fig:tinf}. For comparison we show the short lag timescale for the standard thin disk model in the left panel using \cite{Fausnaugh2016},
\begin{equation}
    \label{eq:short_lag}
    \tau_{\rm c} = \frac{r(\lambda_0)}{c} \left[\left(\frac{\lambda}{\lambda_0}\right)^{4/3}-1\right],
\end{equation}
where $r(\lambda_0)$ is calculated from Equation (\ref{eq:radius}). The remaining three panels are for different values of $h_{\rm W}/r_{\rm W}$: $h_{\rm W}/r_{\rm W}=0.03$, which is similar to its proposed value in the standard thin disk model, and $h_{\rm W}/r_{\rm W}=0.1,0.3$, which are slightly smaller and larger, respectively, than the value found from fitting the lag found in \cite{Yao:2022}. Depending on $h_{\rm W}/r_{\rm W}$, at low redshift the long lag can range from a fraction of a day to dozens of years, for realistic quasar properties.

In addition to the mass and Eddington ratio of the quasar, the long lag also depends strongly on the redshift, $z$, of the quasar. First, time dilation due to increasing redshift should increase lag times by a factor of (1+$z$). Second, the rest frame waveband of a quasar will change by a factor of 1/(1+$z$), i.e. $\lambda_0$ will decrease by a factor of 1/(1+$z$). Since $\tau_0$ scales as $\lambda_0^{4/3(7/2-2\beta)}$ for $\beta \approx 3.6$ overall $\tau_0$ scales as,
\begin{equation}
    \label{eq:zscale}
    \tau_0 \sim (1+z) \left(\frac{1}{1+z}\right)^{4/3(7/2 - 2\beta)} \sim (1+z)^{5.9}.
\end{equation}

The top left panel of Figure \ref{fig:t0z} shows the normalized distribution of $\tau_{\rm long}$ from the \emph{u}- to \emph{y}-band for quasars in the Sloan Digital Sky Survey Data Release 7 \citep[SDSS DR7,][]{Shen_2011} for three different redshift ranges, using the $\beta\approx3.6$ found in \cite{Yao:2022}. $z=1$ is the highest redshift shown for two reasons. First, at $z>1$ the observed-frame \emph{u}-band corresponds to rest frame bands shorter than the shortest UV band in \cite{Yao:2022} and we do not want to extrapolate our models beyond the data available. Second, for $z \gtrsim 1$ the long lag is over three years long for a majority of quasars if $\beta$ is indeed as large as predicted by the model fit in \cite{Yao:2022}. Lags longer than three years would be difficult to detect even with the ten year baseline of LSST. 

In the top right panel of Figure \ref{fig:t0z}, instead of using the exact fit from \cite{Yao:2022} we assume $\beta=2$. For $\beta=2$ the lag timescale would still flatten at longer wavelengths, as in \cite{Yao:2022}, but the aspect ratio would only increase linearly as a function of radius. If the average value of $\beta$ is closer to this value, lags in quasars out to $z \lesssim 2$ should be measurable. 

We now note a few things about the value of $\beta$. First, due to the denominator of $\tau_0$ in Equation (\ref{eq:tau0w}), we restrict $\beta \neq 7/4$. Second, the dependence of the long lag on reference wavelength would be very different for $\beta<7/4$. The term in the brackets in Equation (\ref{eq:func_form}) would become a positive factor and $\tau_0$ would become negative. The increase in $h/r$ becomes less steep than the other radial dependencies of the long lag. Therefore the long lag would increase with reference wavelength as is the case with short lags (see Equation (\ref{eq:short_lag})). 

For example, for $\beta=0$, as in the standard thin disk model, $\tau_{\rm long} \propto \lambda_0^{14/3}$, which is a stronger dependence on $\lambda_0$ than the $\tau_{\rm short} \propto \lambda_0^{4/3}$ for the short lag. As a result, the observed long lag from the \emph{u}- to \emph{y}-band would actually decrease as a function of redshift, because emission from the inner disk where the lag timescale is shorter gets redshifted into longer wavebands. The bottom left panel of Figure \ref{fig:t0z} shows the distribution of $\tau_{\rm long}$ for three different redshift ranges for $\beta = 0$ and $h/r =0.01$, which are typical assumed values for the standard thin disk model. These distributions peak around $10^5 - 10^6$~years. 

As an alternate way of scaling from Fairall 9 to other targets, we can assume that the ratio between short and long lags is preserved between Fairall 9 and other targets. We use this alternative way of scaling to examine how removing our various model assumptions changes the long lag distributions for our different redshift ranges. To calculate these lags we take the redshift of each quasar in SDSS DR7 with $z<1$ and use that to determine the rest-frame wavebands that correspond to the observed frame \emph{u}- and \emph{y}-band for that quasar. We then take the ratio between the long and short lag detected in \cite{Yao:2022} in the rest-frame and multiply it by the short lag (calculated using Equation (\ref{eq:short_lag})) for that quasar. In practice we use the error-weighted mean of this ratio in the rest-frame waveband corresponding to each quasar and both adjacent wavebands. 

Calculated in this way the lags are shifted towards slightly shorter timescales, relative to our model with $\beta=3.6$. This shift is especially the case for $z<0.5$ quasars and $0.75<z<1$ quasars for which the distributions peak around 20 days instead of 40 days and 250 days instead of 400 days, respectively. This shift towards shorter lags would enable the detection of long lags for light curves with shorter baselines than LSST and make it possible to detect long lags out to $z>1$ with LSST. The main reason for the shift to shorter timescales is that this method removes the mass and luminosity dependence of the lag timescale, by scaling all lags to Fairall 9. In the bottom right panel of Figure \ref{fig:t0z} we show the long lag timescale calculated using this approach.

\begin{table}[]
    \centering
    \begin{tabular}{l|cccr}
        Redshift & $\tau_{\rm long}$ & $\tau_{\rm c}$ & \emph{i}-mag & $\tau_{\rm damp}$ \\
        \hline
        $z<$0.5 & 50 days & 7 days &  17.8 mag & 145 days \\
        $0.5<z<0.75$ & 130 days & 9 days &  18.6 mag & 170 days \\
        $0.75<z<1$ &  400 days & 12 days &  18.7 mag & 204 days \\
    \end{tabular}
    \caption{The three redshift ranges we use to bin SDSS DR7 quasars when generating our mock light curves and their corresponding median properties: long lag timescale ($\tau_{\rm long}$), short lag timescale ($\tau_{\rm c}$), \emph{i}-band magnitude, and random walk damping time ($\tau_{\rm damp}$).}
    \label{tab:bins}
\end{table}

\section{Generating Mock Light Curves}

\label{sec:mocks}

\begin{figure*}
    \centering
    \includegraphics[width=\textwidth]{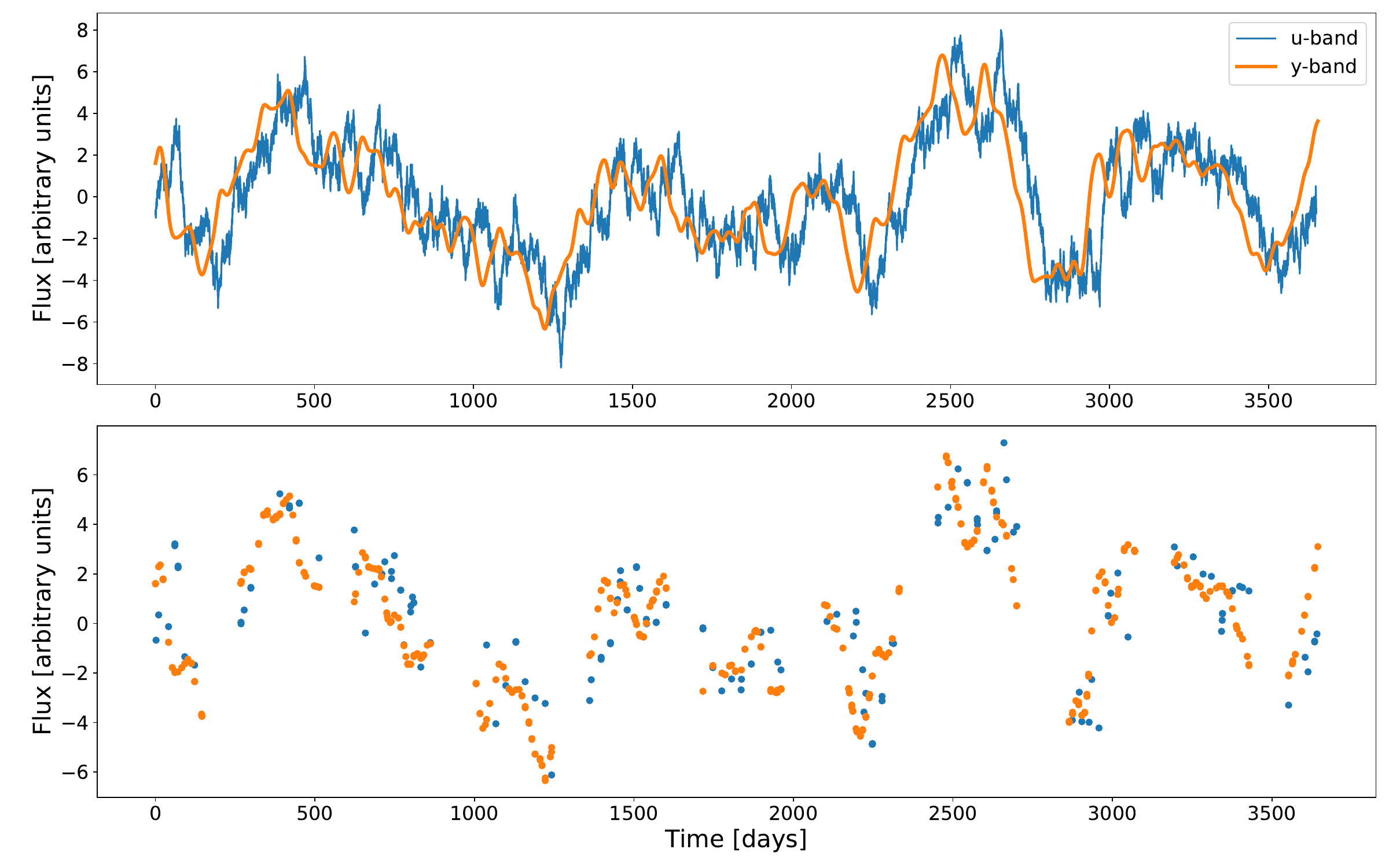}
    \includegraphics[width=0.32\textwidth]{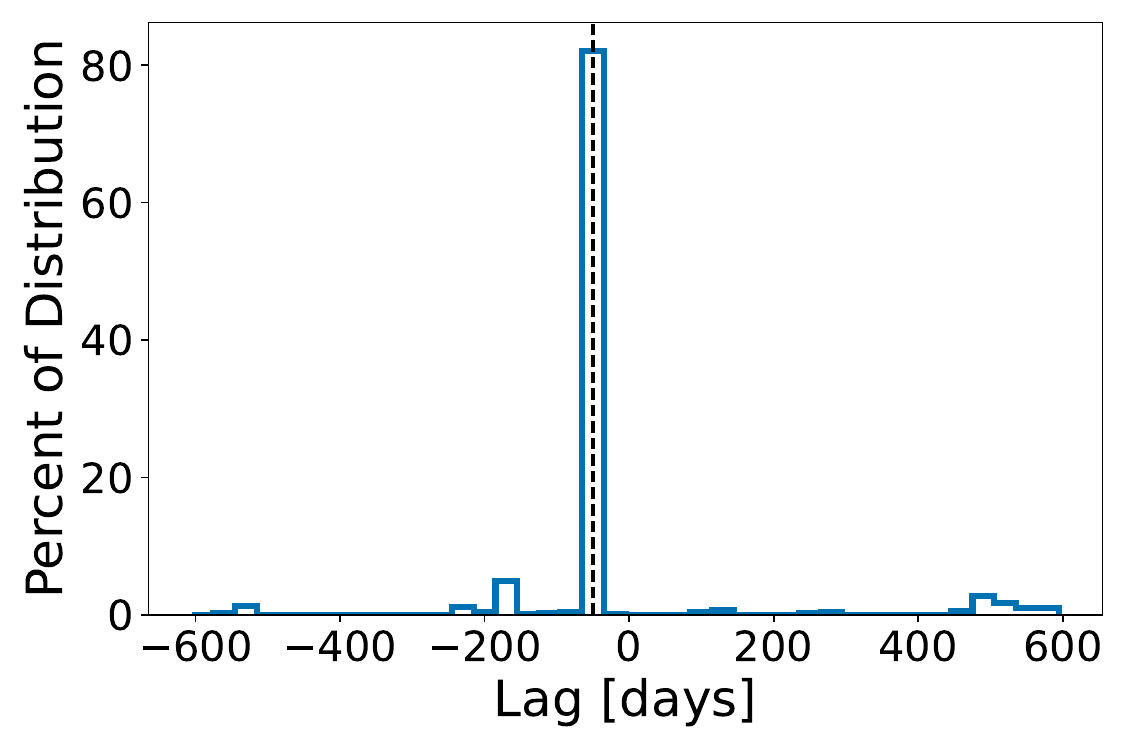}
    \includegraphics[width=0.65\textwidth]{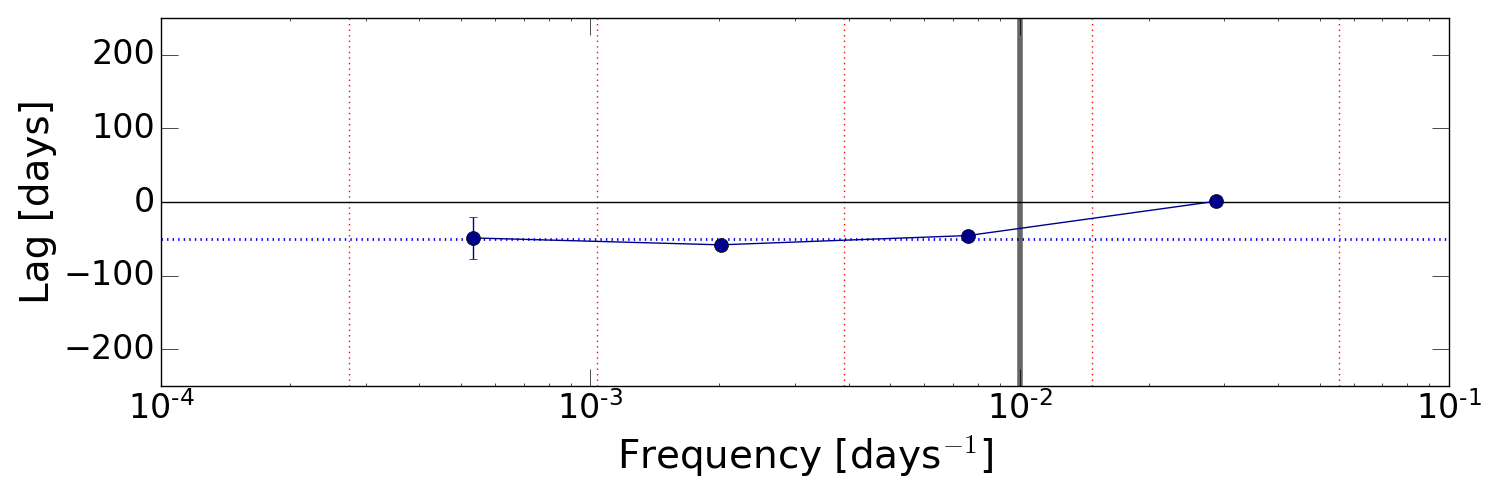}
    \caption{Top panel: Simulated driving \emph{u}-band light curve, in blue, and \emph{y}-band light curve, reprocessed with a long lag of $-50$ days, in orange. Middle panel: The same light curve as in the top panel sub-sampled using the long season LSST cadence. Bottom left: {\sc javelin} probability distribution for the long lag between the light curves above. The long lag is shown as a vertical dashed line. The median of the negative values of the distribution is $-50$~days.
    Bottom right: An example of the lags as a function of frequency recovered by the maximum-likelihood method for a mock light curve with an input long lag of $-50$~days (shown as the horizontal dashed blue line). The points are the lag detected in each bin (shown as the dashed vertical red lines). The vertical lines show the error bars. The lowest frequency bin has the largest error bar. The next two bins are roughly consistent with the input lag and have an error-weighted mean of $-48$~days. Because there is no short lag and it is at frequencies higher than the wrapping frequency (vertical grey line), the highest frequency bin is consistent with zero.}
    \label{fig:ex}
\end{figure*}

With a prescription for calculating lags in hand, we turn now to creating mock multi-band LSST-like light curves incorporating this negative lag. We outline our process for generating mock LSST light curves that have been reprocessed with lags. First, we either generate a driving light curve using a DRW model or use light curves from the radiation MHD simulation in \cite{JiangBlaes2020}. Then, we reprocess these light curves with a Gaussian response function to add lags. Finally, we sub-sample these light curves with proposed LSST cadences.

\subsection{Driving Light Curves}
\label{sec:driving}

\subsubsection{Damped Random Walk Light Curves}
\label{sec:drw}

For light curves analyzed in Sections \ref{sec:long}, \ref{sec:short_long}, and \ref{sec:lco} we start by generating a highly sampled ($\Delta t=0.001$~days) 2000 year-long light curve, using a damped random walk (DRW) model as in \cite{Kelly2009}. A DRW is a stochastic process that goes from a $\nu^{-2}$ power-law, where $\nu$ is frequency, at high-frequency to white noise at low-frequency at some characteristic damping timescale $\tau_{\rm damp}$. The DRW model has been shown to be a decent representation of quasar light curves for timescales of days to years \citep[e.g.][]{Kelly2009, Kozlowski2010, MacLeod2012, Zu2013}. The physical interpretation of this damping timescale is still debated, although \cite{Kelly2009} and \cite{Burkeetal2021} found that it may be related to the thermal timescale of the disk. \cite{Burkeetal2021} found that $\tau_{\rm damp}$ scales with the SMBH mass as,
\begin{equation}
    \label{eq:tdamp}
    \tau_{\rm damp}=107 \left(\frac{M}{10^8 \rm{M}_{\sun}}\right)^{0.38}.
\end{equation}

To determine the properties of our mock light curves we take quasars from SDSS DR7, and put them in three redshift bins: $z<0.5$, $0.5<z<0.75$, and $0.75<z<1$ (see Table \ref{tab:bins}). In this paper we only investigate our ability to detect lags in quasars at $z<1$, because, as can be seen in Figure \ref{fig:t0z}, a majority of quasars at $z \approx 1$ have long lags over 500 days long, which would be difficult to detect with the LSST baseline of 10 years. We use the median quasar mass in each bin to determine $\tau_{\rm damp}$. We assume the structure function at infinity, SF$_{\infty}=2.5$ in flux units, which is a typical value for quasar light curves from the Dark Energy Survey \citep{Yu:2020} and describes the long term variability of the quasar.

We split our 2000 year light curve into 100 individual light curves which we use as underlying driving light curves, $d(t)$. We generate and subsample a 2000 year light curve instead of generating 100, 20 year-long driving light curves because it is important when generating a DRW light curve that the baseline is significantly longer than the damping timescale \citep{Kozlowski:2017,Stone2022}. Done this way our baseline DRW light curve is over 3000 times longer than the largest damping timescale we use.

\subsubsection{Simulated Light Curves}
\label{sec:sim_lcs}

 While the DRW model does a good job describing quasar variability over a range of timescales, it is not based on a physical model of the disk. The DRW model may also not model the PSD properly at timescales less than 1 month \citep{Mushotzky2011,Caplar:2017, Smith2018,Sanchez-Saez:2018}, and be unreliable on the longest timescales no matter how long the modeled DRW is \citep{Stone2022}. Therefore, in Section \ref{sec:light_curves} we test the robustness of our results by using light curves from a simulation performed by \cite{JiangBlaes2020}.

\cite{JiangBlaes2020} used {\sc athena ++} to perform a global radiation MHD simulation of a quasar disk between $30$~$r_{\rm g}$ and $100$~$r_{\rm g}$, where $r_{\rm g}$ is gravitational radii. This region of the disk, which roughly corresponds to the region that emits in the optical, is dominated by the Rosseland mean opacity, which increases around $1.8 \times 10^5$~K due to the iron opacity bump. \cite{JiangBlaes2020} found that the iron opacity bump causes the disk to become convectively unstable leading to turbulence which puffs up and heats the disk. Eventually the puffing up of the disk decreases the opacity enough for the disk to become convectively stable and cool, until it becomes optically thick again. In \cite{JiangBlaes2020} this cycle repeats itself every few years and leads to variability in the luminosity of the disk by factors of around 3-6.

Figure 14 in \cite{JiangBlaes2020} shows the luminosity scaled by the Eddington luminosity emitted from $r\leq 60 r_{\rm g}$ and $r\leq 80 r_{\rm g}$ over roughly 57 years for their simulated $\num{5e8}$~M$_{\sun}$ quasar. Over the 57 years the Eddington ratio can vary from $l/l_{\rm edd} \approx 0.01 - 1$. We split both of these light curves into eight chunks of 20 years each, and linearly interpolate them down to sub-daily cadence from an original sampling of roughly every three days. This interpolation should not have a large impact on our results because the LSST cadence we down-sample the light curve with has a cadence of roughly every 10--30 days (see Section \ref{sec:cadence}).

\subsection{Reprocessed Light Curves}
\label{sec:reprocess}
Next we reprocess our driving light curves with a Gaussian response of the form,
\begin{equation}\label{eq:gaussian}
    \rm \psi(t) = \frac{1}{S\sqrt{2\pi}}exp^{\frac{-(t-C)^2}{2S^2}},
    \end{equation}   
where $C$ is the center of the Gaussian (i.e. the input lag) and $S$ is the width of the Gaussian, set to $S\equiv C/5$, for simplicity. For a negative lag ($C<0$) we replace $t$ with $-t$. We choose $S\equiv C/5$ such that there is some width to the response that scales with the lag duration. However, the exact shape of the response function is unknown, especially for the long lag. 

In cases where we are simulating both a long and a short lag we add two response functions together. We scale the relative strength of the two response functions by fixing the amplitude of the long lag response function to some fraction of the amplitude of the short lag response function. As a result, the integrated Gaussian for the long lag will no longer sum to one, which means the amplitude of the reprocessed light curve will differ from the amplitude of the driving light curve. 

Because we want to examine a broad range of long lag timescales, we use long lags of $-50$, $-130$, and $-400$~days. From Figure \ref{fig:t0z} we can see this range of timescales does a good job representing quasars from SDSS DR7 at $z\lesssim 1$. To match long lag timescales to other quasar light curve properties we estimate that each quasar bin $z<0.5$, $0.5<z<0.75$, and $0.75<z<1$ corresponds to a long lag of $-50$, $-130$, and $-400$ days, respectively. We also calculate the short lag corresponding to each bin using Equation (\ref{eq:short_lag}). The long and short lags for each redshift bin are in Table \ref{tab:bins}.

To get the reprocessed light curve we Fourier transform the driving light curve ($d(t) \rightarrow D(\nu)$) and response function ($\psi(t) \rightarrow \Psi(\nu)$). In a linear reprocessing model the reprocessed light curve is,
\begin{equation}\label{eq:reproccess}
        \rm r(t) = \int_{-\infty}^{\infty}\psi(\tau)d(t-\tau)d\tau.
\end{equation}     
which in Fourier space is equivalent to, 
\begin{equation}\label{eq:freprocess}
        \rm R(\nu) = \Psi(\nu)D(\nu).
\end{equation}  
We can then inverse Fourier transform $R(\nu) \rightarrow r(t)$ to acquire our reprocessed light curve.

\subsection{LSST Cadence}
\label{sec:cadence}

Now that we have a driving and reprocessed light curve, we down-sample their $\Delta t =0.001$ cadence to a variety of LSST cadence possibilities. The LSST cadence is not final. Over the past decade the 
LSST community has conducted a thorough investigation of prospective cadences \citep{Connolly:2014}. Different proposed cadences are available through the LSST Operations Simulator \citep[OpSim,][]{Reuter:2016}\footnote{Code available at: \url{https://github.com/lsst/sims_featureScheduler}}. OpSim simulates the field selection and image acquisition for LSST observations over its proposed 10 year baseline. It takes into account sky brightness, bad weather predicted from localized weather models, seeing, telescope maintenance times, slew times, and filter changes. 

The LSST observing strategy includes two main components, the Wide Fast Deep Survey and the Deep Drilling Fields (DDFs). The Wide Fast Deep Survey will be at least 18000~deg$^2$ where each 9.6 deg$^2$ field has a median of 825 visits summed over all six filters. The DDFs are five relatively small (9.6 deg$^2$) fields which will have deeper coverage and a higher cadence than the Wide Fast Deep Survey. In this paper, we focus on objects observed in the DDFs. Specifically, here we use the proposed cadences for the XMM-LSS DDF. Other DDFs have similar cadences.

In this work, we use the baseline v2.0 DDF cadence, and the DDF cadence with the longest available season length (ddf season length slf0.10 v2), which uses 80\% of the available season, or a total of roughly 210~days. We primary use the \emph{u}- and \emph{y}-band cadences, to take advantage of the largest wavelength range available for LSST. Over ten years the baseline v2.0 cadence has 1156 and 4515 samples for the \emph{u}- and \emph{y}-bands, respectively, while the long season cadence has 842 and 4434 samples for the \emph{u}- and \emph{y}-bands, respectively. We down-sample this number to eliminate visits that occur within 1/100 of a day of each other and end up with 186 (150) and 443 (443) samples for the baseline (long season) cadence, for the \emph{u}- and \emph{y}-bands, respectively. This down-sampling still leaves a large fraction (15--35\% depending on the band and cadence) of visits that occur within roughly 15 minutes of each other. For the baseline cadence, most of the remaining samples are separated by roughly 1 to 3 days. For the long season cadence the separation is typically around 10 to 30 days, with more visit separations on the longer side for the \emph{u}-band. 

We find a longer season length, and as a result shorter season gaps, to be beneficial for detecting long lags in our mock light curves, even though the longer season has lower cadence during the season (see Figure \ref{fig:jav_exB} and discussion in Sections \ref{sec:detect:jav} and \ref{sec:lco}). Therefore, in Sections \ref{sec:long}, \ref{sec:short_long}, \ref{sec:light_curves}, and \ref{sec:i21} we use the long season cadence. In Section \ref{sec:lco}, we discuss the relative performance of our lag detection methods on light curves with the baseline cadence and ways to improve our results if the baseline cadence is selected. 

After down-sampling our driving light curve, we add band-dependent dust extinction and measurement error by calculating the signal to noise ratio (SNR) for each observation using the sky brightness and instrument noise from OpSim. In this calculation we use the average magnitudes in each SDSS quasar bin for quasars with $i<20$~mag (see Table \ref{tab:bins}), to first test our ability to detect the long lag in the most luminous quasars. The median SNR for our mock quasars over all bins is $\sim 150$. We show that we should be able to detect long lags for fainter quasars as well in Section \ref{sec:i21}.

The top panel of Figure \ref{fig:ex} shows an example DRW light curve. The driving light curve is shown in blue, and the same light curve reprocessed with a long lag of $-50$~days is shown in orange. The middle panel shows the same light curves sub-sampled with the long season LSST cadence for the \emph{u}-band and \emph{y}-band, for the driving and reprocessed light curves, respectively.

\section{Lag Detection Methods}
\label{sec:detection}

In this section, we outline several methods developed for detecting short lags and BLR lags that we will apply to the detection of long lags. First, we briefly discuss different cross-correlation methods \citep{Gaskell1987,WhitePeterson1994,Peterson2004}, and the Von-Neumann estimator presented by \cite{Chelouche2017}. Next, we give a more thorough description of the two methods we explore in detail in this paper, {\sc javelin} \citep{Zu2011} and a Fourier method \citep{Zoghbi2013}. We test a variety of methods on our mock light curves because each one has different strengths and drawbacks. These drawbacks mostly arise because these methods need to work for unevenly sampled UV/optical light curves, like those LSST will produce, which will not have an even cadence because of day-time gaps, bad weather, seasonal gaps, shared telescope time, etc.

\subsection{Cross-Correlation Methods and the Von-Neumann Estimator}
\label{sec:alt_methods}

\begin{figure}
    \centering
    \includegraphics[width=0.48\textwidth]{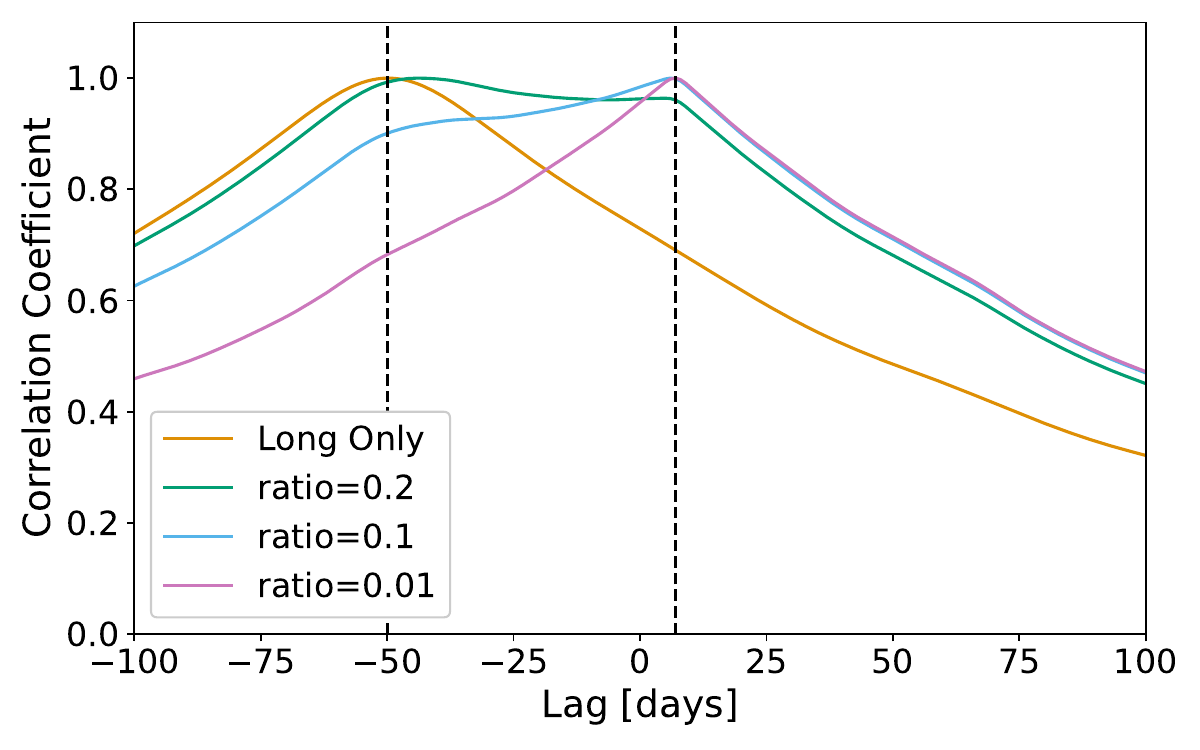}
    \includegraphics[width=0.48\textwidth]{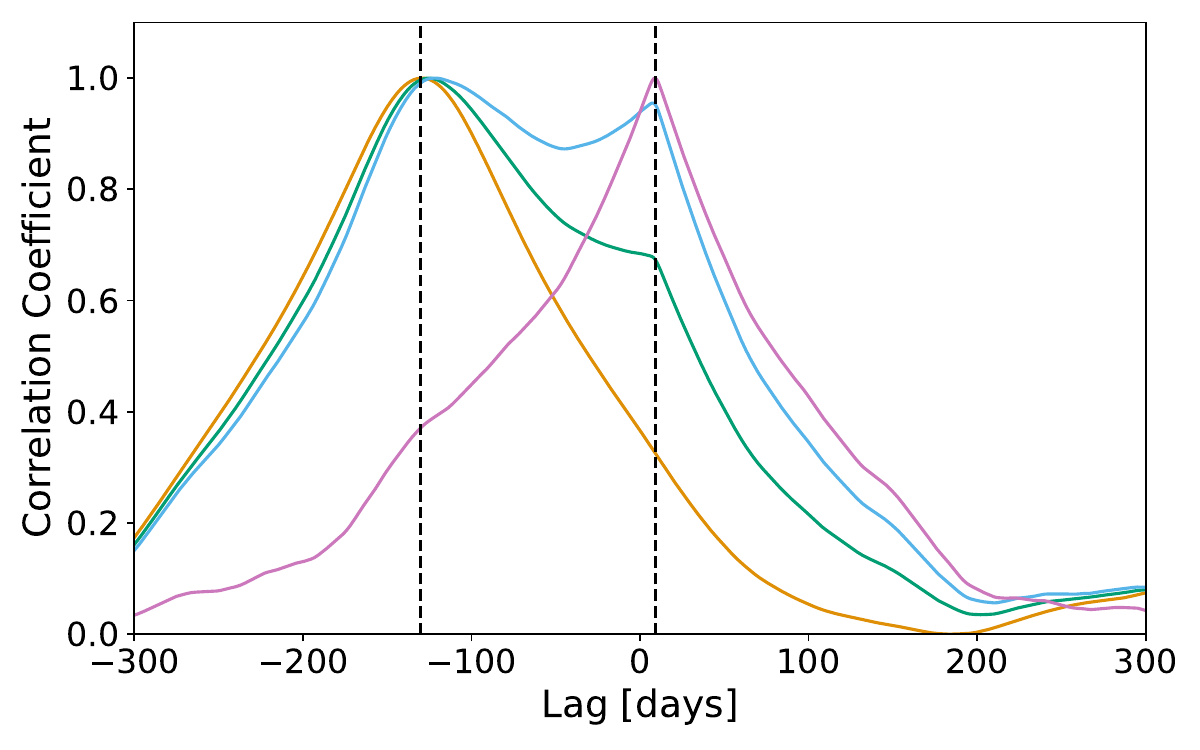}
    \includegraphics[width=0.48\textwidth]{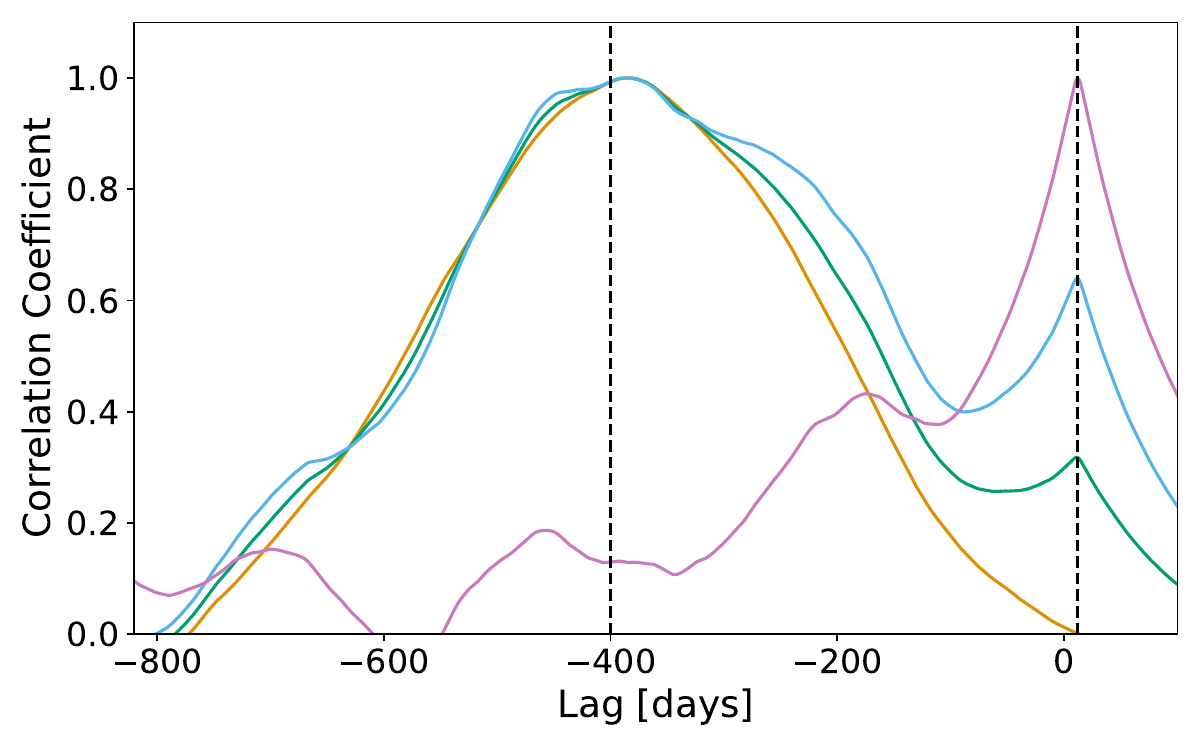}
    \caption{The cross-correlation coefficients of even daily cadence light curves that have been reprocessed with long (short) lags of $-50$ (7), $-130$ (9), and $-400$ (12) days, in the top, center, and bottom panel, respectively. Distributions shown in orange have been reprocessed with only a long lag. The ratios of the amplitude of the long to the amplitude of the short lag for the light curves with distributions shown in green, light blue, and pink, are 0.2, 0.1, and 0.01, respectively. The dashed vertical lines show the input lags.}
    \label{fig:jav_nocad}
\end{figure}

\begin{figure*}
    \centering
    \includegraphics[width=0.32\textwidth]{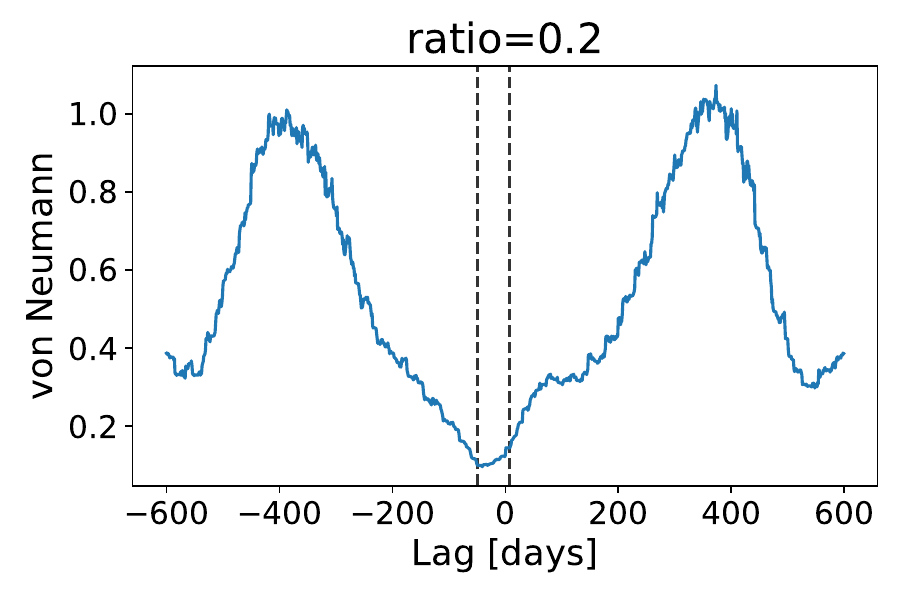}
    \includegraphics[width=0.32\textwidth]{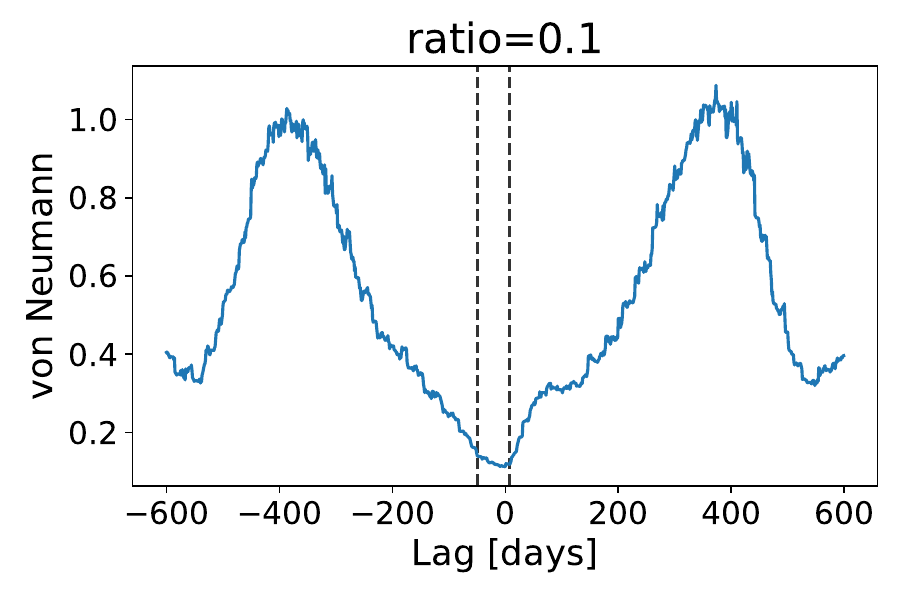}
    \includegraphics[width=0.32\textwidth]{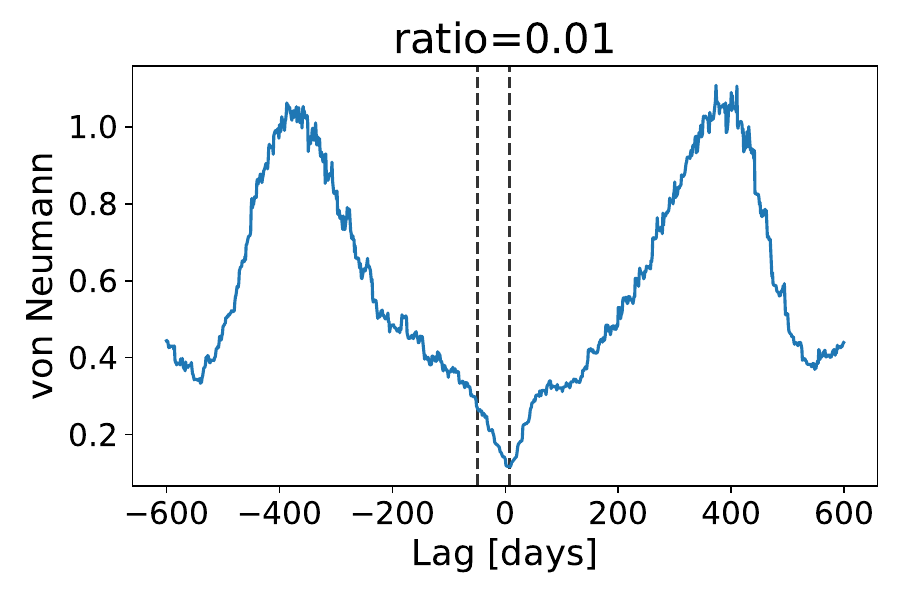} \\
    \includegraphics[width=0.32\textwidth]{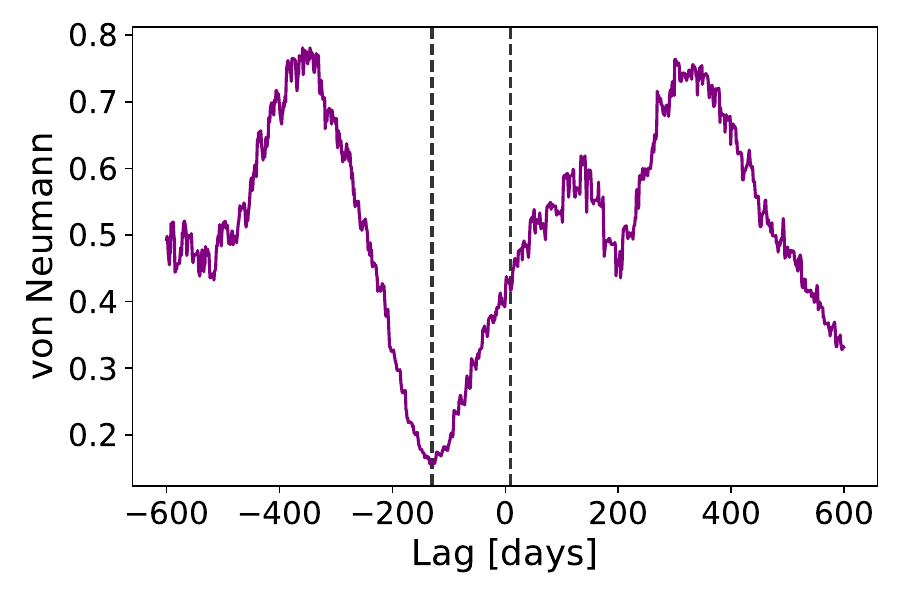}
    \includegraphics[width=0.32\textwidth]{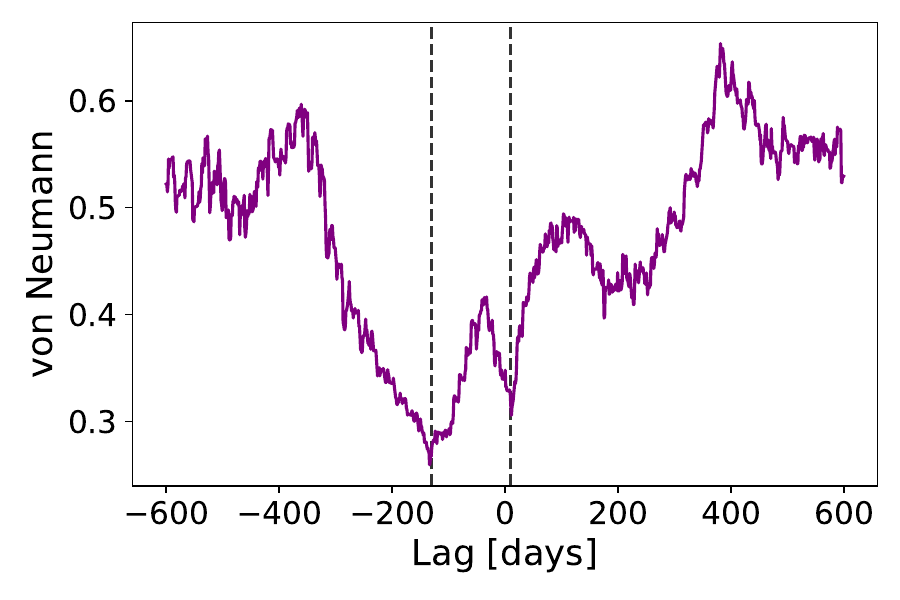}
    \includegraphics[width=0.32\textwidth]{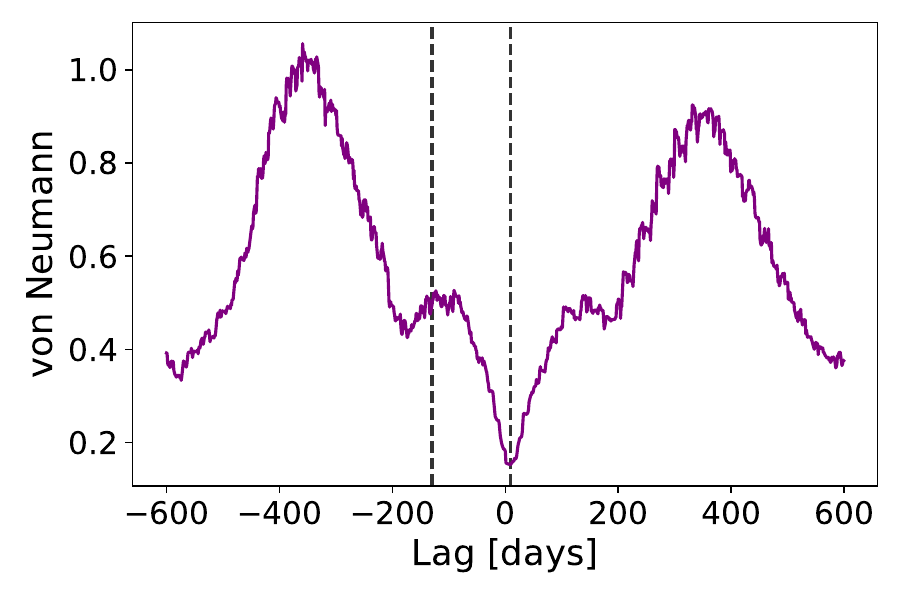} \\
    \includegraphics[width=0.32\textwidth]{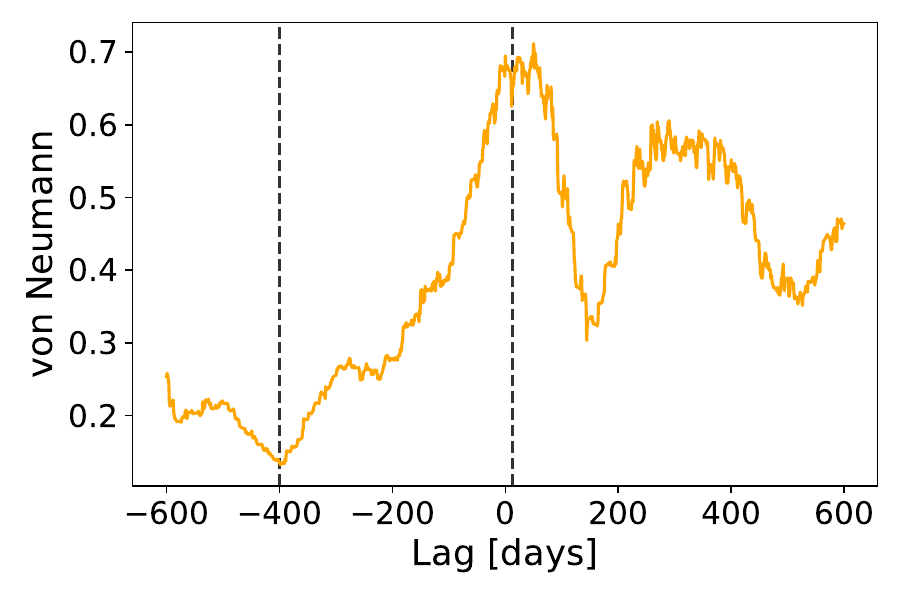}
    \includegraphics[width=0.32\textwidth]{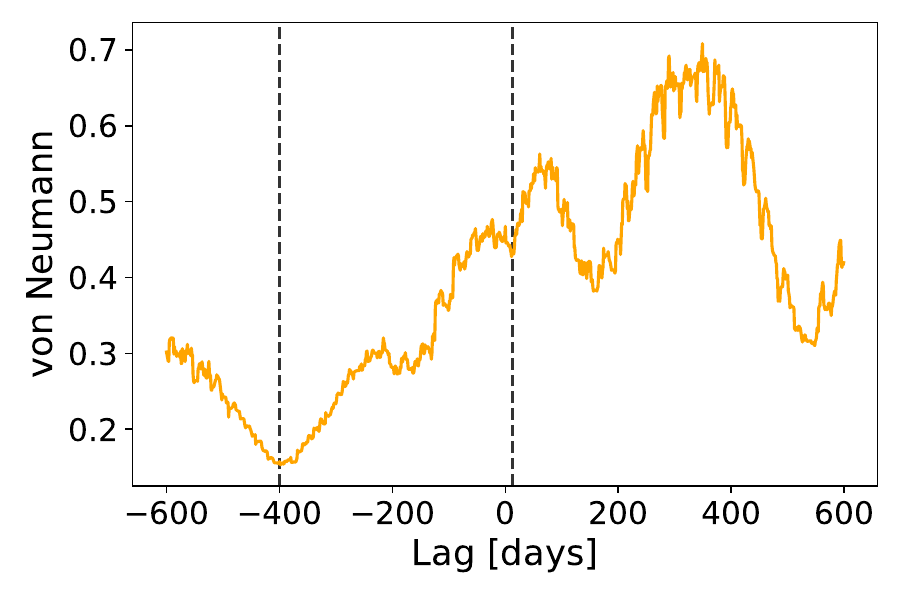}
    \includegraphics[width=0.32\textwidth]{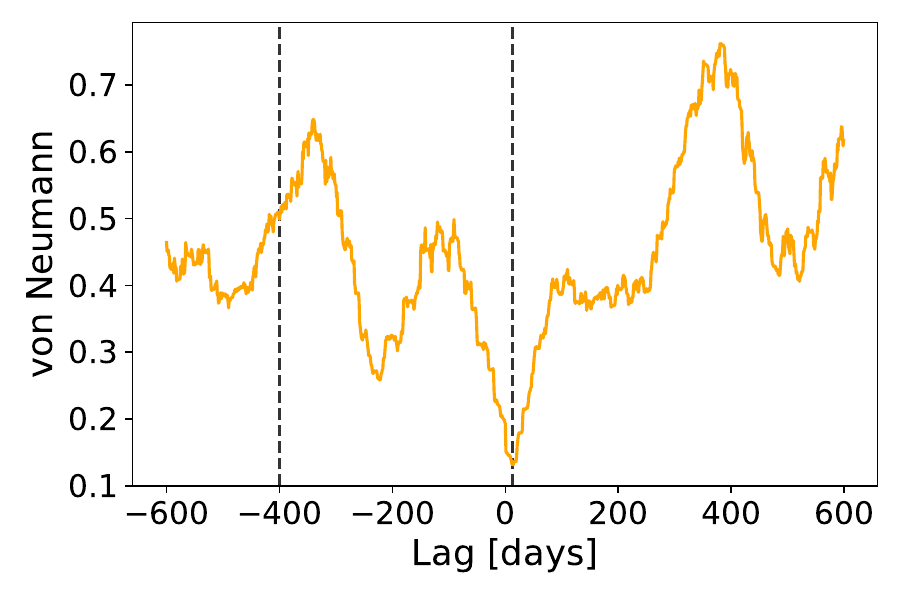} \\
    \caption{The mean of the Von-Neumann value or the mean-square successive difference for 1000 MCMCs as a function of lag for three mock long season LSST light curves with input long lags of $-50$, $-130$, and $-400$~days and short lags of 7, 9, and 12~days in the top, middle and bottom panel, respectively. The ratios of the amplitude of the long lag to the amplitude of the short lag are 0.2 (left panels), 0.1 (middle panels), and 0.01 (right panels). The dashed vertical lines show the input lags.}
    \label{fig:VNexs}
\end{figure*}

With uniformly-sampled light curves, time lags can be estimated by determining the delay time, $\tau$, with the largest cross-correlation coefficient,
\begin{equation}\label{eq:coeff}
    CC_{\rm coeff}(\tau) = \frac{\sum d(t)r(t+\tau)}{\sqrt{\sum d(t)^2 r(t+\tau)^2}}.
\end{equation}
Multiple lags can in principle be detected by finding multiple peaks in the correlation coefficients. 

We show the cross-correlation coefficients as a function of lag for mock DRW light curves with an evenly-sampled, daily cadence and ten-year baseline in Figure \ref{fig:jav_nocad}. The light curves have been reprocessed with long (short) lags of $-50$ (7), $-130$ (9), and $-400$ (12) days, in the top, center, and bottom panel, respectively. As the ratio between the amplitude of the long lag and the amplitude of the short lag decreases the peak around the long lag decreases. When the amplitude ratio is 0.01, the peak around the long lag nearly entirely disappears for input long lags of $-50$ and $-400$~days, although there remains a small peak around the long lag for the light curve with a $-130$~day input long lag.

On the other hand, as the amplitude ratio decreases, the peak around the short lag becomes more pronounced. For an input long lag of $-50$~days it becomes the largest peak when the amplitude ratio is 0.1. For the other two input long lags, the peak around the short lag becomes dominant when the amplitude ratio is 0.01. We show the positive and negative lag values with the highest correlation coefficients for these evenly-sampled light curves in the top four rows of Table \ref{tab:summ}.

To use this cross-correlation technique on unevenly sampled data, such as LSST light curves, we use the interpolated cross-correlation method \citep[ICCF,][]{Gaskell1987}, which performs a linear interpolation to add values to the gaps in the light curves to make them evenly sampled. We use the ICCF on the mock light curves described in Section \ref{sec:mocks} and find that the results are very consistent with what we find when we use cross-correlation on evenly sampled light curves. That is, the uneven cadence does not prevent us from being able to find an accurate lag, but as the amplitude ratio between the long and short lag decreases we are less (more) able to recover the input long (short) lag. We do not try other cross-correlation methods on our light curves, such as the z-transformed discrete correlation function \citep[ZDCF,][]{Alexander:2013}, but we expect these methods to perform similarly to the ICCF.

We do however try the Von-Neumann estimator presented by \cite{Chelouche2017}. This Von-Neumann method shifts two light curves by some time $\tau$ and measures the mean-square successive-difference, which is a measure of the randomness, between the two light curves as a function of $\tau$. The lag is therefore the $\tau$ with the smallest measure of randomness. This method is unique because it involves no model fitting, interpolations, or binning, and is therefore ideal for unevenly sampled light curves.

Figure \ref{fig:VNexs} shows the mean of the mean-square successive differences for 1000 Monte Carlo Markov Chains (MCMCs) for an example mock light curve reprocessed with a long (short) lag of $-50$ (7) days in the top panel, $-130$ (9) days in the center panel, and $-400$ (12) days in the bottom panel. The ratio of the amplitude of the long lag to the amplitude of the short lag is 0.2, 0.1, and 0.01, in the left, center, and right panels, respectively. Because the Von-Neumann method assumes a lag that is constant with frequency it has trouble identifying two lags simultaneously. For the shortest duration, $-50$~day, lag this assumption means the Von-Neumann method combines the two lags into a single wider trough when the amplitude ratio is 0.2. That trough gets skewed towards the short lag when the amplitude ratio is 0.1. For the longer duration lags the assumption of only one lag results in the Von-Neumann method finding only one of either the long or the short lag for most amplitude ratios. Only for the light curve with an input long lag of $-130$~days and an amplitude ratio of 0.1 are both lags detected at all, with significant dips in the distribution near each lag.

As with the cross-correlation method, the single-lag assumption is a drawback of using the Von-Neumann method to detect a multi-lag signal, unless one signal can be filtered out in some way. However, for an amplitude ratio of 0.2, we find that the Von-Neumann method could be a helpful tool to detect long lags in LSST light curves with long seasonal gaps. We discuss this further in Section \ref{sec:lco}.

\subsection{Javelin}
\label{sec:detect:jav}

\begin{figure}
    \centering
    \includegraphics[width=0.45\textwidth]{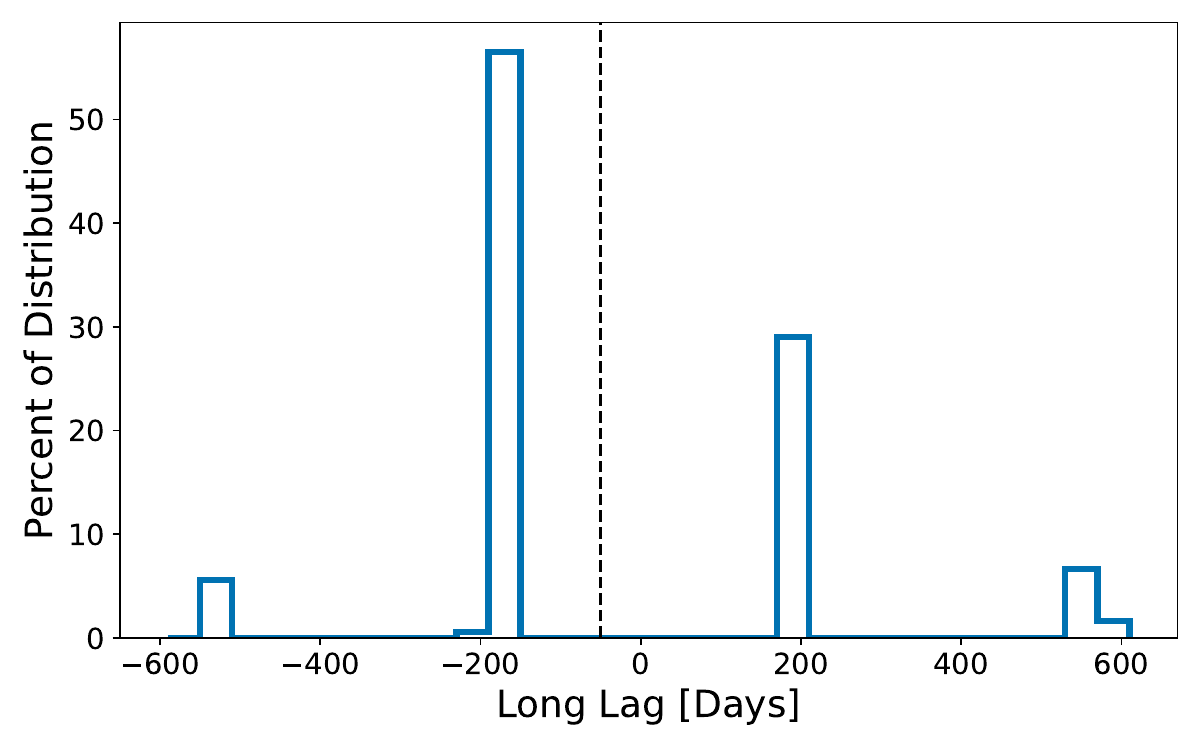}
    \caption{An example {\sc javelin} distribution for a mock light curve with the baseline cadence. While there is a very small peak around the input lag (shown as a dashed line) of $-50$~days, the peaks around $\pm 180$~days are larger.}
    \label{fig:jav_exB}
\end{figure}

Our first main lag detection method is {\sc javelin} \footnote{As presented in \citet{JAVELIN}, code available at: \url{https://github.com/nye17/javelin}}, previously known as {\sc spear}, which is a stochastic process estimator used to look for time lags in spectroscopic \citep{Zu2011} or photometric \citep{Zu2016} light curves. \cite{Li:2019} performed a comprehensive comparison of the performance of {\sc javelin} versus cross-correlation methods over a broad range of light-curve properties for mock SDSS light curves and found that {\sc javelin} performs best at accurately detecting BLR lags. As our lags are of a similar timescale we use {\sc javelin} as one of two main lag detection methods here.

{\sc javelin} works on unevenly sampled light curves by fitting a model to the available data and running an MCMC to find the posterior distributions for the DRW parameters ($\tau_{\rm damp}$ and SF$_{\rm \infty}$) of each light curve. {\sc javelin} then uses these posteriors to statistically interpolate each light curve as it shifts them temporally and uses an MCMC to simultaneously derive the posterior distributions for the DRW parameters and the response function between different light curves. For its default settings, which we use here, {\sc javelin} models the response function as a top hat with a finite width centered at time lag $\tau$.

The DRW statistical interpolation works well for shorter gaps in unevenly sampled light curves but is unable to fit the large seasonal gaps in our data, which becomes a problem for the LSST baseline cadence. Figure \ref{fig:jav_exB} shows a {\sc javelin} distribution for an example mock baseline cadence light curve. The largest peaks are around $\pm 180$~days, which is roughly the length of the gaps in between each observing season. We discuss solutions to this problem in case the baseline cadence is chosen in Section \ref{sec:lco}.

The bottom left panel of Figure \ref{fig:ex} shows the {\sc javelin} distribution for the example light curves in the panels above, which have been sub-sampled with the long season cadence and reprocessed with an input long lag of $-50$~days. The probability distribution peaks very sharply around $-50$~days and the median of the negative values of the distribution is $-50$~days. 

Because the long lag is negative and the short lag is positive, we take the long (short) lag as the median of the negative (positive) values of the {\sc javelin} distribution. This method could bias our values for the short lag towards slightly larger values, because the short lags are near zero and so the peak related to the short lag could extend below zero. However, we find that peaks in {\sc javelin} distributions around the input short lags rarely extend past zero.

While {\sc javelin} is a useful method for detecting lags in quasar light curves it does make assumptions about the light curve that are model dependent. Also, because the second light curve is modeled as a smoothed and shifted version of the first, the lag can depend on which light curve you model first. In addition, because {\sc javelin} uses a DRW model, which is what we use to generate our mock light curves, we could be biasing our comparison to favor {\sc javelin}. We address this potential bias in Section \ref{sec:light_curves}, where we test our methods on light curves from radiation MHD simulations instead of DRW mock light curves.

\subsection{Maximum-Likelihood Fourier Method}
\label{sec:detect:mlm}

\begin{figure}
    \centering
    \includegraphics[width=0.48\textwidth]{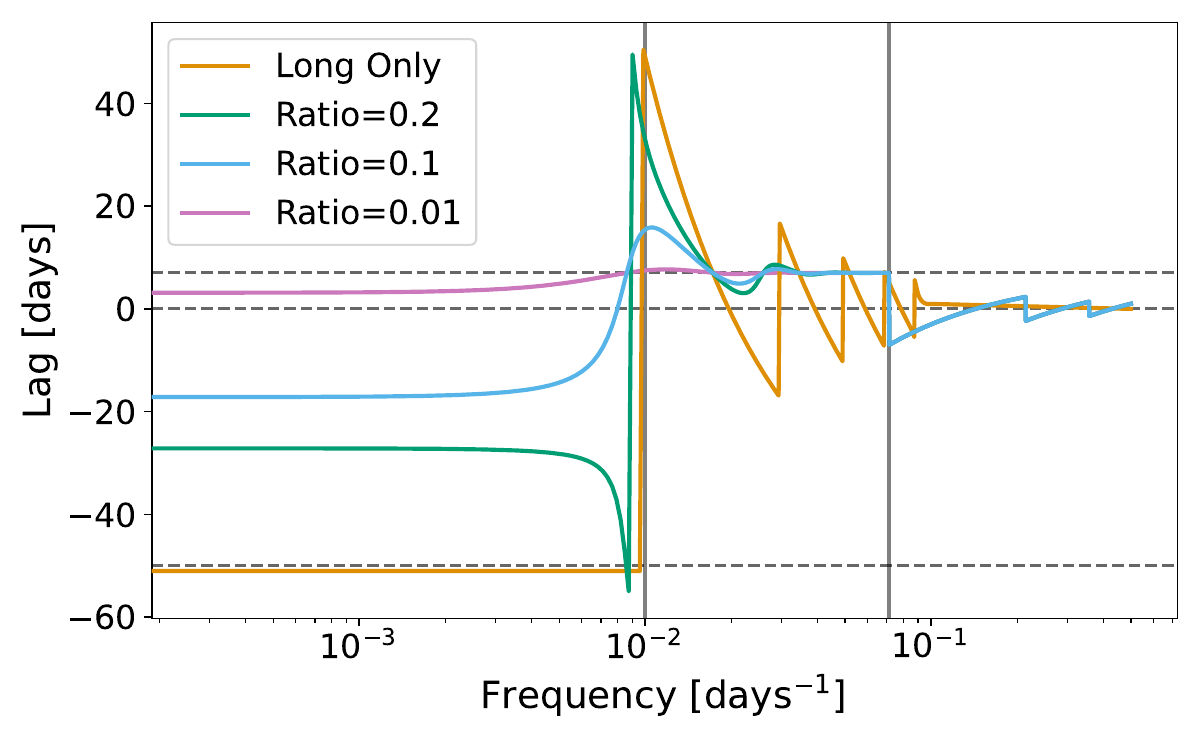}
    \includegraphics[width=0.48\textwidth]{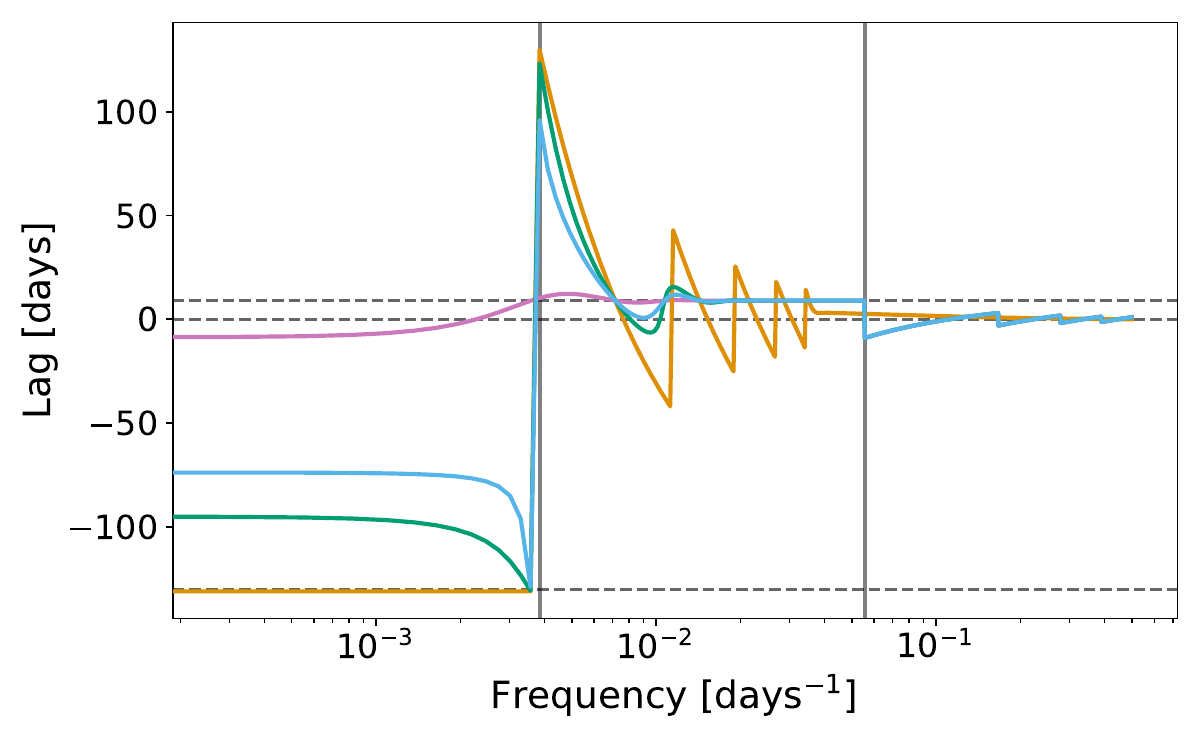}
    \caption{The time lags as a function of frequency between two evenly sampled, daily cadence light curves found using the phase of their complex cross-spectrum and Equation (\ref{eq:mlm_tau}). The input long (short) lags are $-50$ (7) and $-130$ (9) days in the top and bottom panel, respectively. Distributions shown in orange have been reprocessed with only a long lag. The ratio between the amplitudes of the long and short lag for the light curves with distributions shown in green, light blue, and pink, are 0.2, 0.1, and 0.01, respectively. The dashed horizontal lines show the input lags and zero and the vertical grey lines show the wrapping frequencies.}
    \label{fig:mlm_nocad}
\end{figure}

Our second main lag detection method is a Fourier method. Fourier methods are useful because they allow us to separate lags as a function of frequency. They are straightforward as well because calculating the cross-spectrum is a simple multiplication. We can use Equation (\ref{eq:freprocess}) and the power-spectral densities (PSD) of the reprocessed and driving light curve, $\rm |D(\nu)|^2 = D^*(\nu)D(\nu)$ and $\rm |R(\nu)|^2 = R^*(\nu)R(\nu)$, respectively, to derive the cross-spectrum of the two light curves to be,
\begin{equation}
    \label{eq:cross_spec}
    \rm C(\nu) = D^*(\nu)R(\nu) = |D(\nu)|^2\Psi(\nu). 
\end{equation}   
The phase lag, $\phi$, is then just the phase of the complex cross-spectrum, which is related to the time lag as \citep{uttley2014},
\begin{equation}
    \label{eq:mlm_tau}
    \tau = \phi/2\pi \nu.
\end{equation}

We show the lags as a function of frequency recovered by this Fourier method between two 10~year baseline, evenly-sampled, daily cadence light curves in Figure \ref{fig:mlm_nocad}. The input long (short) lags are $-50$ (7) and $-130$ (9) days in the top and bottom panel, respectively. We do not use this method to detect long lags of $-400$~days because the LSST baseline is not long enough to recover enough data points at frequencies $<10^{-3}$~days$^{-1}$. The vertical grey lines show the phase wrapping frequencies, which are equivalent to $\nu_{\rm w}=1/2C$ for a symmetric response function \citep{uttley2014}. Phase wrapping, the sudden flip in the sign of the lag, occurs because phase lags are limited to between $\pm \pi$, so a positive or negative shift of half a wavelength cannot be distinguished.

For light curves with only an input long lag, at frequencies lower than the first phase wrapping the measured lag is $-51$ and $-131$ days for input long lags of $-50$ and $-130$~days, respectively. At frequencies above the long lag's phase wrapping, the lag values oscillate around and then approach zero. When a short lag is added and the ratio between the amplitude of the long lag and the amplitude of the short lag is 0.2, the lags measured at low frequencies become roughly $-27$ and $-95$ days for input long lags of $-50$ and $-130$~days, respectively. For an amplitude ratio of 0.1 the duration of these lags decrease again to $-17$ and $-74$~days for input long lags of $-50$ and $-130$~days, respectively. For an amplitude ratio of 0.01 the lag measured at the lowest frequencies becomes positive for an input long lag of $-50$~days and is only $-8$~days for an input long lag of $-130$~days. 

The decrease in the duration of the recovered long lags as the amplitude ratio decreases is due to the growing influence of the short lag on lower frequencies. While the short lag has the strongest impact at frequencies around the inverse of the short lag timescale, it is also applied to all frequencies below that. Therefore it reduces the value of the long lag.

On the other hand, because the long lag only acts at lower frequencies, the short lag is easier to isolate in Figure \ref{fig:mlm_nocad}. For light curves with two input lags, at frequencies higher than the long lag phase wrapping frequency the lag values oscillate before quickly converging on the short lag value for all amplitude ratios. The smaller the amplitude ratio the sooner they converge. The lag values stay at the input short lag, until the short lag wrapping frequency is reached, when the lag values again oscillate and then in this case approach zero. We summarize the long and short lag values in Figure \ref{fig:mlm_nocad} in the top four rows of Table \ref{tab:summ}.

This Fourier method is very useful for recovering the shape of a response function for evenly sampled X-ray data \citep[e.g.][]{Papadakis2001, McHardy2007, Arevalo2008, Fabian2009, Zoghbi2010, DeMarco2013, Cackett2013, uttley2014}, but requires a work-around for unevenly sampled data, which can not be fast Fourier transformed. Here we use a maximum-likelihood method described in \cite{Zoghbi2013} and first presented by \cite{Miller2010}. 

\begin{figure}
    \centering
    \includegraphics[width=\columnwidth]{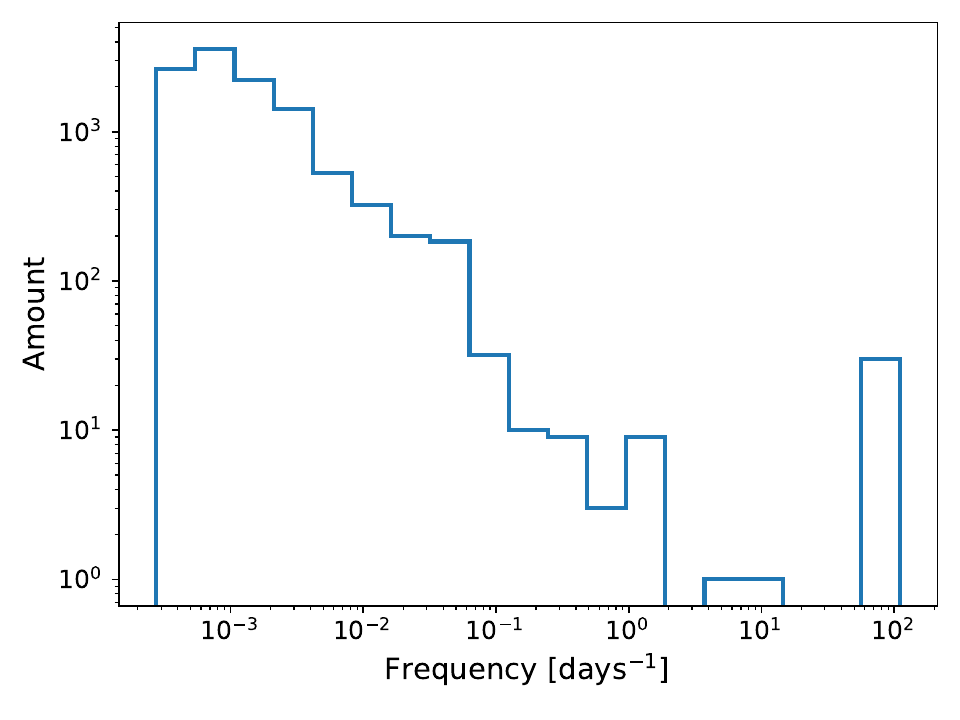}
    \caption{The distribution of the inverse of the time between each observation in the LSST long season cadence.}
    \label{fig:freqs}
\end{figure}

First, we create $N=4$ bins in frequency space ranging from $f_{\rm min} = 1/T$, where $T=10$~years is the length of the light curve, to $f_{\rm max}=0.055$~day$^{-1}$. As in \cite{Zoghbi2013} we also add a bin on either end to be discarded later to reduce bias in the minimum and maximum bins. $f_{\rm max}$ is chosen based on the maximum frequency for which there is significant power for an LSST cadence. Figure \ref{fig:freqs} shows the distribution of the inverse of the time between each observation for the LSST long season cadence. Because the length of the light curves is 10 years, frequencies less than $10^{-2}$~days$^{-1}$ are common, but because the cadence is roughly every ten days, frequencies greater than a few times $10^{-2}$~days$^{-1}$ are rarer.

It is easier to obtain accurate results with fewer bins. Previous testing shows that errors for each bin are smaller when the width of bins are at least a few times $1/T$. This requirement makes it difficult to detect $-400$~day lags, and so we do not attempt to detect that lag with this method. Otherwise, four bins are sufficient for our purpose here: detecting lags from two Gaussian response functions representing the long and short lag, which occur at order of magnitude different frequencies. However, additional bins would be useful for detecting the shape of more complicated response functions. 

Using these bins we fit a piece-wise PSD to each light curve and calculate the likelihood using the autocovariance function,
\begin{equation}\label{eq:autoc}
    \rm  \mathcal{A}(\tau) = \int_{-\infty}^{\infty}|D(\nu)|^2cos(2\pi\nu\tau) d\nu,
\end{equation}   
where $|D(\nu)|^2 = \sum_i D_i$ is the PSD as a piecewise sum over a range of frequency bins. We then maximize this likelihood to find the PSD for each light curve. We use these PSD as input parameters for an MCMC, which gives us a distribution for the power in each bin. We take the median of the distribution for each bin as the power for that bin and combine them to form a piecewise PSD over the full range of frequencies.

Next, we maximize the log-likelihood of the cross-covariance,
\begin{equation}
\begin{split}
                \rm \chi(\tau) &= \int_{-\infty}^{\infty}  \textit{C}(\nu)cos(2\pi \nu \tau+\phi(\tau)) df \\
                            &= \int_{-\infty}^{\infty}|D(\nu)|^2|\Psi(\nu)|cos(2\pi\nu\tau-\phi(\nu)) d\nu,
\end{split}
\end{equation}          
to obtain the best fit values of $\Psi(\nu) = \sum_i \Psi_i$ and $\phi(\nu) = \sum_i \phi_i$ as a function of frequency using the PSD from the MCMC distributions. We again use the best fit parameters as input for an MCMC, which gives us the distribution of $\phi(\nu)$. For each bin we use the median of the distribution as that bin's $\phi(\nu)$ and the standard deviation as the error of $\phi(\nu)$. Finally, we can use $\phi(\nu)$ to determine the lag for each frequency bin using Equation (\ref{eq:mlm_tau}). 

We show an example of the lag as a function of frequency detected using this maximum-likelihood method in the bottom right panel of Figure \ref{fig:ex}, for the same light curve with an input long lag of $-50$~days shown above. The vertical grey line shows the phase wrapping frequency for $-50$ days. In Figure \ref{fig:ex} the lowest frequency bin has the largest error-bar and is  consistent with zero. In general for our mock light curves this bin tends to have the largest error and to be the least consistent with the input long lag. For these reasons we neglect the lag found in this bin, and then calculate the long lag as the error-weighted mean of all recovered lags $<-10$~days. We apply a cut-off of $-10$~days, because lags will approach zero at the wrapping frequency, and we only care about negative lag values when searching for the long lag. In Figure \ref{fig:ex} the two bins below the wrapping frequency have an error-weighted mean value of $-49$~days, which is similar to the input lag. The bin above the wrapping frequency is consistent with zero for this light curve, because there is no input short lag. For light curves with input short lags we also use the maximum-likelihood method to detect the short lag. To avoid interference from a long lag signal, we follow \cite{Yao:2022} and calculate a short lag by taking the lag value from the one bin centered between $0.01 < \nu < 0.1$. 

\section{Using LSST to Detect Long Lags}
\label{sec:detect}

\begin{deluxetable*}{ccccccccc}
\tablecolumns{9}
\tablewidth{0pt}
\label{tab:summ}
\tablehead{
& \colhead{Cadence} & \colhead{Ratio}  &\multicolumn{2}{c}{Input Lag} & \multicolumn{2}{c}{Cross-Correlation/{\sc javelin}}& \multicolumn{2}{c}{Fourier} \\
& & & \colhead{Long} & \colhead{Short}&\colhead{Long} &\colhead{Short} &\colhead{Long} &\colhead{Short} 
}
\startdata
\multirow{3}{*}{(1)} & \multirow{3}{*}{\begin{tabular}[c]{@{}c@{}}Even\\ Daily\end{tabular}} & \multirow{3}{*}{Long Only} & $-50$ & -- & $-50$ & 1 & $-51$ & $-1$ \\
 &  &  & $-130$ & -- & $-129$ & 1 & $-131$ & 3 \\
 &  &  & $-400$ & -- & $-385$ & 518 & -- & -- \\
\multirow{3}{*}{(2)} & \multirow{3}{*}{\begin{tabular}[c]{@{}c@{}}Even \\ Daily\end{tabular}} & \multirow{3}{*}{0.2} & $-50$ & 7 & $-44$ & 5 & $-27$ & 7 \\
 &  &  & $-130$ & 9 & $-125$ & 1 & $-96$ & 9 \\
 &  &  & $-400$ & 12 & $-386$ & 500 & -- & -- \\
\multirow{3}{*}{(3)} & \multirow{3}{*}{\begin{tabular}[c]{@{}c@{}}Even\\ Daily\end{tabular}} & \multirow{3}{*}{0.1} & $-50$ & 7 & $-1$ & 6 & $-17$ & 7 \\
 &  &  & $-130$ & 9 & $-122$ & 8 & $-74$ & 9 \\
 &  &  & $-400$ & 12 & $-390$ & 12 & -- & -- \\
\multirow{3}{*}{(4)} & \multirow{3}{*}{\begin{tabular}[c]{@{}c@{}}Even\\ Daily\end{tabular}} & \multirow{3}{*}{0.01} & $-50$ & 7 & $-1$ & 7 & 3 & 7 \\
 &  &  & $-130$ & 9 & $-1$ & 9 & $-8$ & 9 \\
 &  &  & $-400$ & 12 & $-1$ & 12 & -- & -- \\ \hline
\multirow{3}{*}{(5)} & \multirow{3}{*}{\begin{tabular}[c]{@{}c@{}}Long \\ Season\end{tabular}} & \multirow{3}{*}{Long Only} & $-50$ & -- & $-50 ^{+1}_{-1}$ & $460^{+60}_{-300}$ & $-49^{+5}_{-4}$ & $0^{+4}_{-4}$ \\
 &  &  & $-130$ & -- & $-130^{+0}_{-0}$ & $490^{+50}_{-300}$ & $-130^{+100}_{-30}$ & $0^{+4}_{-4}$ \\
 &  &  & $-400$ & -- & $-410^{+10}_{-20}$ & $270^{+260}_{-100}$ & -- & -- \\
\multirow{3}{*}{(6)} & \multirow{3}{*}{\begin{tabular}[c]{@{}c@{}}Long \\ Season\end{tabular}} & \multirow{3}{*}{0.2} & $-50$ & 7 & $-48^{+1}_{-4}$ & $190^{+300}_{-90}$ & $-25^{+7}_{-7}$ & $6^{+2}_{-2}$ \\
 &  &  & $-130$ & 9 & $-130^{+0}_{-0}$ & $250^{+250}_{-120}$ & $-98^{+43}_{-28}$ & $8^{+2}_{-5}$ \\
 &  &  & $-400$ & 12 & $-400^{+30}_{-40}$ & $170^{+260}_{-60}$ & -- & -- \\
\multirow{3}{*}{(7)} & \multirow{3}{*}{\begin{tabular}[c]{@{}c@{}}Long \\ Season\end{tabular}} & \multirow{3}{*}{0.1} & $-50$ & 7 & $-48^{+3}_{-7}$ & $14^{+120}_{-1}$ & $-20^{+7}_{-4}$ & $5^{+2}_{-2}$ \\
 &  &  & $-130$ & 9 & $-130^{+0}_{-0}$ & $240^{+250}_{-220}$ & $-62^{+15}_{-27}$ & $8^{+2}_{-4}$ \\
 &  &  & $-400$ & 12 & $-410^{+40}_{-40}$ & $160^{+310}_{-110}$ & -- & -- \\
\multirow{3}{*}{(8)} & \multirow{3}{*}{\begin{tabular}[c]{@{}c@{}}Long \\ Season\end{tabular}} & \multirow{3}{*}{0.01} & $-50$ & 7 & $-140^{+30}_{-170}$ & $9^{+5}_{-1}$ & $-26^{+3}_{-32}$ & $5^{+2}_{-2}$ \\
 &  &  & $-130$ & 9 & $-130^{+30}_{-80}$ & $11^{+2}_{-2}$ & $-53^{+38}_{-6}$ & $8^{+2}_{-3}$ \\
 &  &  & $-400$ & 12 & $-250^{+90}_{-230}$ & $14^{+2}_{-1}$ & -- & -- \\
\multirow{3}{*}{(9)} & \multirow{3}{*}{\begin{tabular}[c]{@{}c@{}}Baseline \\ u-y\end{tabular}} & \multirow{3}{*}{0.2} & $-50$ & 7 & $-180^{+0}_{-10}$ & $190^{+10}_{-10}$ & $-31^{+12}_{-17}$ & $9^{+2}_{-3}$ \\
 &  &  & $-130$ & 9 & $-180^{+10}_{-0}$ & $190^{+10}_{-10}$ & $-64^{+21}_{-34}$ & $9^{+2}_{-14}$ \\
 &  &  & $-400$ & 12 & $-180^{+10}_{-350}$ & $190^{+10}_{-0}$ & -- & -- \\
\multirow{3}{*}{(10)} & \multirow{3}{*}{\begin{tabular}[c]{@{}c@{}}Baseline \\ g-z\end{tabular}} & \multirow{3}{*}{0.2} & $-50$ & 7 & $-180^{+30}_{-10}$ & $180^{+350}_{-10}$ & $-27^{+8}_{-12}$ & $7^{+2}_{-3}$ \\
 &  &  & $-130$ & 9 & $-170^{+20}_{-10}$ & $190^{+350}_{-10}$ & $-98^{+50}_{-30}$ & $9^{+2}_{-14}$ \\
 &  &  & $-400$ & 12 & $-190^{+10}_{-340}$ & $190^{+350}_{-10}$ & -- & -- \\
\multirow{3}{*}{(11)} & \multirow{3}{*}{\begin{tabular}[c]{@{}c@{}}Baseline \\ + LCO\end{tabular}} & \multirow{3}{*}{0.2} & $-50$ & 7 & $-51^{+2}_{-9}$ & $450^{+100}_{-280}$ & $-22^{+4}_{-5}$ & $6^{+1}_{-1}$ \\
 &  &  & $-130$ & 9 & $-130^{+0}_{-0}$ & $370^{+170}_{-200}$ & $-85^{+35}_{-28}$ & $8^{+1}_{-1}$ \\
 &  &  & $-400$ & 12 & $-390^{+3}_{-20}$ & $170^{+280}_{-80}$ & -- & -- \\
\multirow{3}{*}{(12)} & \multirow{3}{*}{\begin{tabular}[c]{@{}c@{}}Long Season\\ (Non-DRW)\end{tabular}} & \multirow{3}{*}{0.2} & $-50$ & 7 & $-53^{+3}_{-1}$ & $150^{+370}_{-50}$ & $-21^{+1}_{-3}$ & $6^{+0}_{-0}$ \\
 &  &  & $-130$ & 9 & $-130^{+0}_{-10}$ & $130^{+350}_{-20}$ & $-95^{+6}_{-14}$ & $9^{+1}_{-2}$ \\
 &  &  & $-400$ & 12 & $-430^{+40}_{-20}$ & $190^{+30}_{-20}$ & -- & -- \\
\multirow{3}{*}{(13)} & \multirow{3}{*}{\begin{tabular}[c]{@{}c@{}}Long Season\\ (i=21 mag)\end{tabular}} & \multirow{3}{*}{0.2} & $-50$ & 7 & $-50^{+2}_{-4}$ & $480^{+40}_{-350}$ & $-26^{+7}_{-7}$ & $5^{+2}_{-2}$ \\
 &  &  & $-130$ & 9 & $-130^{+10}_{-10}$ & $480^{+40}_{-350}$ & $-98^{+34}_{-30}$ & $8^{+2}_{-8}$ \\
 &  &  & $-400$ & 12 & $-410^{+30}_{-30}$ & $170^{+320}_{-40}$ & -- & --
\enddata
\caption{Above divider: The peaks in the cross-correlation coefficient and the long and short lags detected by the Fourier method for the even-cadence light curves with different input lags and amplitude ratios (between the long and short lag) shown in Figures \ref{fig:jav_nocad} and \ref{fig:mlm_nocad}. Below divider: The median long and short lags detected by {\sc javelin} and the maximum-likelihood method for 100 mock light curves with each cadence, amplitude ratio, and input long and short lag used throughout the paper. The error ranges represent one standard deviation.}
\end{deluxetable*}

\subsection{Detecting Long Lags Alone}
\label{sec:long}

\begin{figure}
    \centering
    \includegraphics[width=\columnwidth]{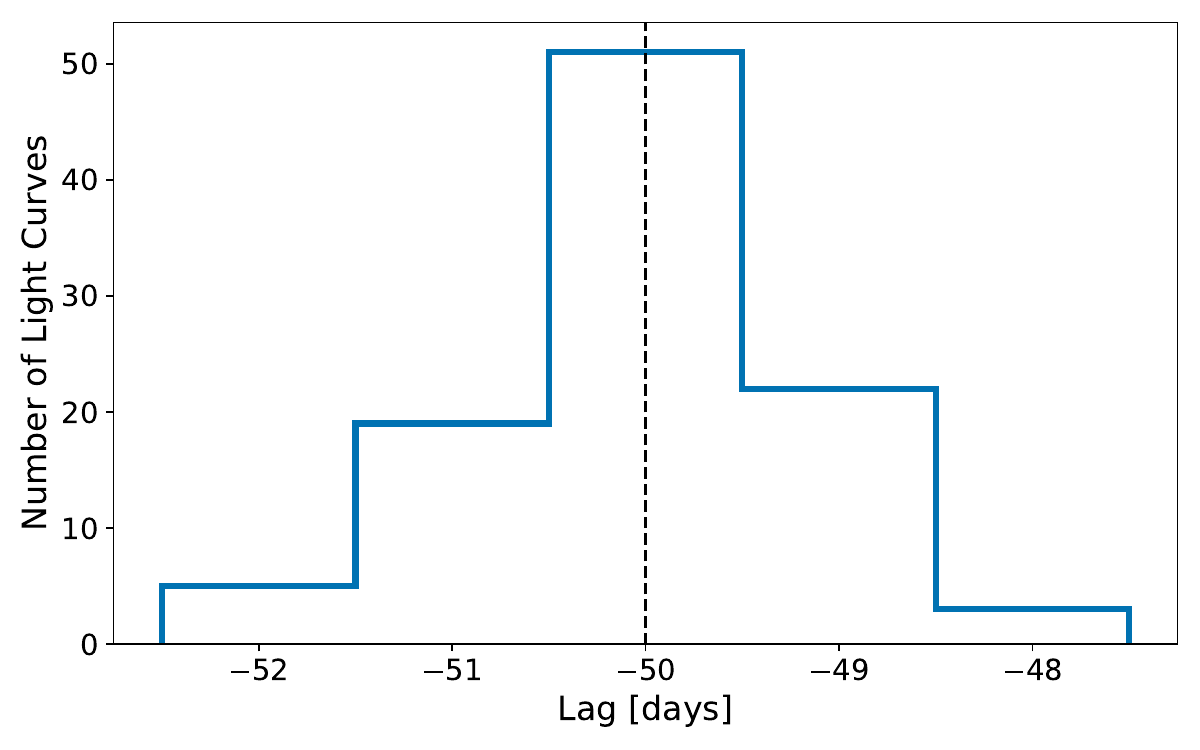}
    \\
    \includegraphics[width=\columnwidth]{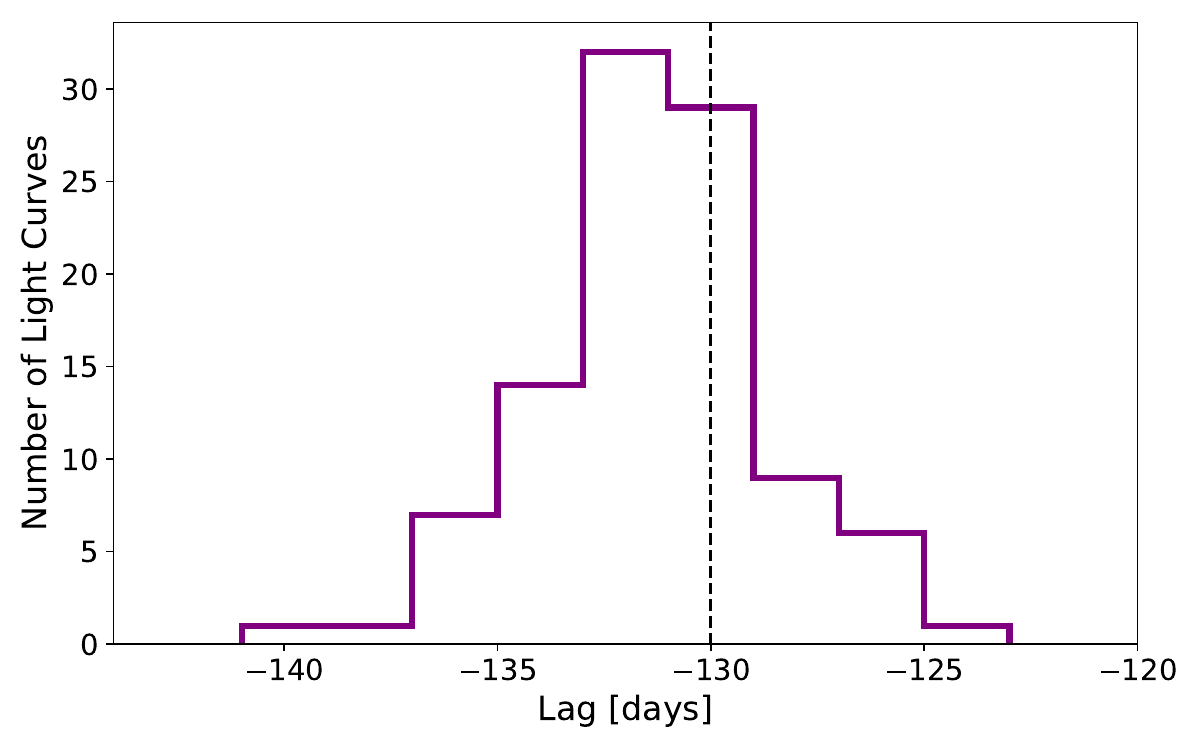}
    \\
    \includegraphics[width=\columnwidth]{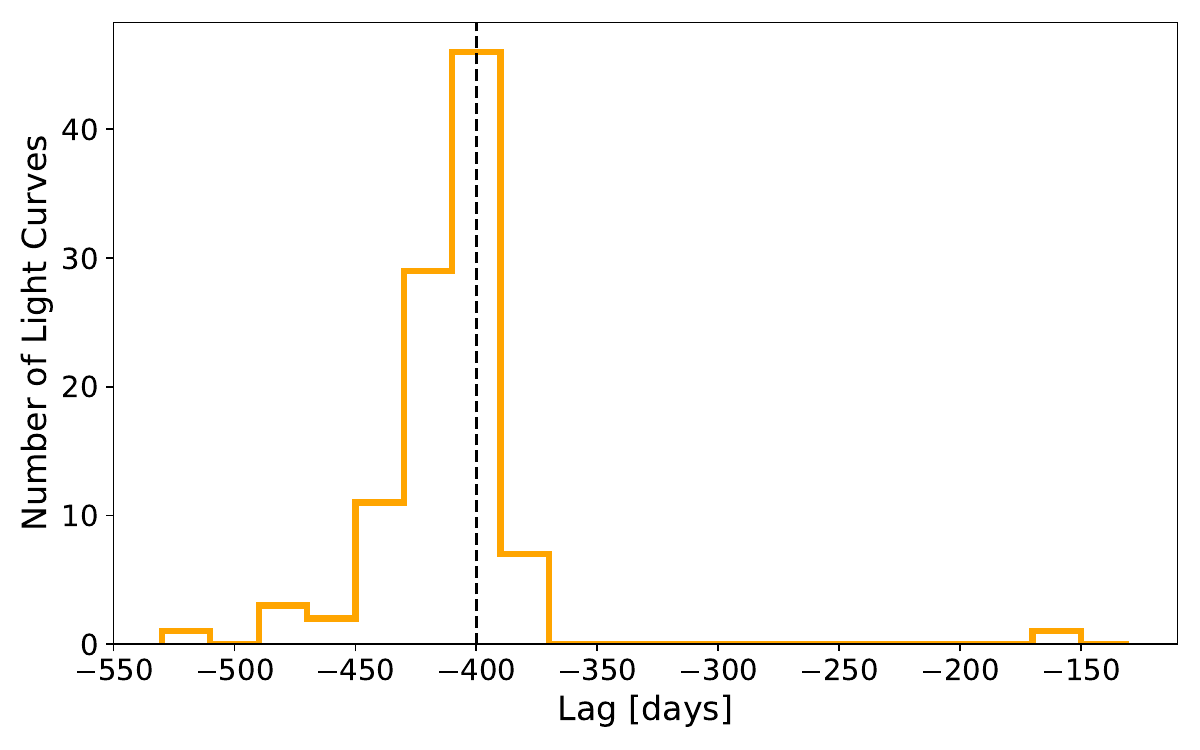}
    \caption{The distribution of the long lags detected by {\sc javelin} for 100 mock long season cadence LSST light curves with input long lags of $-50$, $-130$, and $-400$~days (shown as dashed black lines) in the top, center, and bottom panel, respectively.}
    \label{fig:jav_single}
\end{figure}

\begin{figure}
    \centering
    \includegraphics[width=\columnwidth]{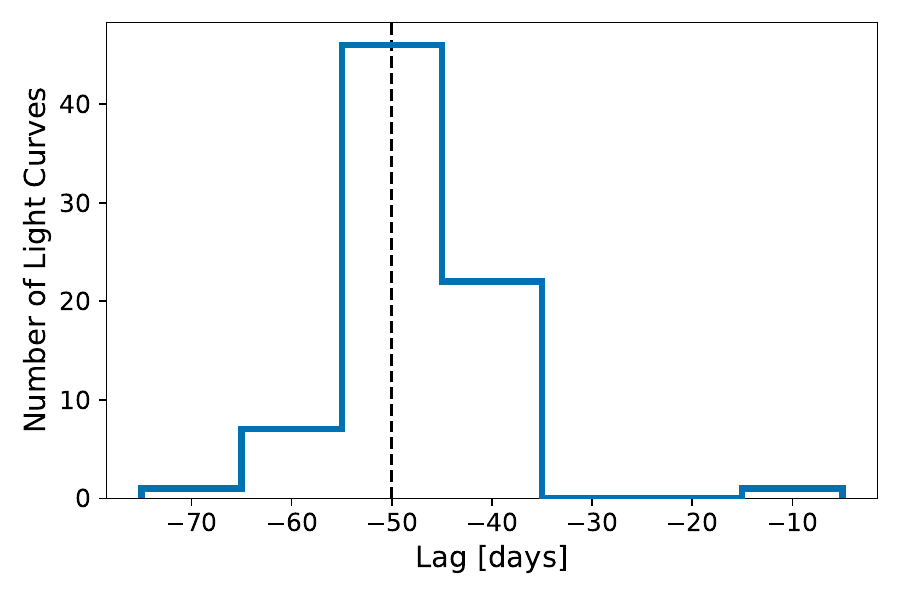} \\
    \includegraphics[width=\columnwidth]{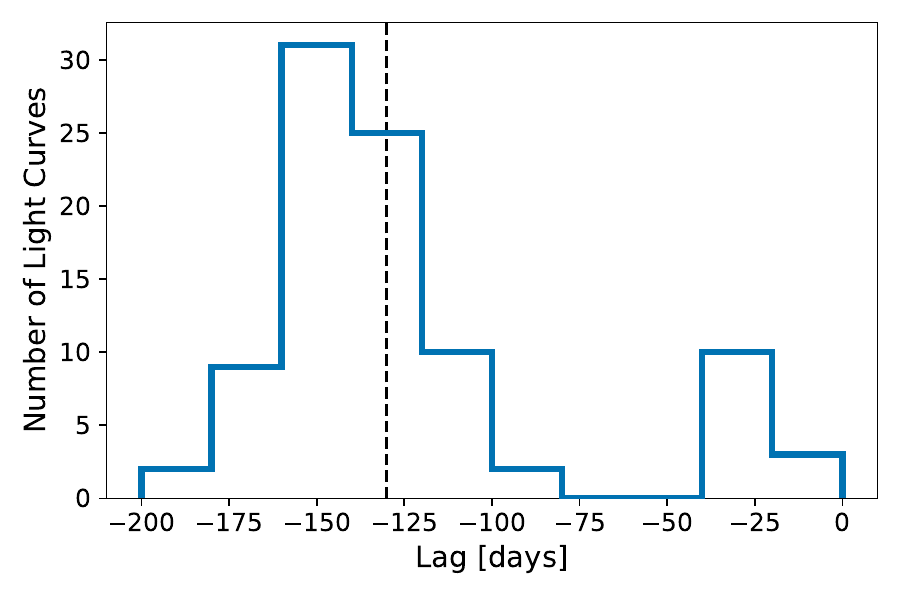} \\
    \caption{The distribution of long lags recovered by the maximum-likelihood method for 100 mock long season cadence LSST light curves with input long lags (shown as dashed black lines) of $-50$~days (top panel) and $-130$~days (bottom panel).}
    \label{fig:mlm_single}
\end{figure}

First we examine the performance of {\sc javelin} and the maximum-likelihood method on light curves with the long season LSST cadence (see Section \ref{sec:cadence}) that have been reprocessed with a long lag only. Figure \ref{fig:jav_single} shows the distributions of the medians of the negative values of the {\sc javelin} distributions for 100 mock light curves with input long lags of $-50$~days (top panel), $-130$~days (middle panel), and $-400$ days (bottom panel). The lag detected by {\sc javelin} agrees within 10\% for all light curves with input lags of $-50$ days and $-130$ days and 88\% of light curves with input lags of $-400$. 

Figure \ref{fig:mlm_single} shows the distribution of lags recovered by the maximum-likelihood method for the same mock light curves excluding those with input long lags of $-400$~days. A long lag of $-400$~days is very difficult to detect with the maximum-likelihood method for light curves with the LSST baseline of ten years. For input lags of $-50$ and $-130$~days the maximum-likelihood method is less accurate than {\sc javelin}. However, the maximum-likelihood method is able to detect the long lag to within 30\% of the input lag for around 70\% of light curves. We show the medians and standard deviations of the {\sc javelin} and maximum-likelihood method distributions for these mock light curves in row (5) of Table \ref{tab:summ}.

\subsection{Detecting Short and Long Lags Together}
\label{sec:short_long}

\begin{figure*}
    \centering
    \includegraphics[width=\textwidth]{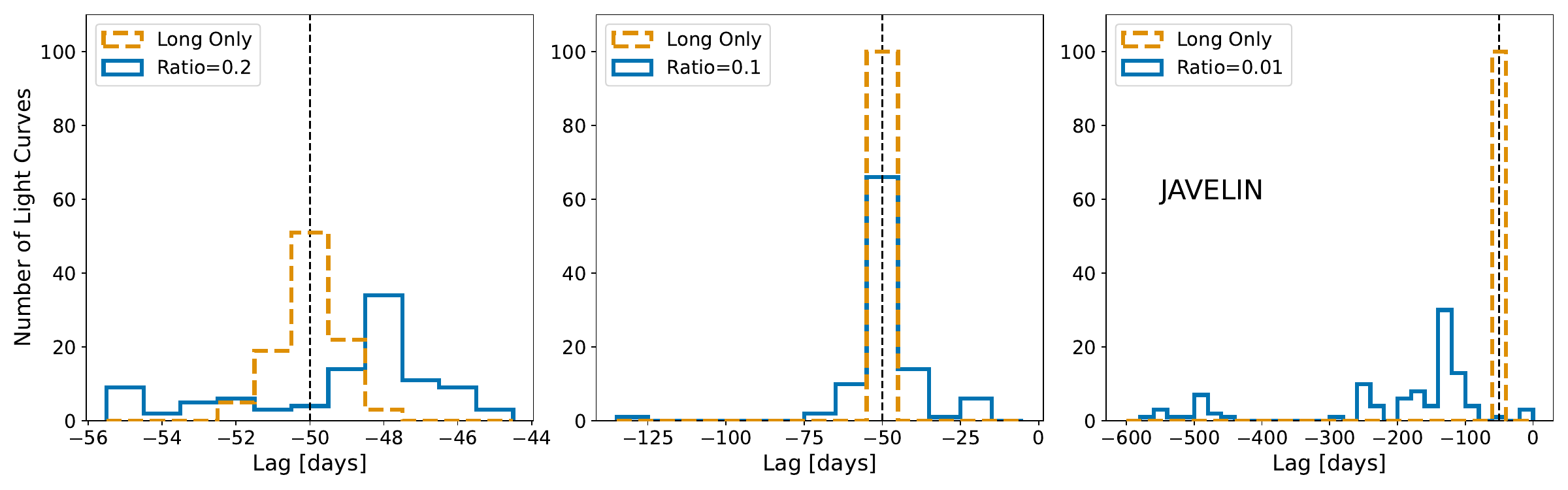} 
    \includegraphics[width=\textwidth]{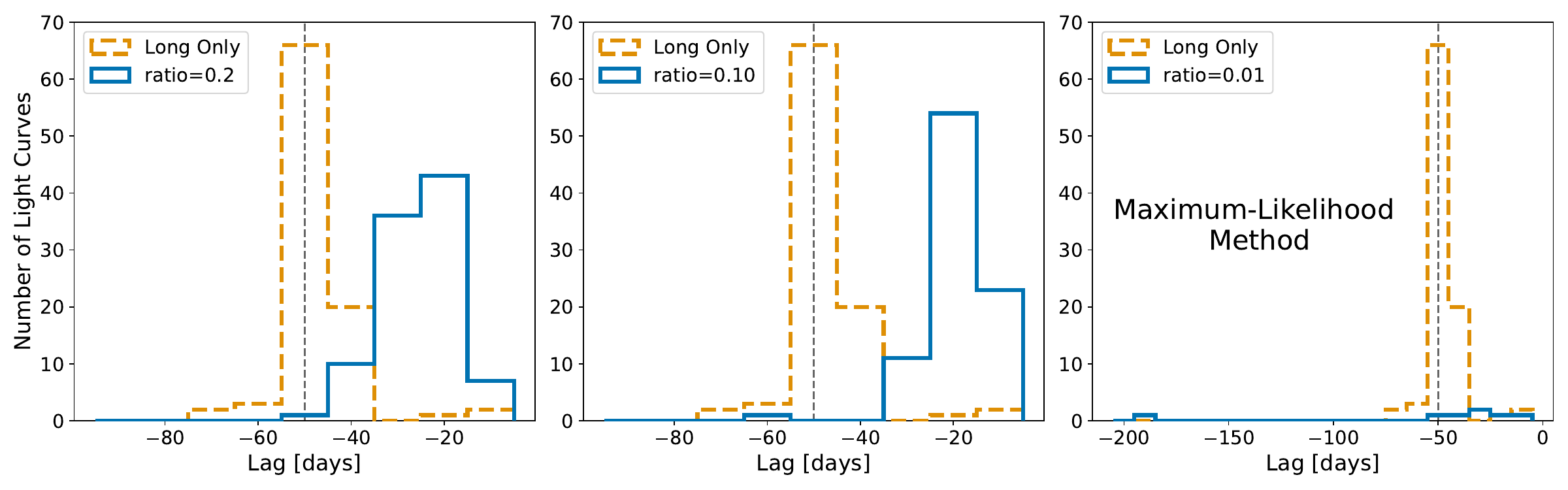}
    \caption{The distributions of the long lags detected by {\sc javelin} (top panel) and the maximum-likelihood method (bottom panel) for 100 mock long season cadence LSST light curves with an input long lag of $-50$ days (shown as a dashed black line). Distributions shown in blue are for light curves that also have a short lag of 7 days. The legends show the ratio of the amplitude of the long lag to the amplitude of the short lag for each blue distribution.}
    \label{fig:both}
\end{figure*}

\begin{figure}
    \centering
    \includegraphics[width=\columnwidth]{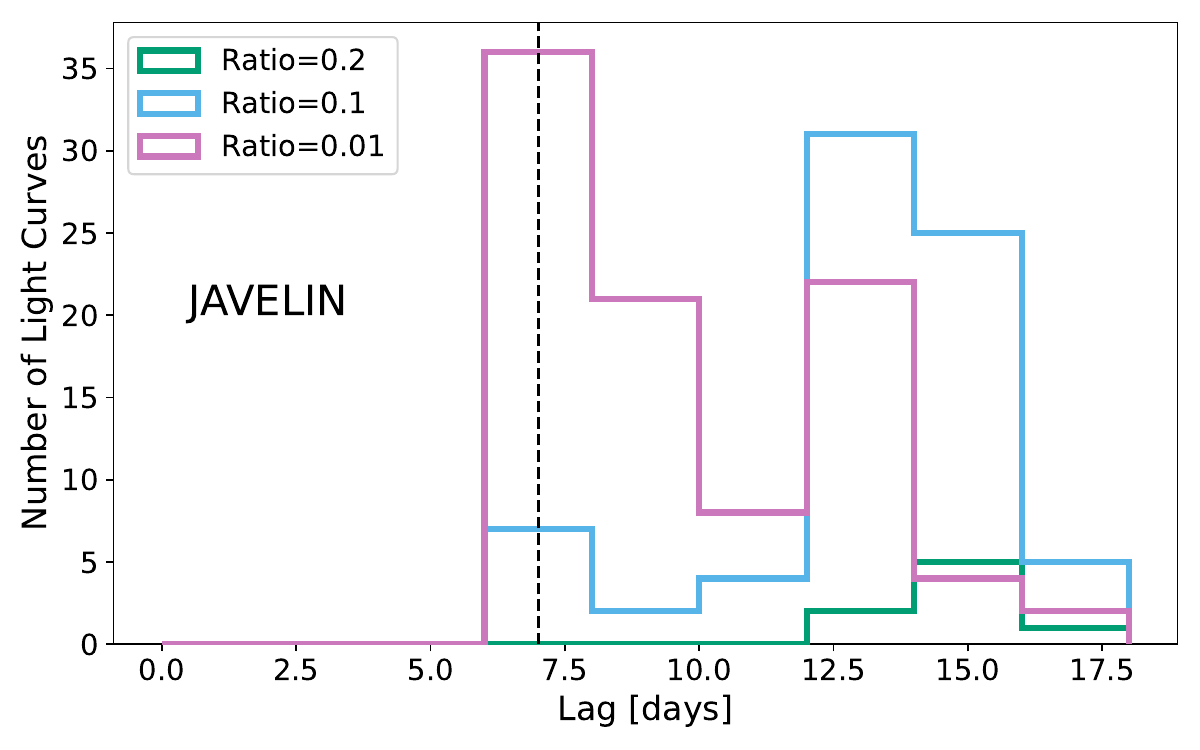} \\
    \includegraphics[width=\columnwidth]{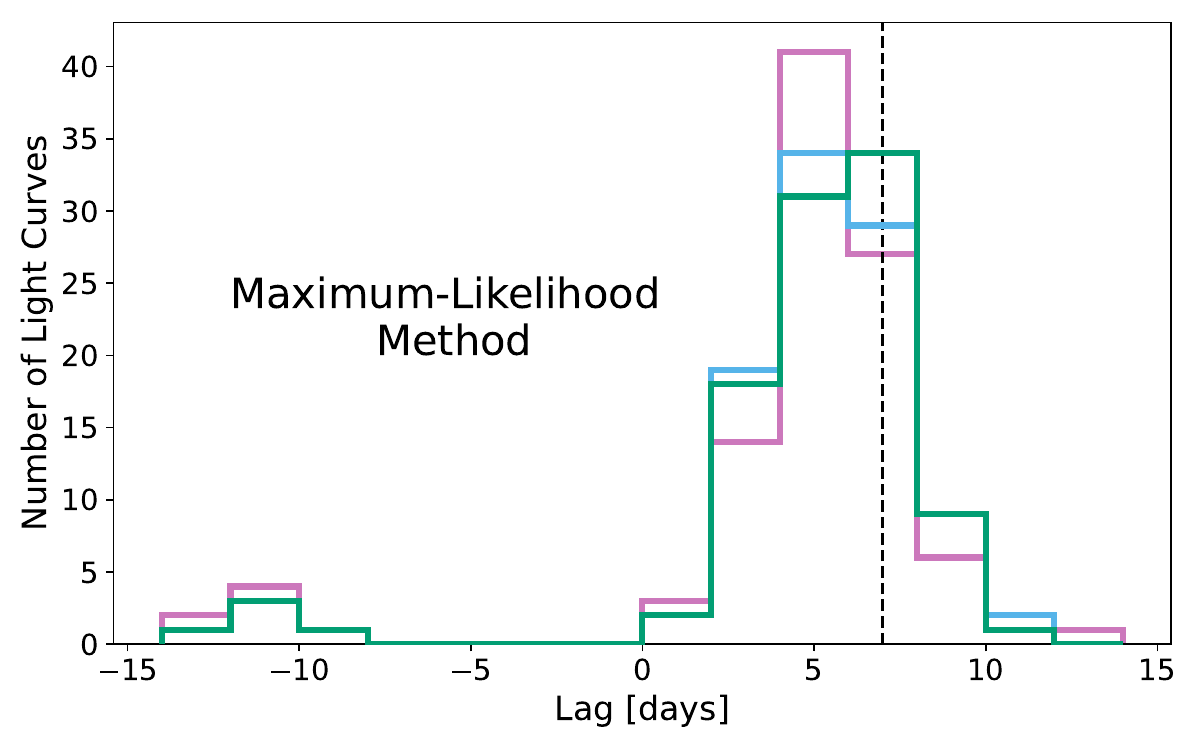}
    \caption{The distribution of the short lags detected by {\sc javelin}  (top panel) and the maximum-likelihood method (bottom panel) for 100 mock long season cadence LSST light curves with different amplitude ratios (shown in the legend). In the top panel we zoom in on the portion of the distribution between 0 and 20 days to better show the accuracy of the short lag when detected. The input short lag, shown as the vertical dashed line, is 7~days.}
    \label{fig:short}
\end{figure}

We now add a short lag to our mock light curves with the long season LSST cadence to examine how multiple lag signals can affect our ability to detect either lag. Our fiducial value for the ratio of the amplitude of the long lag to the amplitude of the short lag is 0.2, as we expect the short lag signal to be stronger. We also examine our ability to detect lags when this ratio is 0.1 and 0.01. For readability, in this section we only show figures for light curves with an input long lag of $-50$~days and an input short lag of 7~days. Unless otherwise noted, the input lag duration does not have an impact on the accuracy of our lag detection methods. We repeat the analysis in this section for light curves reprocessed with long (short) lags of $-130$ (9) and $-400$ (12) days in Appendix \ref{app:long_short}.

In the top panel of Figure \ref{fig:both} we compare the {\sc javelin} results for light curves reprocessed only with a long lag (in orange) to {\sc javelin} results for light curves reprocessed with a long and short lag (in blue). For light curves with both lags we take the median of the negative (positive) values in the {\sc javelin} distribution as the long (short) lag. When the amplitude ratio between the long and short lag is 0.2 or 0.1 the distributions of long lags detected by {\sc javelin} are consistent with the distribution of long lags detected for light curves without a short lag. Adding in a short lag only makes the distributions somewhat broader. Comparing these results to the correlation coefficients for evenly-sampled daily cadence light curves, shown in Figure \ref{fig:jav_nocad}, the long lag tends to be more prominent than we might expect, in particular for an amplitude ratio of 0.1. It may be that the roughly every 10--30 day cadence of the long season LSST cadence, helps us to detect the long lag, because it hides some of the signal from the short lag.

On the other hand, when the ratio is decreased to 0.01, we fail to detect the correct long lag for nearly all mock light curves with an input long lag of $-50$~days. In Figure \ref{fig:jav_both} in Appendix \ref{app:long_short}, for an input long lag of $-130$ days there is still a peak in the distribution around $-130$ days. It is unclear whether this peak is tied to the cadence of the light curve or the injected long lag. For one, there is also a peak around $-130$~days in the lag distributions for input long lags of $-50$ and $-400$~days when the amplitude ratio is 0.01. There is also some evidence for a peak around $-130$~days in our null tests using light curves that have not been reprocessed with a lag (see top left panel of Figure \ref{fig:app:jav} in Appendix \ref{app:jav}). {\sc javelin} detects a long lag of around $-130$~days for roughly 10\% of these light curves with no input lag. However, in the right middle panel of Figure \ref{fig:jav_both} {\sc javelin} accurately recovers a long lag of around $-130$~days for a significantly higher percentage ($\sim 50\%$) of light curves. Furthermore, in Figure \ref{fig:jav_nocad} there is a slight peak in the cross-correlation coefficient for the evenly-sampled light curve with a $-130$~day input long lag and an amplitude ratio of 0.01, while there is no corresponding peak in the correlation coefficient for other input lags when the amplitude ratios are 0.01. Therefore it is unclear if the peak in the distribution in the right middle panel in Figure \ref{fig:jav_both} is due to a recovered lag or not.

We show the medians of the positive values of the {\sc javelin} distribution for 100 mock light curves with input lags of $-50$ and 7~days in the top panel of Figure \ref{fig:short}. We zoom in to the portion of the distribution falling between 0 and 20 days to better show the accuracy of the short lag when it is detected. When the ratio of the amplitude of the long lag to the amplitude of the short lag is 0.01 {\sc javelin} is able to detect the short lag, although the distribution is skewed towards larger values than the input short lag. In previous studies, {\sc javelin} has detected short lags that appear to be skewed towards longer frequencies due to diffuse BLR emission \citep[e.g.,][]{Cackett:2018}. Something similar may be happening here due to the presence of both a long and short lag.

For an amplitude ratio of 0.1, {\sc javelin} detects a short lag of around 13--16 days for roughly 60\% of light curves with input lags of $-50$ and 7~days. The null tests do not show a peak around 13 days for this cadence (see top right panel of Figure \ref{fig:app:jav} in Appendix \ref{app:jav}). In addition, for the even cadence light curve with an input long lag of $-50$~days shown in the top panel of Figure \ref{fig:jav_nocad}, the peak in the correlation coefficient around the short lag is larger than the peak around the long lag when the amplitude ratio is 0.1. The fact that this peak around the short lag is already larger at this amplitude ratio suggests that it might be easier to detect the short lag with {\sc javelin} in this particular case. Therefore it is possible that the light blue peak in the top panel of Figure \ref{fig:short} is due to the input short lag, but heavily skewed towards longer lags because of the presence of the long lag. For other input long lags {\sc javelin} is unable to detect a short lag for most light curves unless the amplitude ratio is 0.01 (see Figure \ref{fig:jav_short} in Appendix \ref{app:long_short}.)

The bottom panel of Figure \ref{fig:both} is the same as the top panel, but instead shows the long lag detected by the maximum-likelihood method. As with the even cadence light curves in Figure \ref{fig:mlm_nocad}, when a short lag is added, the long lag is shifted towards shorter timescales. The median of the distribution for light curves with an input long lag of $-50$~days goes from $-49$~days for light curves with no short lag, to $-25$~days and $-20$~days for light curves with amplitude ratios of 0.2 and 0.1, respectively. When the amplitude ratio is 0.01, for light curves with input long lags of $-50$~days there are very few negative lags detected at all and for light curves with an input long lag of $-130$~days there are only about 20 negative lags detected, mostly between $-10$ and $-65$~days (see Figure \ref{fig:mlm_both} in Appendix \ref{app:full}). It is clear that like {\sc javelin} the maximum-likelihood method is unable to isolate the long lag when the amplitude ratio is 0.01.

The bottom panel of Figure \ref{fig:short} shows the distribution of short lags detected by the maximum-likelihood method. Unlike {\sc javelin}, the maximum-likelihood method is able to detect short lags that are accurate to within a few days of the input lag for over 60\% of light curves with amplitude ratios of 0.2 and 0.1. Detecting short lags more accurately for a wider range of amplitude ratios is therefore one benefit to using the maximum-likelihood method in addition to {\sc javelin}.

The accuracy of the maximum-likelihood method on light curves where we include two input lags appears to be consistent with Figure \ref{fig:mlm_nocad}, where we use the Fourier method on uniform 1-day cadence, 10~year-long, DRW light curves. It is promising that the maximum-likelihood method is able to perform nearly as well on light curves with the long season LSST cadence as the Fourier method performs on light curves with an even cadence. However, it will be difficult with this method to disentangle the length of the long lag versus the amplitude ratio between the long and short lag. 

\begin{figure}
    \centering
    \includegraphics[width=\columnwidth]{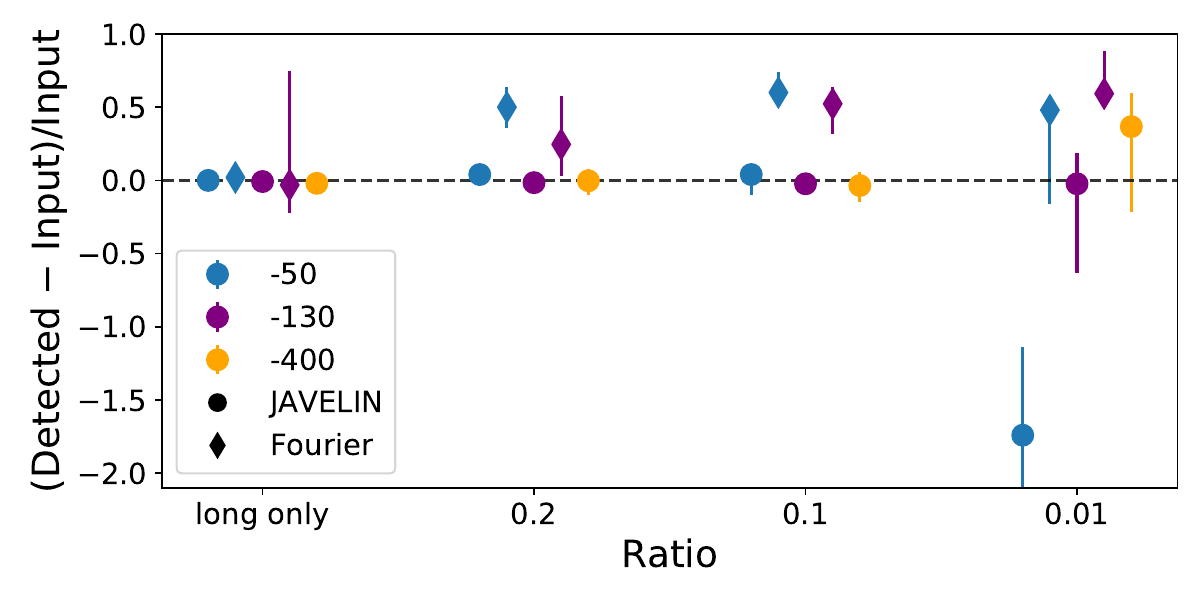}
    \caption{The fractional error of the median of the distribution of long lags recovered by {\sc javelin} (points) and the Fourier method (diamonds) for 100 mock long season cadence LSST light curves as a function of the ratio of the amplitude of the long lag to the amplitude of the short lag for input long lags of $-50$, $-130$, and $-400$~days in blue, purple, and orange, respectively. The error bars represent the fractional error of one standard deviation for each distribution and the horizontal dashed line shows zero. A slight scatter is added in the x-direction to make the individual points easier to view.}
    \label{fig:summ}
\end{figure}

We show the medians and standard deviations of the distributions of lags recovered by {\sc javelin} and the maximum-likelihood method for our 100 mock light curves in rows (6), (7), and (8) in Table \ref{tab:summ} for light curves with amplitude ratios between their input long and short lags of 0.2, 0.1, and 0.01, respectively. In Figure \ref{fig:summ} we show the fractional error of the medians of the distributions of lags recovered by {\sc javelin} and the maximum-likelihood method for the 100 mock light curves as a function of these amplitude ratios for all input long lags. The error-bars show the fractional error of one standard deviation for each distribution. The accuracy of the recovered lags is similar regardless of the durations of the input lags.

\subsection{Improving the Baseline Cadence}
\label{sec:lco}

\begin{figure}
    \centering
    \includegraphics[width=0.48\textwidth]{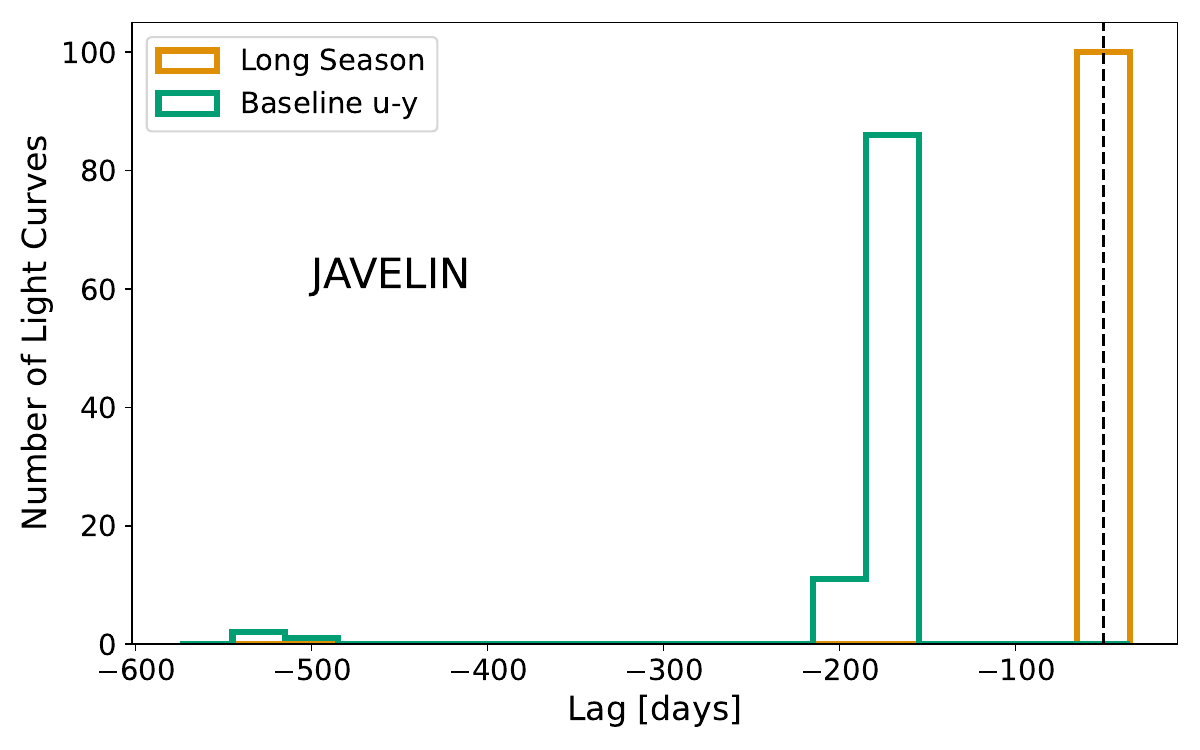}
    \\
    \includegraphics[width=0.48\textwidth]{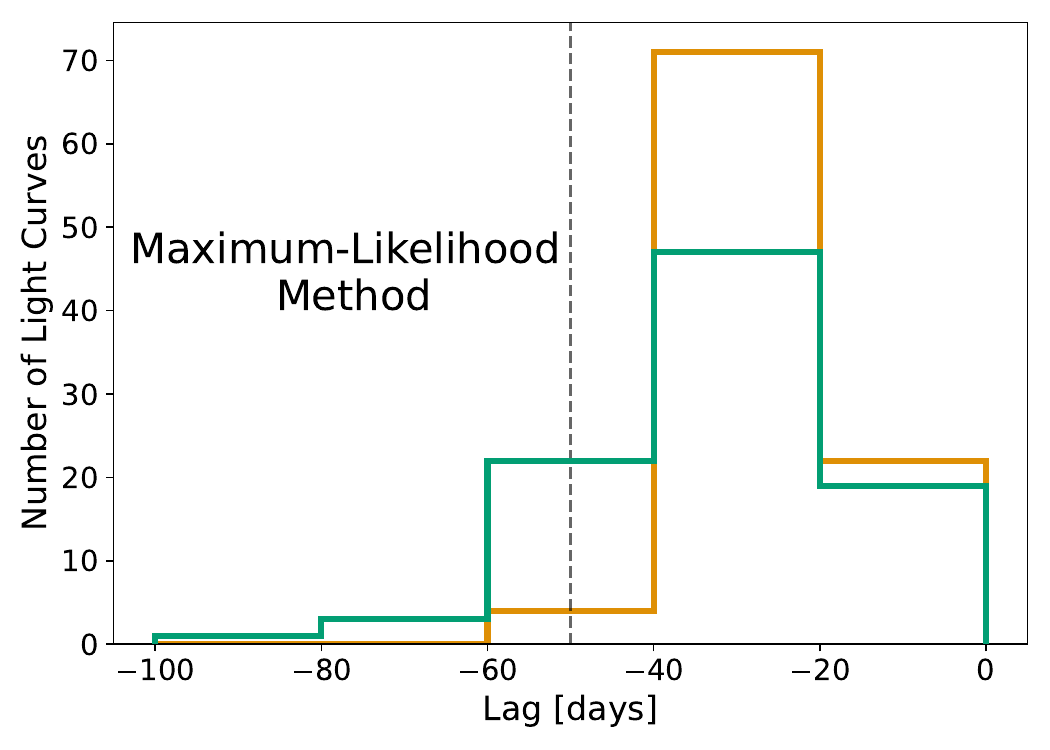}
    \caption{The distributions of the long lags recovered by {\sc javelin} (top panel) and the maximum-likelihood method (bottom panel) for 100 light curves with input lags of $-50$ and 7 days. The green and orange distributions are for mock light curves with the long season and baseline v2.0 LSST cadences, respectively, for the \emph{u}- and \emph{y}-band. The dashed vertical line shows the input long lag.}
    \label{fig:base}
\end{figure}

\begin{figure}
    \centering
    \includegraphics[width=0.48\textwidth]{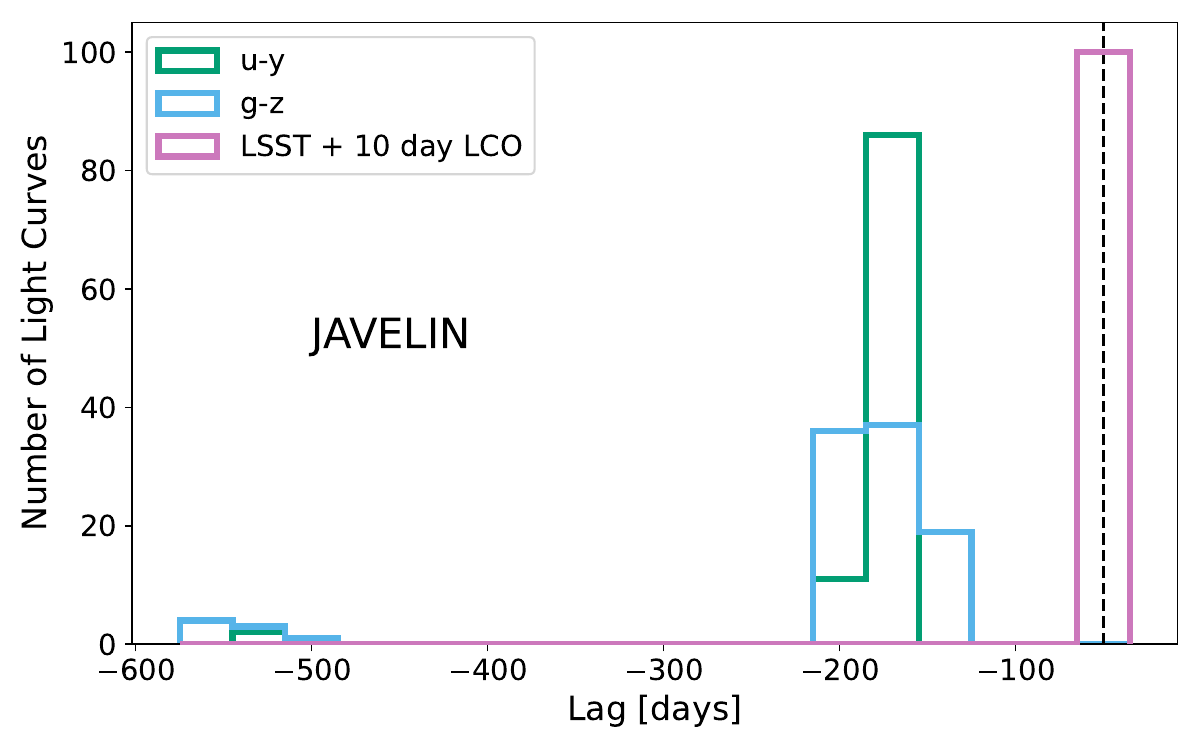} \\
    \includegraphics[width=0.48\textwidth]{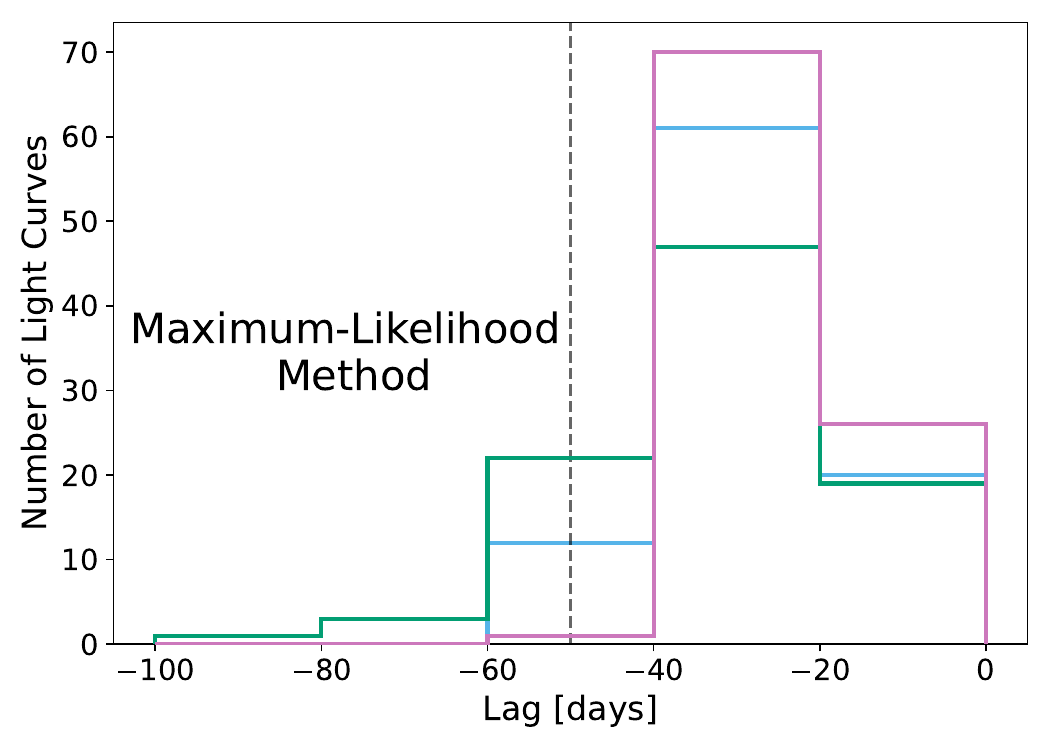}
    \caption{Distribution of the long lags recovered by {\sc javelin} (top panel) and the maximum-likelihood method (bottom panel) for 100 mock light curves with input lags of $-50$ and 7 days (top panel). The dashed vertical line shows the input long lag. The green and light blue distributions are for mock light curves with the baseline v2.0 LSST cadence for the \emph{u}- and \emph{y}-band and \emph{g}- and \emph{z}-band, respectively. The distributions shown in pink are for light curves with the baseline \emph{g}- and \emph{z}-band cadence plus an LCO observation taken roughly every ten days in each LSST observing gap. These LCO observations amount to around 18 observations in each gap, or per year, for a total of around 180 observations.}
    \label{fig:lco}
\end{figure}

\begin{figure}
    \centering
    \includegraphics[width=0.48\textwidth]{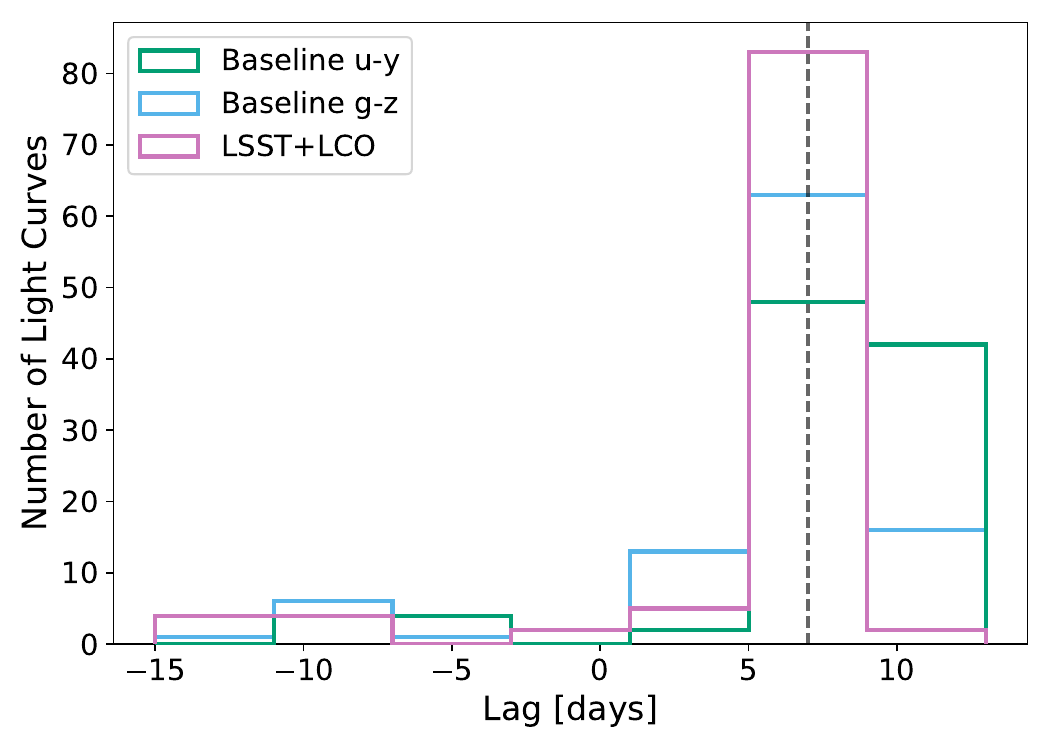}
    \caption{Distribution of the short lag recovered by the maximum-likelihood method for 100 mock light curves with input lags of $-50$ and 7 days. The dashed vertical line shows the input short lag. The green and light blue distributions are for mock light curves with the baseline v2.0 LSST cadence for the \emph{u}- and \emph{y}-band and \emph{g}- and \emph{z}-band, respectively. The distributions shown in pink are for light curves with the baseline \emph{g}- and \emph{z}-band cadence plus an LCO observation taken roughly every ten days in each LSST observing gap. These LCO observations amount to around 18 observations in each gap, or per year, for a total of around 180 observations.}
    \label{fig:lco_short}
\end{figure}

\begin{figure}
    \centering
    \includegraphics[width=\columnwidth]{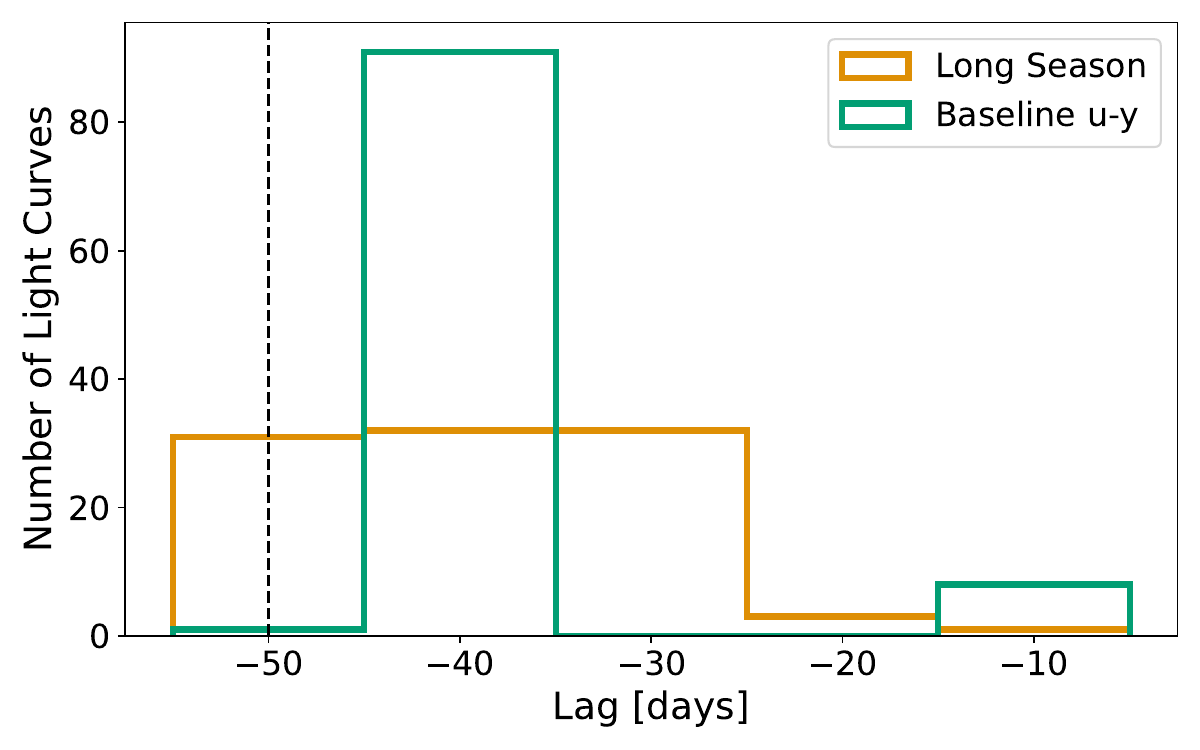}
    \caption{Distribution of the long lag recovered by the Von-Neumann method for 100 mock light curves with input lags of $-50$ and $7$ days. The dashed vertical line shows the input long lag. The orange (green) distribution is for mock light curves with the long season (baseline \emph{u}- and \emph{y}-band) cadence.}
    \label{fig:vnm_baseline50}
\end{figure}

The LSST collaboration is still evaluating different possible cadences. We have evaluated and compared the ability of our lag detection methods to accurately pick out long lags for several cadences. Above we asserted that the most significant factor for improving our ability to detect lags is increased season length. In the previous sections we use the long season LSST cadence. In this section we compare the detected lags for light curves with this long season cadence to light curves with the baseline v2.0 cadence. We again only show results here for light curves with input lags of $-50$ and 7~days.  We show results for all three lag bins in Table \ref{tab:bins} in Appendix \ref{app:baseline}. Unless otherwise noted, our results for light curves reprocessed with a long lag of $-50$~days are consistent with results for light curves reprocessed with other input long lags.

The top panel of Figure \ref{fig:base} shows the long lags detected by {\sc javelin} for 100 mock light curves sub-sampled with the long season and baseline cadences, in orange and green, respectively. Unfortunately, when using the baseline cadence {\sc javelin} fails to accurately recover any long lags. Instead, {\sc javelin} primarily recovers long lags of around $-180$ days. The second panel from the top in Figure \ref{fig:app:jav} in Appendix \ref{app:jav} shows the distributions of the medians of the negative and positive values of the {\sc javelin} distributions for 100 light curves with the baseline cadence but with no input lag, in the left and right panel, respectively. The positive and negative distributions peak strongly around $\pm 180$~days, like the distribution shown in green in top panel of Figure \ref{fig:base}, suggesting that the peak around $\pm 180$~days is due to the cadence, not a lag.

The maximum-likelihood method, on the other hand, still performs reasonably well on the light curves sub-sampled with the baseline cadence. Because the maximum-likelihood method separates the signal out in frequency space it is easier to separate out the lag signal from the cadence signal. The bottom panel of Figure \ref{fig:base} shows the distribution of lags recovered by the maximum-likelihood method for light curves with the long season or baseline cadence in orange and green, respectively. The distribution of long lags recovered by the maximum-likelihood method is slightly less skewed towards shorter timescales, with a median of $-31$~days. In Figure \ref{fig:mlm_base} in Appendix \ref{app:full} the distribution of long lags for light curves with an input long lag of $-130$~days is more skewed towards shorter timescales, with a median of $-58$~days. It could be that the presence of a gap at longer timescales actually helps to skew the long lag back towards longer timescales for light curves with input lags significantly shorter than the observing gaps ($\sim -50$~days), while longer lags ($\sim -130$~days) closer to the gap length are affected differently.

In order to improve the performance of {\sc javelin} on light curves with large seasonal gaps, we can add observations from other telescopes to fill in the 180 day gap. Here, we explore the effectiveness of filling these gaps with observations from LCO. LCO has been used for numerous RM campaigns and has wavebands ranging from the \emph{u}--\emph{y}-band. Here we experiment with adding observations in the \emph{g}- and \emph{z}-bands to each seasonal gap on roughly a ten day cadence. We use the \emph{g}- and \emph{z}-bands instead of the \emph{u}- and \emph{y}-bands to increase the SNR. We determine the additional LCO cadence in the gaps using the LCO \emph{g}- and \emph{z}-band cadences from the \cite{F92020} Fairall 9 observing campaign, which are roughly daily. We randomize these Fairall 9 observing times over a uniform distribution with a range of half the distance to the following observation, and then down-sample to every tenth observation. 

For each lag bin in Table \ref{tab:bins}, we make 100 mock light curves sub-sampled at times corresponding to the LSST baseline cadence for the \emph{g}- and \emph{z}-band and these additional LCO observations. The LCO observations add roughly 18 observations per year or 180 total. We add measurement errors corresponding to the SNR for LSST and LCO. The SNR for our mock LCO observations were calculated using the LCO Exposure Time Calculator\footnote{\url{https://exposure-time-calculator.lco.global/}} and range from 19 to 77, depending on the band and redshift bin.

We show the long lags recovered by {\sc javelin} for 100 of these mock light curves with input lags of $-50$ and 7~days in pink in the top panel of Figure \ref{fig:lco}. For comparison we show the lags recovered by {\sc javelin} for 100 mock light curves with the same input lags and the \emph{u}- and \emph{y}-band LSST baseline cadences (in green) and \emph{g}- and \emph{z}-band LSST baseline cadences (in light blue). Light curves with the \emph{g}- and \emph{z}-band LSST baseline cadences experience the same cadence-related issues as light curves with the \emph{u}- and \emph{y}-band LSST baseline cadences (see also Appendix \ref{app:jav}). However, adding LCO observations eliminates most of the false lag detections from the large gaps in the baseline cadence. {\sc javelin} recovers a long lag within 10\% of the input lag for 74\% (100\% are within 30\%) of light curves with input long lags of $-50$~days. The decline in false lag detections with this LSST+LCO cadence can also be seen in our null tests in Appendix \ref{app:jav}. In the bottom panel of Figure \ref{fig:app:jav}, the distributions of lags detected by {\sc javelin} for light curves with this LSST+LCO cadence and no input lag peak around zero, as expected.

The bottom panel of Figure \ref{fig:lco} is the same as the top panel except for the maximum-likelihood method. The distribution of long lags detected for light curves with the LSST+LCO cadence is similar to the distribution for light curves with the long season cadence. The median long lag detected for light curves with the LSST+LCO cadence is $-22$~days for an input long lag of $-50$~days. This median is similar to, although skewed slightly shorter than, the median found for light curves with the long season cadence ($-25$~days). 

We show the distributions of short lags detected by the maximum-likelihood method for the same light curves as above in Figure \ref{fig:lco_short}. For all cadences the distributions peak around the input short lag. However, for the baseline cadences only around 50\% of short lags detected are within 2 days of the input short lag, whereas when the LCO observations are added around 80\% of the short lags detected are within 2 days of the input short lag, which is slightly more accurate than with the long season cadence. For both {\sc javelin} and the maximum-likelihood method it is clear that adding just 18 LCO observations a year in the gaps of the baseline cadence should be sufficient to erase the gap issue and detect lags at rates consistent with those for the long season cadence. We show the medians and standard deviations of the distributions of lags recovered by {\sc javelin} and the maximum-likelihood method for light curves with the baseline \emph{u}- and \emph{y}-band cadences, baseline \emph{g}- and \emph{z}-band cadences, and baseline+LCO cadence in rows (9), (10), and (11) of Table \ref{tab:summ}, respectively.

Finally, we find that the Von-Neumann method (see Section \ref{sec:alt_methods}) is more effective at detecting long lags for mock light curves with the baseline LSST cadence than {\sc javelin} or the Fourier method. The Von-Neumann method is more accurate than these other methods in this case because it does not involve any interpolation or binning.

Figure \ref{fig:vnm_baseline50} shows the distributions of long lags recovered by the Von-Neumann method for 100 mock light curves with input lags of $-50$ and 7~days and either the long season cadence (in orange) or the baseline v2.0 cadence for the \emph{u}- and \emph{y}-band (in green). Unlike with {\sc javelin}, the Von-Neumann method is still able to detect an accurate long lag for light curves with the baseline cadence. The median long lag for both distributions are $-40^{+8}_{-7}$~days and $-40^{+1}_{-2}$~days for an input long lag of $-50$~days for the long season and baseline cadences, respectively. Therefore, the Von-Neumann method appears to be a very useful tool for recovering the long lag if the baseline cadence is selected.

\subsection{Beyond DRW Light Curves}
\label{sec:light_curves}

\begin{figure}
    \centering
    \includegraphics[width=\columnwidth]{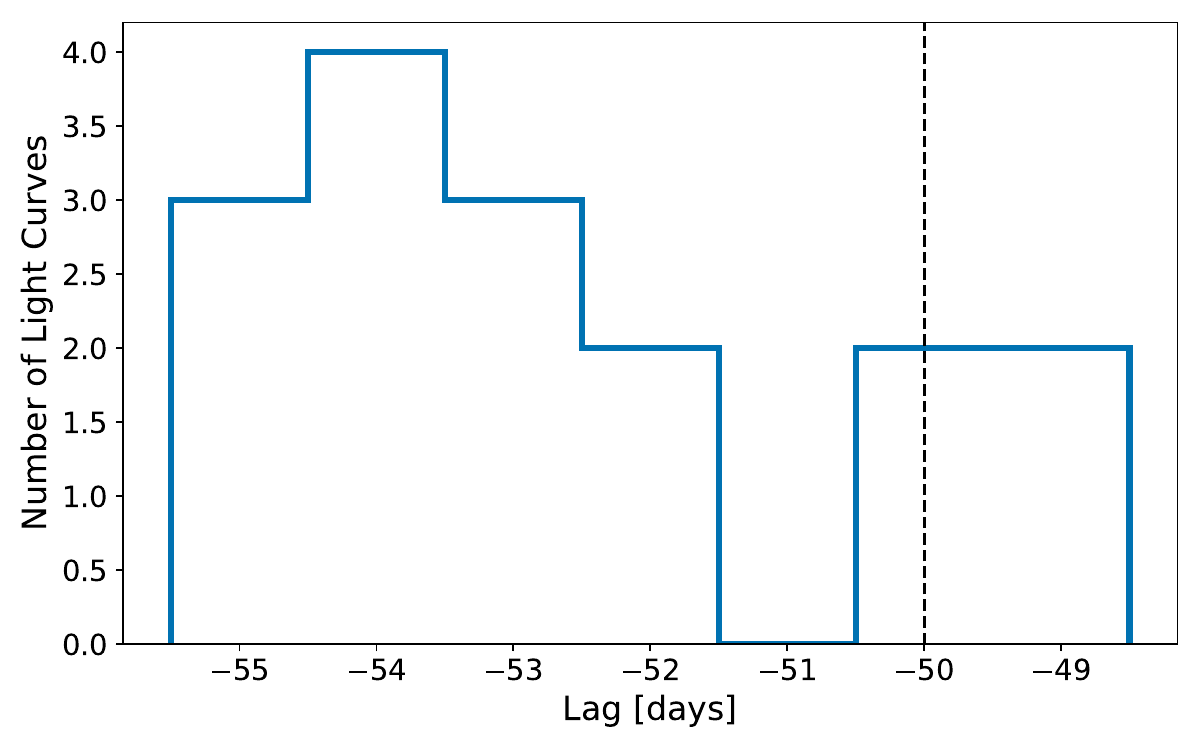}
    \\
    \includegraphics[width=\columnwidth]{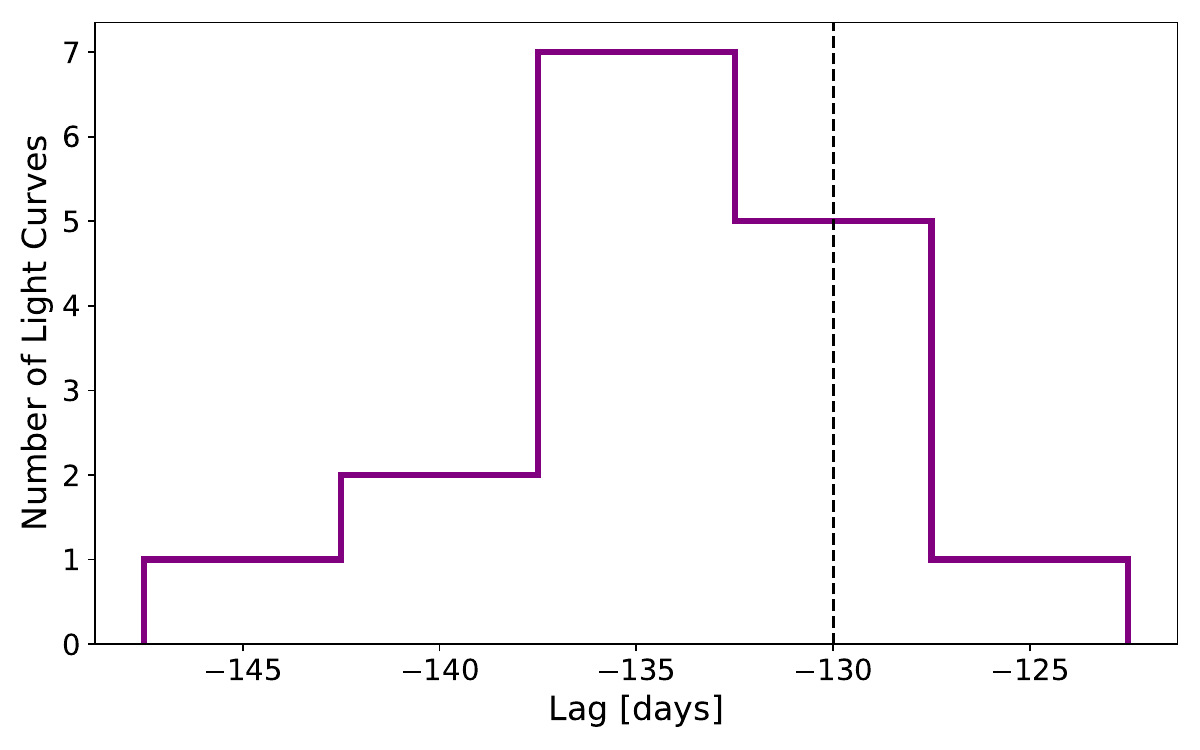}
    \\
    \includegraphics[width=\columnwidth]{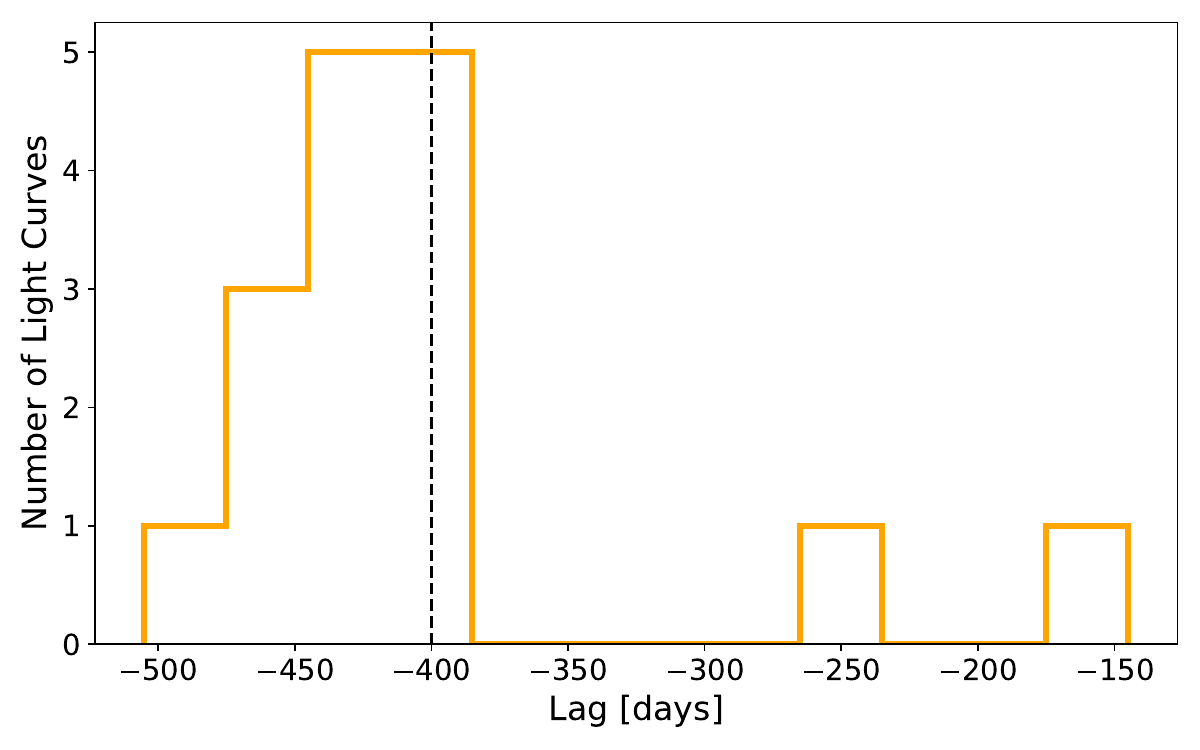}
    \caption{Distribution of the long lags recovered by {\sc javelin} for 16 mock long season cadence LSST light curves made using light curves from the simulation in \cite{JiangBlaes2020} for input lags of $-50$ and 7~days (top panel), $-130$ and 9~days (middle panel), and $-400$ and 12~days (bottom panel). The long lags are shown as the dashed vertical line.}
    \label{fig:jyf}
\end{figure}

\begin{figure}
    \centering
    \includegraphics[width=0.48\textwidth]{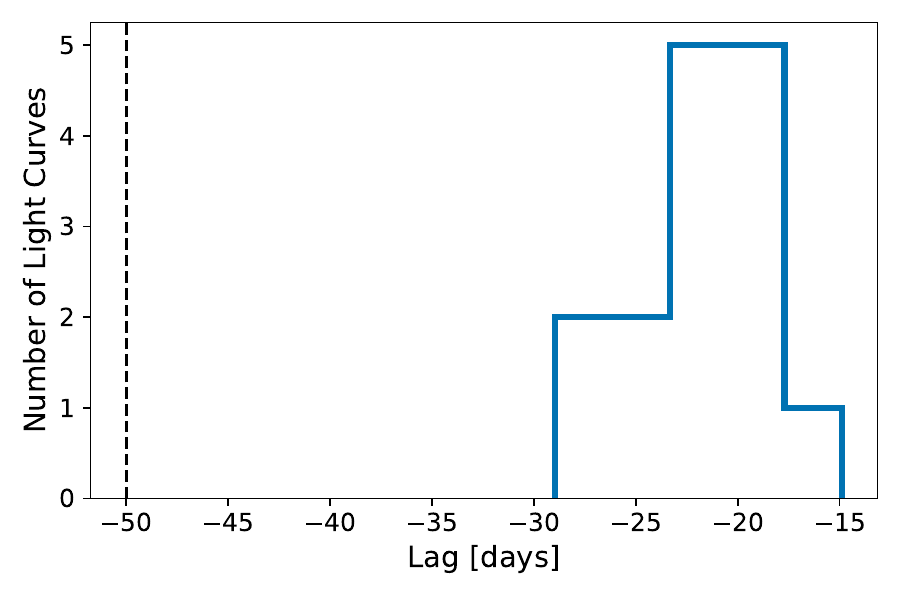}
    \\
    \includegraphics[width=0.48\textwidth]{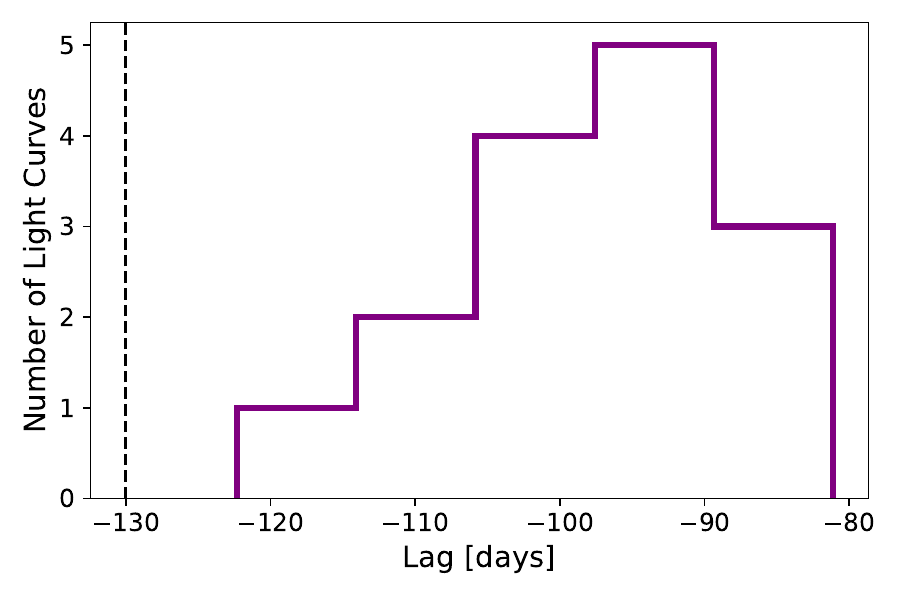}
    \caption{Distribution of the long lags detected by the maximum-likelihood method for 16 mock long season cadence LSST light curves made using light curves from the simulation in \cite{JiangBlaes2020} for input lags of $-50$ and 7~days (top panel) and $-130$ and 9~days (bottom panel). The long lags are shown as the dashed vertical line.}
    \label{fig:myf}
\end{figure}

\begin{figure}
    \centering
    \includegraphics[width=0.48\textwidth]{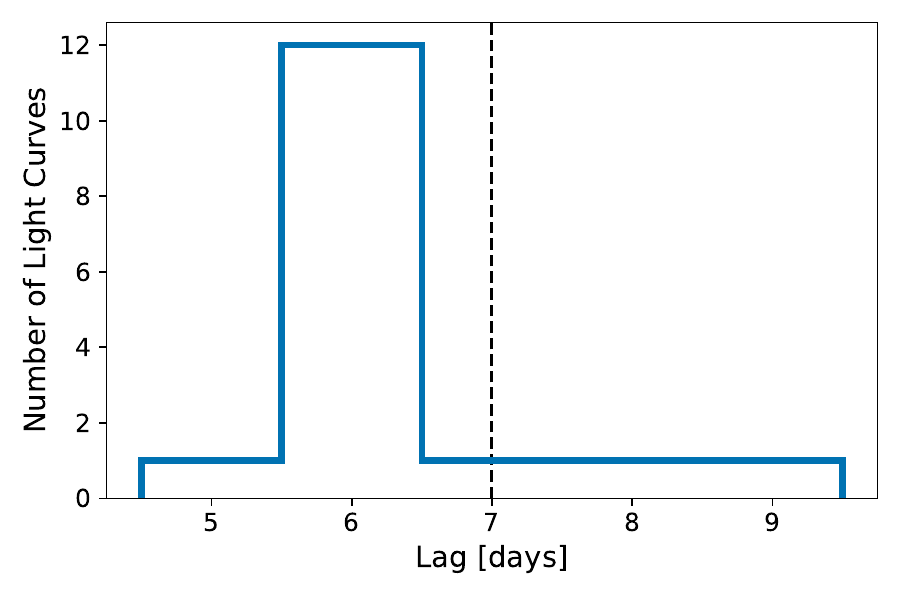}
    \\
    \includegraphics[width=0.48\textwidth]{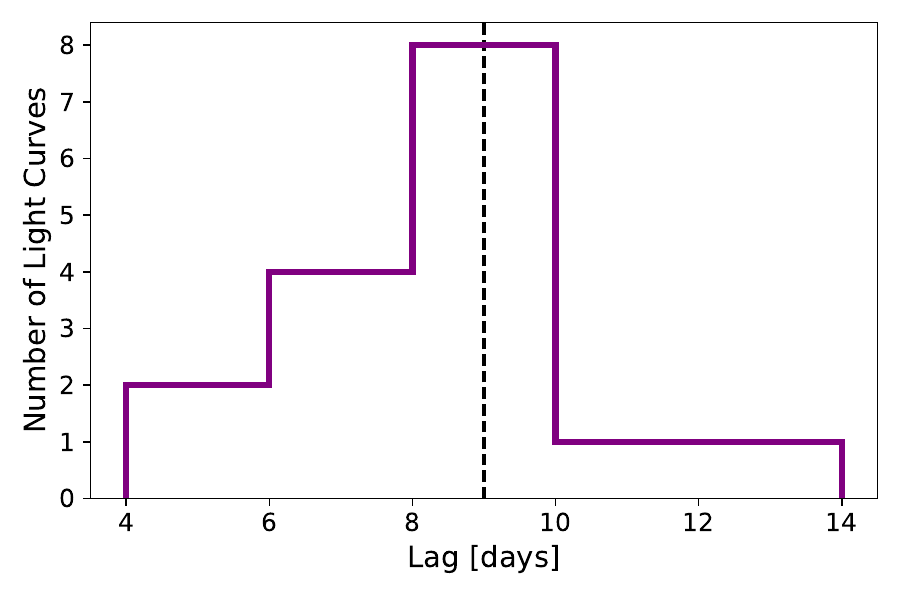}
    \caption{Distribution of the short lags detected by the maximum-likelihood method for 16 mock long season cadence LSST light curves made using light curves from the simulation in \cite{JiangBlaes2020} for input lags of $-50$ and 7~days (top panel) and $-130$ and 9~days (bottom panel). The short lags are shown as the dashed vertical line.}
    \label{fig:myfS}
\end{figure}

Although the DRW model has been shown to be a good representation of quasar light curves \citep[e.g.,][]{MacLeod2012}, the DRW model is not physically motivated and may be unreliable at the smallest and largest scales. Furthermore, since {\sc javelin} uses a DRW model to interpolate unevenly-sampled light curves, it is prudent to investigate its performance on light curves that are not generated using a DRW. Therefore, in this Section we use the light curves from the radiation MHD simulation in \cite{JiangBlaes2020} (see Section \ref{sec:sim_lcs}). These mock light curves have the long season LSST cadence and the fiducial amplitude ratio of 0.2. The simulation in \cite{JiangBlaes2020} is for a $\num{5e8}$~M$_{\sun}$ SMBH and the light curves vary in Eddington ratio from $l/l_{\rm edd}\approx 0.01 - 1$. A DRW can be fit to the light curves with $\tau_{\rm damp}=200$, although this fit is not unique.

Figure \ref{fig:jyf} shows the distribution of the median of the negative values of the {\sc javelin} distribution for the 16 mock light curves made using the simulated light curves from \cite{JiangBlaes2020}. For an input long lag of $-50$ and $-130$~days all medians fall within $\sim 10\%$ of the input lag. 14 out of 16 lags detected for light curves with input long lags of $-400$~days are within 20\% of the input long lag, with only two outliers. Both of these outliers are from light curves made using only the inner radius light curve (see Section \ref{sec:sim_lcs}), and may be from a portion of the light curve with lower variability, making it more difficult to detect the lag. Nonetheless, despite using a DRW model to interpolate light curves, {\sc javelin} still performs well on light curves that are not generated as a DRW.

We show the lags detected by the maximum-likelihood method for the 16 mock light curves with input long lags of $-50$ and $-130$ days in Figure \ref{fig:myf}. We again see that the distribution of long lags is skewed towards shorter timescales as was the case for light curves reprocessed with a long and short lag in Section \ref{sec:short_long}. The shifts in these distributions are mildly larger. The maximum-likelihood method finds long lags of $-20$~days (40\% of the input lag) versus $-25$~days (50\% of the input lag) and $-95$~days (73\% of the input lag) versus $-98$~days (75\% of the input lag) for light curves with input long lags of $-50$ and $-130$, respectively.

As with DRW mock light curves where the ratio of the amplitude of the long lag to the amplitude of the short lag is 0.2, {\sc javelin} is unable to accurately detect the short lag. The maximum-likelihood method, however, is able to detect the short lag to within 50\% accuracy for all light curves and to within a day for a majority of light curves.

 We show the medians and standard deviations of the distributions of lags recovered by {\sc javelin} and the maximum-likelihood method for these non-DRW modeled light curves in row (12) of Table \ref{tab:summ}. Using light curves from the radiation MHD simulation in \cite{JiangBlaes2020} instead of DRW light curves appears to have little affect on our ability to detect long and short lags. Because we derive similar success rates for more physically motivated light curves, we are hopeful that our results for light curves generated with a DRW model will hold for future observed LSST light curves.
 
\subsection{Dimmer Magnitude Quasars}
\label{sec:i21}

\begin{figure}
    \centering
    \includegraphics[width=\columnwidth]{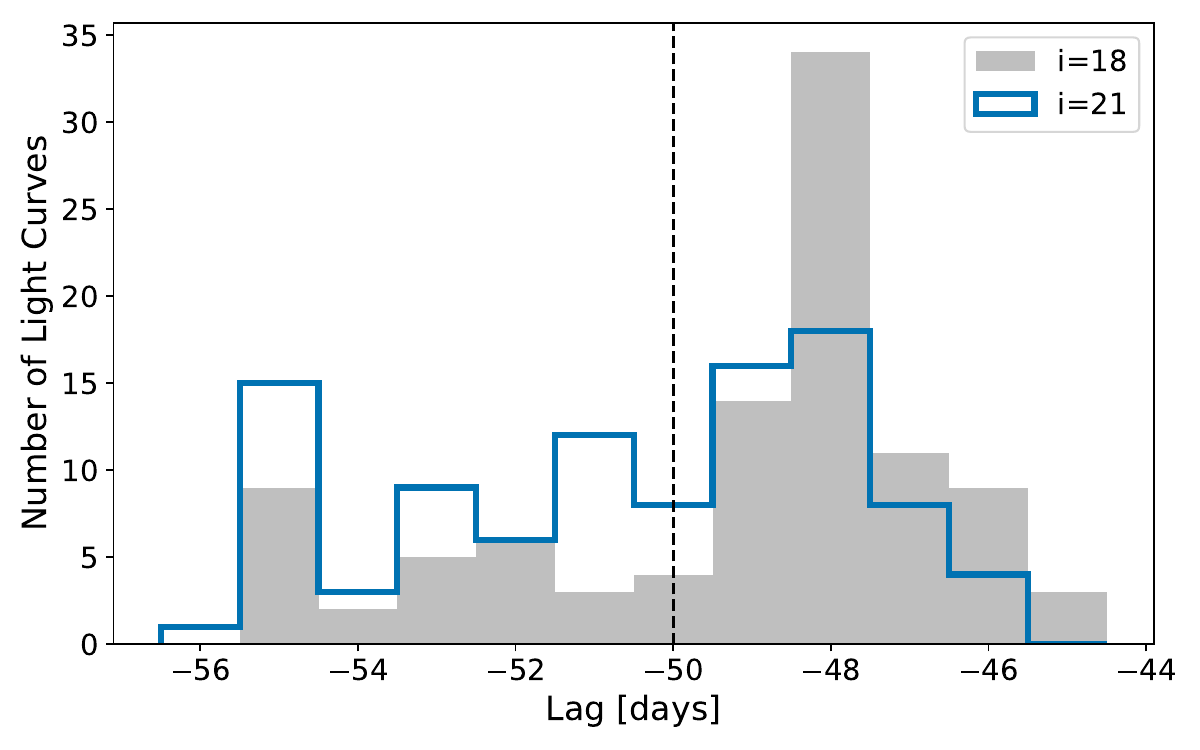} \\
    \includegraphics[width=\columnwidth]{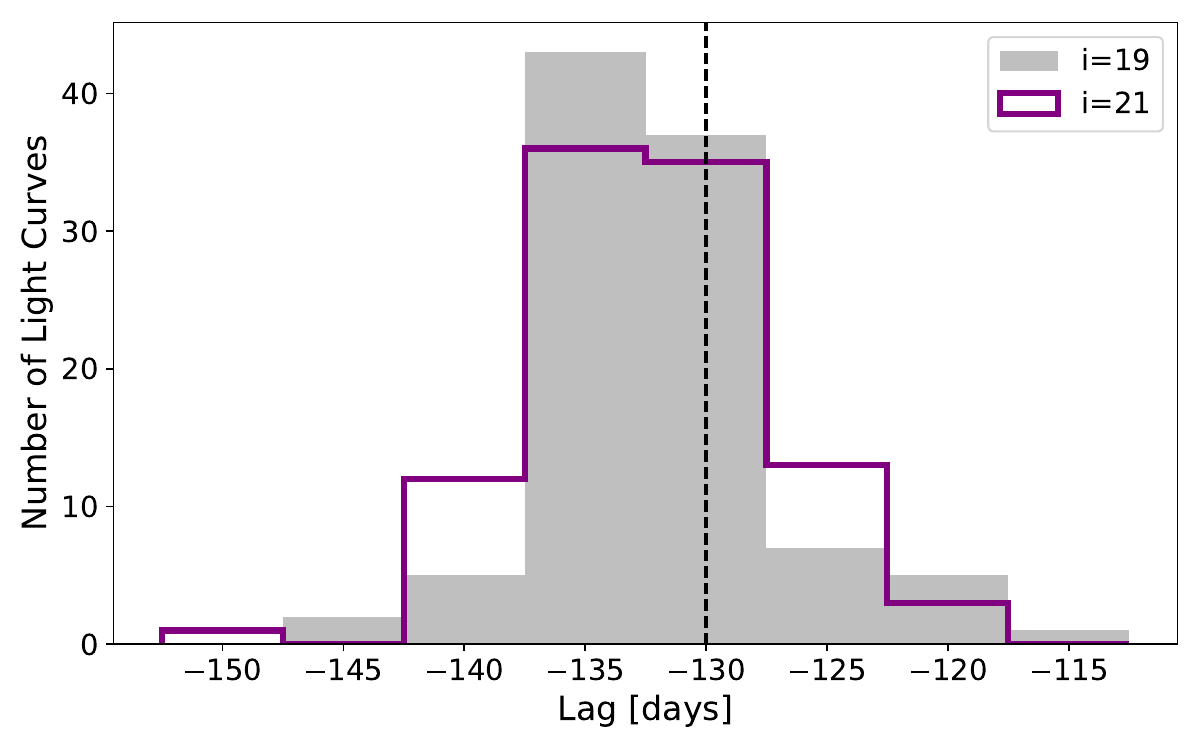} \\
    \includegraphics[width=\columnwidth]{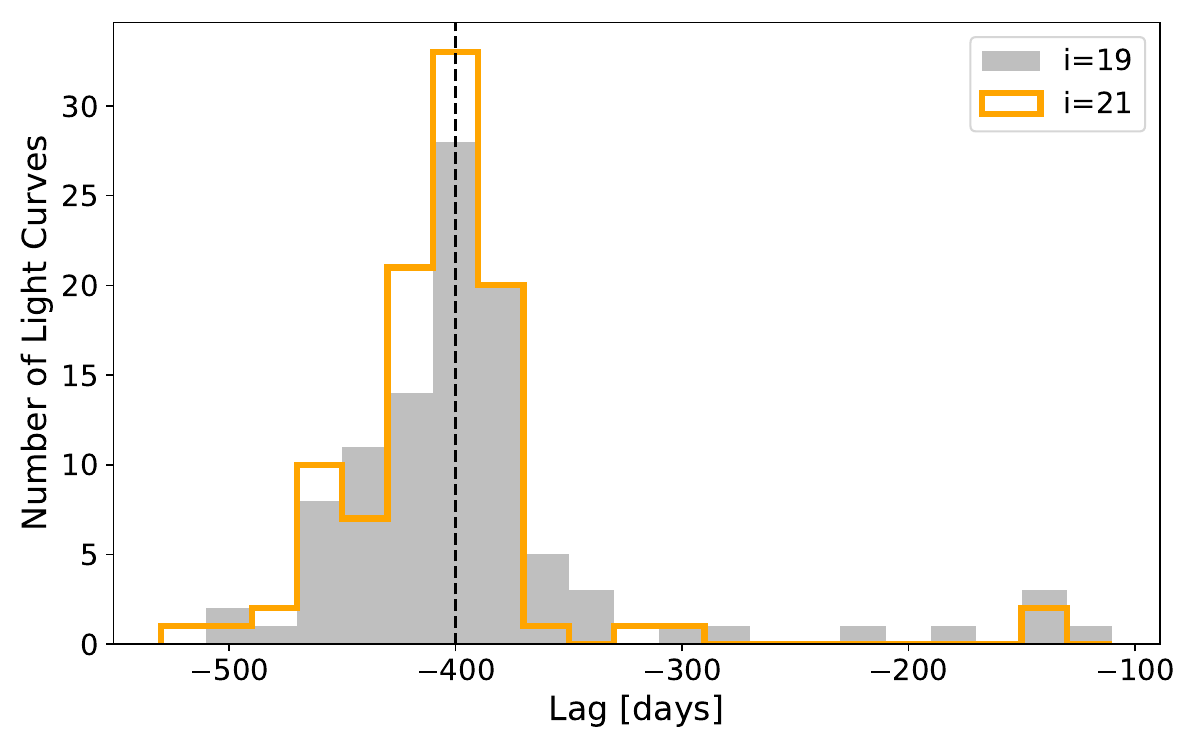}
    \caption{The distributions of the long lags recovered by {\sc javelin} for 100 mock light curves with the long season LSST cadence and input lags of $-50$ and 7 days (top panel), $-130$ and 9 days (middle panel), and $-400$ and 12~days (bottom panel). The vertical dashed lines show the input long lag.  The colored lines are the distributions for light curves where we set $i=21$ to explore how a lower SNR changes our ability to detect lags. The distributions of these lags are very similar to the distributions shown in gray which are for light curves with the \emph{i}-band magnitudes in Table \ref{tab:bins}.}
    \label{fig:jav_snr}
\end{figure}

\begin{figure}
    \centering
    \includegraphics[width=\columnwidth]{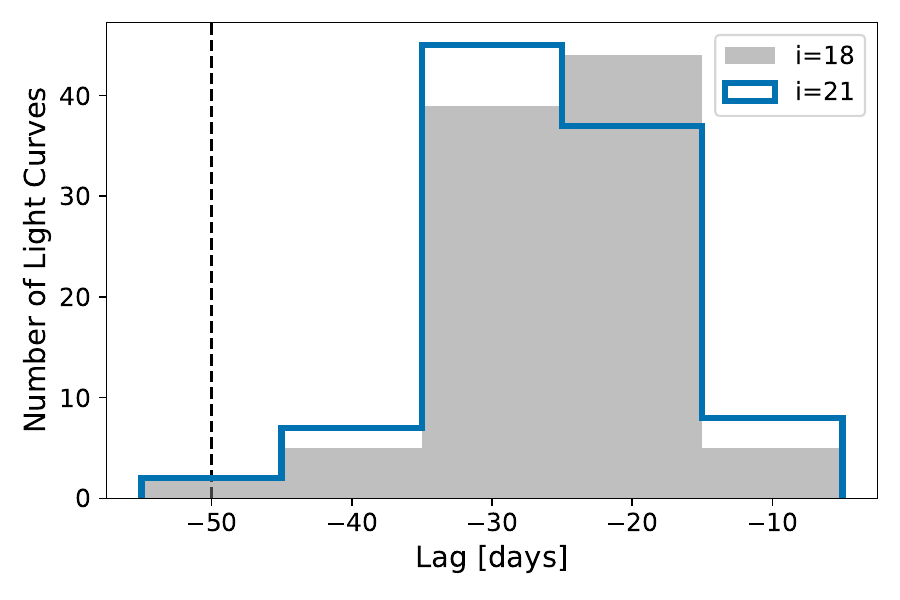} \\
    \includegraphics[width=\columnwidth]{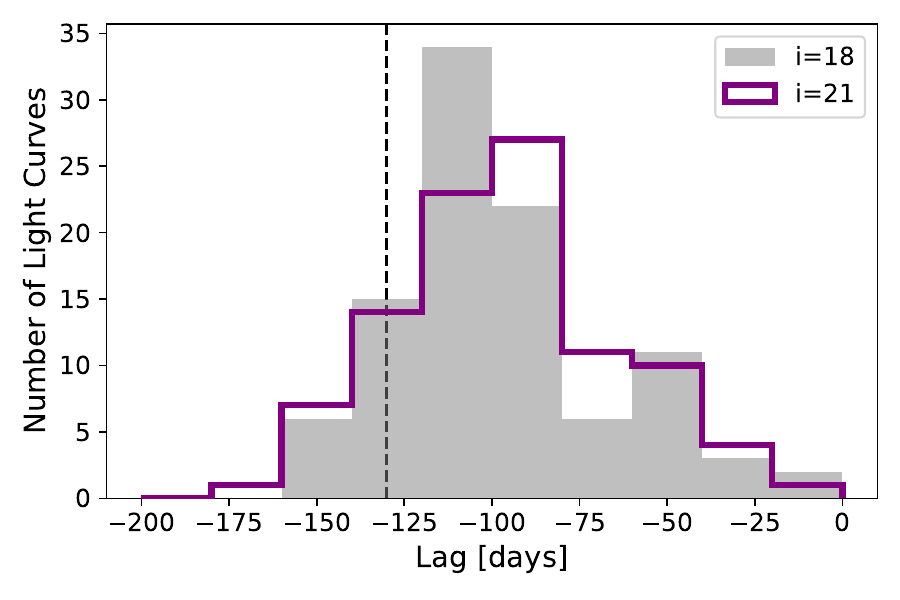}
    \caption{The distributions of the long lags recovered by the maximum-likelihood method for 100 mock light curves with the long season LSST cadence and input lags of $-50$ and 7 days (top panel) and  $-130$ and 9 days (bottom panel). The vertical dashed lines show the input long lag. The colored lines are the distributions for light curves where we set $i=21$ to explore how a lower SNR changes our ability to detect lags. The distributions of these lags are very similar to the distributions shown in gray which are for light curves with the \emph{i}-band magnitudes in Table \ref{tab:bins}.}
    \label{fig:mlm_snr}
\end{figure}

In the previous sections we test our lag recovery methods on light curves with $i<19$~mag, which have SNR of around 150. To test if we can detect lags at lower SNR, we generate mock long season cadence LSST light curves with the same input long and short lags as in the previous sections and a ratio of 0.2 between the long and short lag amplitudes, but set the band magnitudes to $i=21$ for all input lags. For LSST this band magnitude gives an SNR of around 30.

The colored lines in Figures \ref{fig:jav_snr} and \ref{fig:mlm_snr} show the distributions of lags recovered by {\sc javelin} and the maximum-likelihood method, respectively, for these mock light curves. These distributions are very similar to the distributions, shown in gray, of lags recovered by both of these methods for light curves with the band magnitudes in Table \ref{tab:bins}. The median long lags recovered by {\sc javelin} are $-50$, $-130$, and $-410$~days for input long lags of $-50$, $-130$, and $-400$~days, respectively. 

The median long lags recovered by the maximum-likelihood method for these light curves, $-26$ and $-98$~days for input long lags of $-50$ and $-130$~days, respectively, are also very similar to the median long lags recovered for light curves with brighter magnitudes. We show these median values, their standard deviations, and these values for the short lags in the bottom row of Table \ref{tab:summ}. Because the results for these dimmer magnitude quasars are nearly identical to the results in Section \ref{sec:short_long}, we can expect to be able to detect long lags for quasars brighter than $i=21$~mag, and suggest it might be possible to look for lags for even fainter quasars.

 \subsection{Summary of Results}
 \label{sec:result summary}

We now provide a summary of Section \ref{sec:detect}. First, in Section \ref{sec:long} we find that {\sc javelin} and the maximum-likelihood method both detect long lags with high accuracy in mock light curves with the LSST long season cadence that only have an input long lag. When we also reprocess these mock light curves with a short lag in Section \ref{sec:short_long}, {\sc javelin} is still able to accurately recover the long lag as long as the ratio of the amplitude of the long lag to the amplitude of the short lag is 0.2 or 0.1. The long lags recovered by the maximum-likelihood method get skewed to shorter and shorter durations as this amplitude ratio decreases, a trend we also see in our tests of the Fourier method on even cadence light curves (see Figure \ref{fig:mlm_nocad}). This trend could potentially help us to determine the amplitude ratio between the long and short lag in cases where we can make a robust long lag prediction from an alternative method. Both detection methods fail to recover the long lag for most light curves when the amplitude ratio is 0.01, although {\sc javelin} may in some cases be able to detect a long lag for this ratio when the input long lag is $-130$~days.

{\sc javelin} is unable to detect a short lag for an amplitude ratio of 0.2 for most light curves. For an amplitude ratio of 0.1, {\sc javelin} is only able to detect the short lag for a significant amount of light curves for light curves with input lags of $-50$ and 7~days. When the amplitude ratio is 0.01 {\sc javelin} is able to detect a short lag for a majority of light curves, although the short lags detected are often skewed towards longer durations. The maximum-likelihood method, on the other hand, is always able to detect the short lag, which is a real strength of the method.

In Section \ref{sec:lco} we find that {\sc javelin} performs poorly on light curves with the LSST baseline v2.0 cadence, identifying the season gaps instead of the long or short lag. The maximum-likelihood method and the Von-Neumann method perform better on light curves with the baseline cadence, seeming to be only minorly affected by the change in cadence. We show that adding just 18 observations a year from LCO in the gaps in the baseline cadence fixes the cadence-issues for {\sc javelin}, resulting in highly accurate lag detections. Adding these additional LCO observations also improves the accuracy of the short lags recovered by the maximum-likelihood method.

In Section \ref{sec:light_curves} we find that both {\sc javelin} and the maximum-likelihood method detect lags with roughly the same level of accuracy for the light curves from the radiation MHD simulation in \cite{JiangBlaes2020} as for our DRW light curves. That our methods perform this consistently on more physically motivated light curves suggests that the results in this section should be applicable to real LSST light curves in the future. Finally, in Section \ref{sec:i21} we find that when we generate dimmer mock light curves with a magnitude of $i=21$, which lowers their SNR, the distributions of lags detected by {\sc javelin} and the maximum-likelihood method are nearly identical to the distributions for the brighter magnitude mock light curves ($i<19$) in Section \ref{sec:short_long}.

\section{Number of Detectable Long Lags}
\label{sec:number}

We now provide an estimate for the number of long lags we may expect to measure with LSST. To do this estimate we will use quasar properties from SDSS and metrics developed for the robustness of BLR and short lag detections. 

First, we select only quasars at $z<1$, because for our models based on Fairall 9 Figure \ref{fig:t0z} suggests that we should be able to detect the long lags for a majority of SDSS quasars at $z<1$. Note that if $\beta=0$ and $h/r=0.01$, as is often assumed for the standard thin disk model, the long lag will be at least on the order of 100 years for bright SDSS quasars, and therefore not detectable using LSST. Next, we select only quasars with an \emph{i}-mag brighter than 21. In Section \ref{sec:i21} we show that our lag detection methods are accurate down to this magnitude.

Whether a long lag is detectable or not also depends on the variability of the quasar. When looking for BLR lags in SDSS quasar light curves, \cite{Grier:2017} defined a criterion SNR2 based on the continuum and line light curve rms variability. The rms variability is the intrinsic variability of the light curve about a fitted linear trend, divided by the uncertainty of the estimated intrinsic variability.

To detect C IV lags, \cite{Grier:2019} require an SNR2 $>20$. They found that 348 out of 492 ($\sim 71\%$) quasars in their sample from the SDSS reverberation mapping project that fall within $1.35 < z < 4.32$ met that criterion. \cite{Homayouni:2020} look for Mg II lags for 453 quasar light curves with redshifts more similar to our target redshifts here, $0.35 < z <1.7$, and find that 198, or $\sim 44\%$, have an SNR2 $>20$. The number of quasars with SNR2 $>20$ is fewer in this case because Mg II tends to be less variable. 

A second criterion for determining lag significance for both BLR and short continuum lags is setting a minimum threshold for $r_{\rm max}$, the maximum ICCF correlation coefficient (see Section \ref{sec:alt_methods}). Setting a minimum threshold for $r_{\rm max}$ ensures that the light curves are well correlated. \cite{Guo:2022} require $r_{\rm max}>0.8$ for their sample of $z<0.8$ quasar light curves from the Zwicky Transient Facility. Of 4255 quasars, 559 (13\%) met that criterion. 

SDSS DR14 has $\sim \num{5e5}$ quasars over 9376~deg$^2$ \citep{paris2018}. Limiting ourselves to  $z<1$ and $i<21$~mag quasars we are left with $\sim \num{9.5e4}$ quasars. Accounting for the five planned 9.6~deg$^2$ DDF, we can expect to detect around 500 of these quasars within the DDFs. If we take the more conservative fraction of quasars with SNR2 $>20$ from \cite{Homayouni:2022} we find that around 200 quasars should have detectable lags. If we assume only 13\% of light curves will have $r_{\rm max}>0.8$ as in \cite{Guo:2022} we would be able to detect long lags for 65 quasars within the DDFs.

Here, we only look at $z<1$ quasars because the long lag gets very long at $z>1$ and LSST has a ten year baseline. However, it is possible LSST will operate for 20 years. In addition, at higher redshifts we could potentially use a narrower range of wavebands (for example, \emph{g}-\emph{i}) to detect lags with lengths similar to those tested here for $z<1$. Finally, we could also expand on the 10 year LSST baseline by using lower cadence archival data. Future work should examine if such a cadence enables lag detection. If we use the same \emph{i}-band magnitude cut-off for SDSS DR14 quasars out to $z<2$ we would roughly double the number of detectable long lags. 

The Wide-field Infrared Survey Explorer and eROSITA provide other catalogues of quasars not included in SDSS. In particular these catalogues could be useful for studying lower luminosity, and thus lower mass, or redder, more dust-obscured quasars. Since $\tau_{\rm long} \propto M^{1/2}$, the lag timescales for these quasars will be shorter. Shorter lag timescales could help us detect a higher yield of longs lags for lower mass quasars, although this exact yield is difficult to estimate.

Even with our conservative estimates, we should be able to detect long lags for dozens to hundreds of new quasars over the ten year lifetime of LSST. In addition, \cite{Yao:2022} were able to potentially detect a lag of $\sim 70$~days using light curves with a less than 1 year baseline, so we speculate that we should be able to detect long lags in some fraction of nearby quasars with durations $\lesssim 50$~days within the first 1--3 years of LSST. Roughly a third of SDSS quasars with $i<21$ and $z<1$ are within $z<0.5$, where we expect lags to have durations of $\lesssim 50$~days. Therefore it could be possible to detect long lags for roughly 20--100 quasars within the first couple of years of LSST.

\section{Discussion and Summary}
\label{sec:discuss}

In this Section we discuss caveats related to our simulated light curves (Section \ref{sec:discuss:lcs}) and lag detection methods (Section \ref{sec:discuss:detect}) that we plan to address in future work. We then provide a summary of the paper in Section \ref{sec:discuss:summary}.

\subsection{Improved Simulated Light Curves}
\label{sec:discuss:lcs}

In this paper we create mock observations of our driving and reprocessed light curves with both the proposed LSST baseline v2.0 and longest season cadence from OpSim. While OpSim models SNR and observing gaps by taking into account sky brightness, bad weather predicted from localized weather models, seeing, telescope maintenance times, slew times, and filter changes, the cadences and SNR measurements may still be optimistic. However, small changes in cadence should not have a large impact on our ability to detect lags on the order of months. In addition, our tests in Section \ref{sec:i21} using SNR corresponding to $\emph{i}=21$~mag light curves show that even if the SNR corresponding to the magnitudes in Table \ref{tab:bins} is too high by an order of magnitude, we should still be able to recover lags with similar accuracy.

We use two types of light curves, self-generated DRW light curves and light curves from the simulation in \cite{JiangBlaes2020}. While often used to model quasar light curves, the DRW models are not physically motivated and may fail to model the quasar PSD properly at short and long timescales. In particular, PSD from highly sampled light curves from the Kepler mission appear to have a steeper slope at short timescales \citep{Mushotsky:2011,Kasliwal:2015,Smith2018} and the length of the light curve may have an effect on the PSD at long timescales \citep{Stone2022}. On the other hand, the simulated light curves from \cite{JiangBlaes2020} are not intended to capture quasar variability for a wide range of quasar parameters. Therefore, future work to simulate light curves for a wider range of quasar parameters is needed to better understand our ability to detect long lags in their variability.

We reprocess our light curves with a Gaussian response function. However, the shape of the response function for a long lag is unknown. For this paper we set the width of the Gaussian to $S\equiv C/5$ (see Equation (\ref{eq:gaussian})), but preliminary tests suggest that {\sc javelin} has more difficulty accurately recovering lags for response functions with larger widths ($S \gtrsim C/2$). In addition, using other functional forms for the response function, such as the log normal distribution, may make it more difficult to detect a long lag using {\sc javelin}. Therefore the width and shape of the response function are important parameters. Here, we use the default {\sc javelin} response function model, a top hat with no priors on the width. It is possible that adjusting the model or priors for the width of the response function could improve the accuracy of {\sc javelin}. Still, {\sc javelin} may not be the best tool if detecting long lags relies on finding lags in variability that is reduced or smeared as it moves inwards. The maximum-likelihood method is better designed to capture the shape of the response function and should therefore be able to handle broader or more complex response functions. 

Our response function is centered on frequencies corresponding to the timescale of the long lag. However, if the perturbations that are accreted inward leading to the long lag are on the thermal time scale it may be that, unlike the short lag timescale, the frequencies most affected by the long lag do not correspond exactly to the inverse of the long lag timescale. The thermal timescale is similar to the viscous timescale we use to determine the long lag timescale, for the large aspect ratios in our model. However, a future improvement would be to experiment with the recoverability of long lags depending on the frequencies most affected by the long lag.

Another simplification is that we model the radial dependence and value of the aspect ratio as independent of SMBH mass or Eddington ratio, when that is very likely not the case. Most likely, as the Eddington ratio of a quasar increases, its aspect ratio will increase, which will shorten its long lag timescale.

Ultimately, simulations of the reprocessing of high frequency light in the UV/optical emitting regions of turbulent quasar disks for a wide range of quasar parameters are needed to better model a more accurate response function for the long lag. To perform these simulations it will be necessary to use a multi-band radiation MHD code, such as the code developed by \cite{jiang2022} for {\sc athena ++}. Radiation MHD is necessary to capture the Maxwell stress, magnetic pressure, radiative viscosity, and radiation pressure, all of which may play a role in locally generated turbulence that can be accreted inwards on the viscous timescale \citep{JiangBlaes2020}. The multi-band capabilities will be useful because it will allow us to inject high-frequency radiation and compare it to the radiation reprocessed and emitted by the local disk.

\subsection{Improved Lag Detection Methods}
\label{sec:discuss:detect}

Here we use two main lag detection methods. The first, {\sc javelin}, very accurately detects long lags when the ratio between the amplitude of the long lag and the amplitude of the short lag is $\gtrsim 0.1$. On the other hand, it performs very poorly when this amplitude ratio is 0.01. {\sc javelin} has been shown previously to be biased in the presence of multiple lag signals \citep[e.g.,][]{Cackett:2018}. 

We can potentially improve the performance of {\sc javelin} by simultaneously fitting several reprocessed light curves, instead of just fitting one driving and one reprocessed light curve. Fitting several reprocessed light curves will be possible for LSST, which will have six different wavebands. With multiple bands, {\sc javelin} will be less likely to misidentify seasonal gaps as lags when the lags of the different wavebands are significantly different from each other \citep{Zu2016}, which would be especially helpful if the baseline v2.0 cadence is selected. However, in our model for the long lag, the lag timescales between longer wavelength bands becomes very short. As a result, fitting multiple wavebands simultaneously may not be as helpful as it is when fitting BLR lags, which differ significantly from band to band. Another potential improvement would be to use a custom model for the response function in {\sc javelin} that includes a positive and negative lag, instead of using the default model which is just a single top hat function.

The maximum-likelihood method is less accurate overall than {\sc javelin} at detecting the long lag. The long lags also become increasingly skewed towards zero as the amplitude of the short lag becomes stronger relative to the long lag. Figure \ref{fig:mlm_nocad} shows that this skewing towards zero is due to the shape of the combined long and short lag response function. However, because of this bias, it will be difficult to disentangle the exact value of the long lag versus the ratio of the long lag amplitude to the short lag amplitude. Utilizing the full frequency dependent information instead of summarizing the long lag with a single average value may help disentangle the two and reveal more details about the shape of the response function. More details could also be obtained by increasing the number of bins. In this paper, we only use four frequency bins for the Fourier method, because of the relatively low sampling rate of LSST, and because it is easier to detect a single lag with fewer bins.

There are several ways that using all of the frequency dependent lag information can provide additional information on the shape of the response function. For one, regardless of the ratios between the long and short lags, the wrapping frequency seems to consistently occur very near $1/2\tau_{\rm long}$ in Figure \ref{fig:mlm_nocad}. Therefore, we may be able to more effectively exploit the location of the wrapping frequency to derive the lag. In addition, as can be seen in Figure \ref{fig:mlm_nocad}, for larger amplitude ratios even if the value of the long lag at lower frequencies is skewed towards zero, right before the lag wraps it drops to roughly the value of the input long lag. If, on the other hand, the lags increase towards zero or positive values as the lag approaches the wrapping frequency, as is the case for smaller amplitude ratios, that would provide information on the possible presence of a long lag with a small amplitude relative to the short lag.

Overall, there is room for improvement in the lag recovery techniques used here, especially if we want to detect long lags with amplitudes less than 10\% of the amplitude of the short lag. We have mentioned several ways our employment of {\sc javelin} and the maximum-likelihood method could be improved. It is possible that alternative methods used for detecting lags in quasar light curves may be more accurate. The Von-Neumann method, which we also test on our mock light curves, seems like a particularly promising alternative, especially if the baseline cadence is chosen for LSST (see Section \ref{sec:lco}). Ultimately, the greatest improvement may be to create a new method, directly tailored to detecting the long lag.

\subsection{Summary}
\label{sec:discuss:summary}

Inspired by the recent potential detection of a long lag in Fairall 9, we have laid out a model for the timescale of negative long lags, which occur due to turbulence in the outer disk and propagate inward on the viscous timescale. Because the viscous timescale depends on the aspect ratio of the disk, measuring long lags can provide information on the vertical structure of the disk. Based on the measured long lags in \cite{Yao:2022} for Fairall 9, we find that the scale height depends on the radial disk position as $h \propto r^{3.59}$. However, additional long lags need to be detected to properly constrain our initial model for the radial dependence of the scale height. We therefore hope to use LSST to detect these additional long lags.

To study whether LSST will be able to detect long lags, we simulate mock LSST light curves using both DRW modeled light curves and light curves from the simulation in \cite{JiangBlaes2020}, which we then mock observe with several LSST cadences from OpSim. We reprocess the lagging light curve with a Gaussian response function. We then test several lag detection methods, including the ICCF \citep{Gaskell1987,WhitePeterson1994,Peterson2004}, a Von-Neumann estimator presented by \cite{Chelouche2017}, {\sc javelin} \citep{Zu2011}, and a maximum-likelihood Fourier method \citep{Zoghbi2013}, on our mock light curves. We primarily focus on two main lag detection methods. The first is the commonly used {\sc javelin} code. We find that when only the long lag is included (Section \ref{sec:long}), {\sc javelin} is accurate to within 10\% of the input lag for all light curves with input long lags of $-50$ and $-130$~days, and accurate to within 10\% of the input lag for 79\% of light curves with input lags of $-400$~days in our DRW modeled light curves. 

When we include a short lag as well (Section \ref{sec:short_long}) {\sc javelin} remains very accurate, unless the ratio of the amplitude of the long lag to the amplitude of the short lag is around 0.01, at which point the accuracy decreases dramatically. Conversely, the short lag is not detected by {\sc javelin} unless this ratio $\le 0.1$ (for an input long lag of $-50$~days) or $\le 0.01$ (for input long lags of $-130$ and $-400$~days). {\sc javelin} detects long lags with a similar level of accuracy even when using a DRW model to interpolate the non-DRW modeled light curves from the simulation in \cite{JiangBlaes2020} in Section \ref{sec:light_curves}.

Because {\sc javelin} is not aimed at detecting two lags simultaneously we also use a maximum-likelihood Fourier method, which separates lags in frequency space. Separating lags in frequency space should make it easier to detect multiple lags. However, as the amplitude of the long lag decreases relative to the amplitude of the short lag, the influence of the short lag on lower frequencies increases, skewing the recovered long lag towards shorter durations. When the amplitude ratio between the long and short lags is 0.01 the maximum-likelihood method is unable to identify the long lag at all for most light curves, just like {\sc javelin}. On the other hand, the maximum-likelihood method provides a significant advantage when it comes to detecting the short lag, which it is able to accurately detect for all amplitude ratios. Finally, the performance of this method on the light curves from \cite{JiangBlaes2020} is comparable to its performance on DRW light curves.

Based on the results in Section \ref{sec:lco} and other cadence tests we performed, we believe a long season cadence\footnote{including strong rolling or weak accordion cadences which both have long seasons} is helpful for detecting long lags. A long season cadence is particularly beneficial when using {\sc javelin} to detect long lags, whereas the Von-Neumann method and the maximum-likelihood method for shorter duration long lags ($\sim -50$~days) are able to detect long lags with similar accuracy regardless of the cadence. However, adding just 18 LCO observations annually to mock light curves sub-sampled with the baseline v2.0 cadence, leads to a comparable or higher lag detection rate than for the long season cadence, for both our main methods for all lag durations. Additional LCO observations or the Von-Neumann method may also be effective tools for finding long lags in the Wide Fast Deep Survey, which will cover a much larger area of the sky at lower cadence.

Even with our conservative estimates in Section \ref{sec:number}, we should be able to detect long lags for dozens to hundreds of new quasars over ten years. These detections will significantly improve our ability to model the long lag as a function of wavelength, SMBH mass, and Eddington ratio. These models will help us constrain the vertical structure of quasar disks for a range of quasar parameters.

\acknowledgements
 The authors would like to thank both anonymous referees for increasing the clarity and thoroughness of the paper. The authors would also like to thank Ashley Villar for help and advice on using OpSim cadences and Francisco Pozo Nu{\~n}ez for his help providing a python version of the Von-Neumann method. A.S. is supported by a fellowship from the NSF Graduate Research Fellowship Program under Grant No. DGE-1656466. JEG is supported in part by NSF grants AST1007052 and AST1007094. The Center for Computational Astrophysics at the Flatiron Institute is supported by the Simons Foundation. Part of the work was done when YFJ was attending the Binary22 program in KITP, which was supported in part by the National Science Foundation under Grant No. NSF PHY-1748958. AZ is supported by NASA under award number 80GSFC21M0002

\bibliography{agn_lag.bib}{}
\bibliographystyle{aasjournal}

\appendix

\section{Full Tests}
\label{app:full}
In Sections \ref{sec:short_long} and \ref{sec:lco} we show the distributions of lags detected by our various methods only for light curves with input lags of $-50$ and 7~days, for the sake of brevity. Here we present the same results, but for all lag bins in Table \ref{tab:bins}.

\subsection{Long and Short Lags}
\label{app:long_short}
\begin{figure*}[h]
    \centering
    \includegraphics[width=\textwidth]{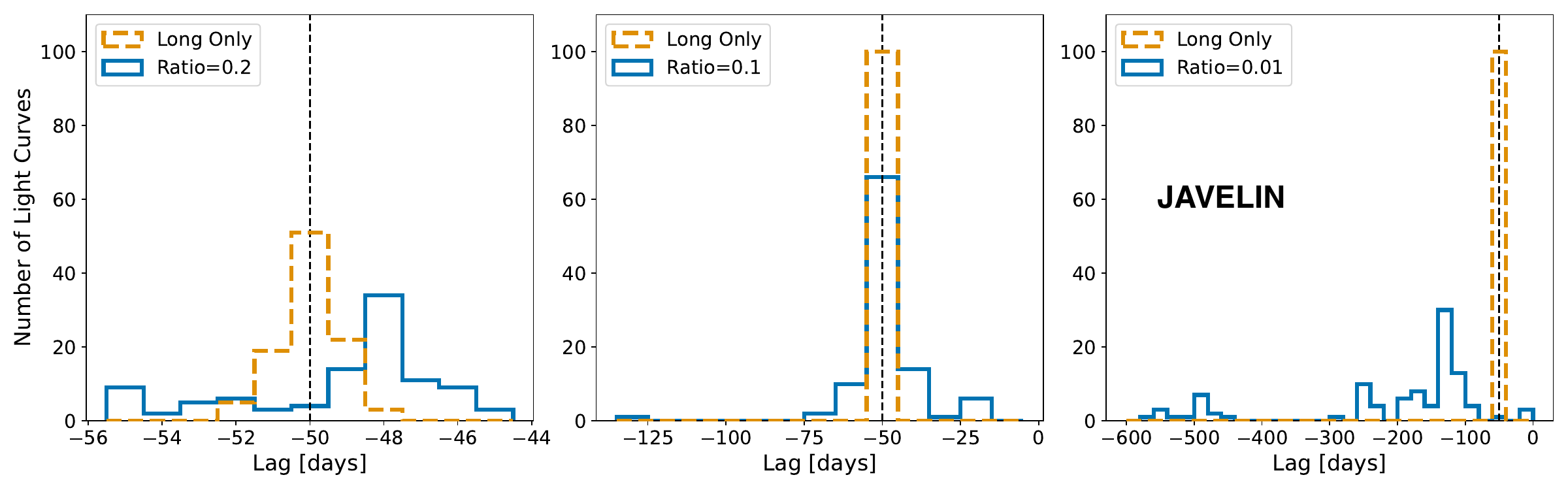}
    \\
    \includegraphics[width=\textwidth]{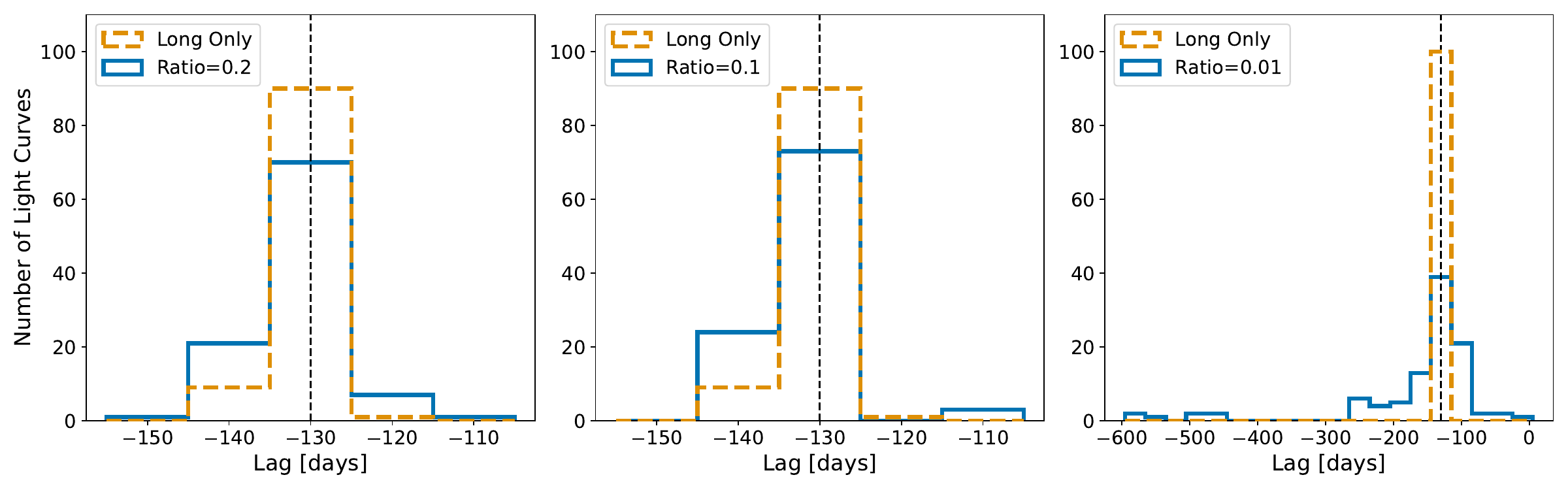}
    \\
    \includegraphics[width=\textwidth]{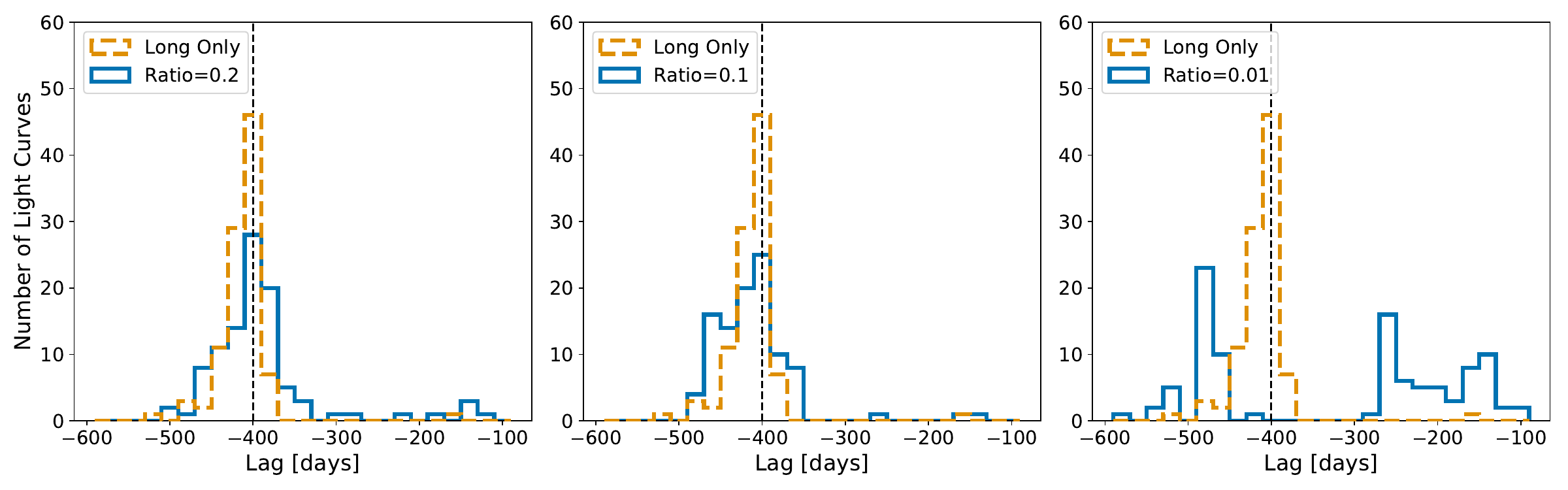}
    \caption{The distribution of the long lags detected by {\sc javelin} for 100 mock long season cadence LSST light curves with input long lags, shown by the dashed vertical lines, of $-50$ days (top panel), $-130$ days (middle panel), and $-400$ days (bottom panel). Distributions shown in blue also include a short lag of 7 days (top panel), 9 days (middle panel), and 14 days (bottom panel). The ratio of the amplitude of the long lag to the amplitude of the short lag is shown in the legends for each blue distribution.}
    \label{fig:jav_both}
\end{figure*}

 We show the distributions of the long lags detected by {\sc javelin} for 100 mock long season cadence LSST light curves with input long lags of $-50$~days (top panel), $-130$~days (middle panel), and $-400$~days (bottom panel) in Figure \ref{fig:jav_both}. Distributions shown in orange have only long lags, while distributions shown in blue have long and short lags with varying ratios between the amplitude of the long and short lag. The accuracy of the distributions for different input long lags is mostly consistent, with the exception of the distribution for light curves with input long lags of $-130$~days and an amplitude ratio of 0.01, which appears to be more accurate than the other distributions (see Section \ref{sec:short_long} for more discussion).

\begin{figure*}
    \centering
    \includegraphics[width=\textwidth]{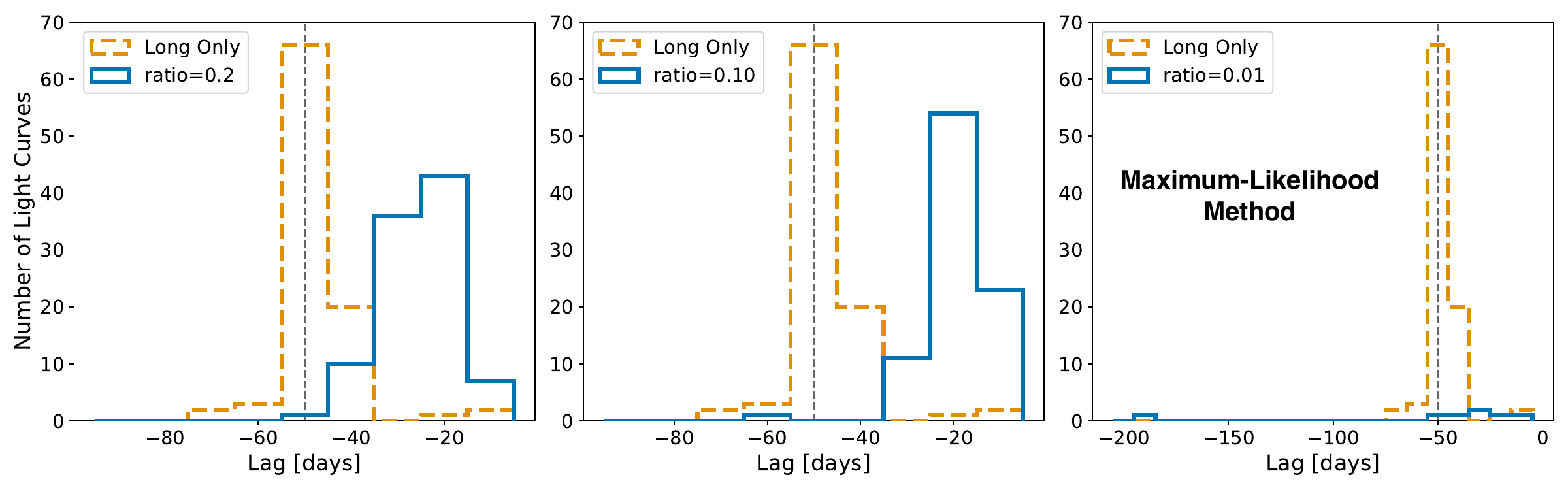} \\
    \includegraphics[width=\textwidth]{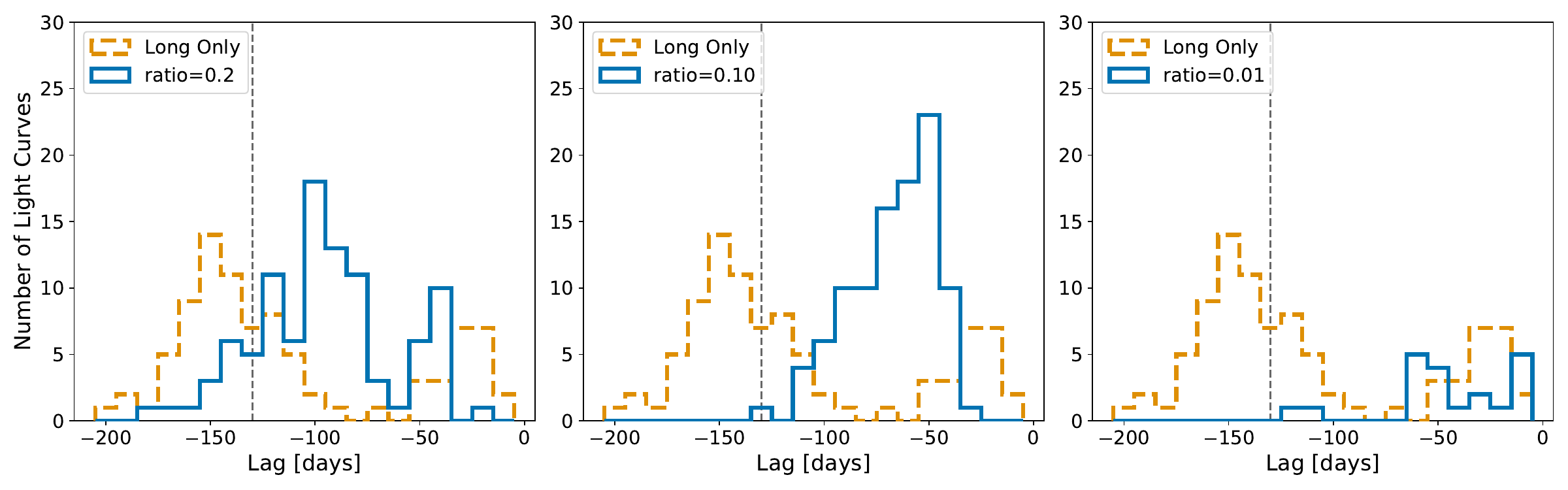}
    \caption{Distribution of the long lags recovered by the maximum-likelihood method for 100 mock long season cadence LSST light curves with input long lags, shown by the dashed vertical lines, of $-50$ days (top panel) and $-130$ days (bottom panel). Distributions shown in blue also include a short lag of 7 days (top panel) and 9 days (bottom panel). The ratio of the amplitude of the long lag to the amplitude of the short lag is shown in the legends for each blue distribution.}
    \label{fig:mlm_both}
\end{figure*}

Figure \ref{fig:mlm_both} is the same as Figure \ref{fig:jav_both} but for the maximum-likelihood method. The results for light curves with an input long lag of $-130$~days are mostly consistent with the results for an input long lag of $-50$~days. The maximum-likelihood does do slightly better at detecting some negative lags for an amplitude ratio of 0.01 for $-130$~day lags. This improvement is expected based on our tests of the Fourier method with even-cadence light curves shown in Figure \ref{fig:mlm_nocad}. It appears that because the long lag is greater for light curves with $-130$~day input lags, it is not skewed all the way positive by the short lag.

\begin{figure}
    \centering
    \includegraphics[width=0.32\textwidth]{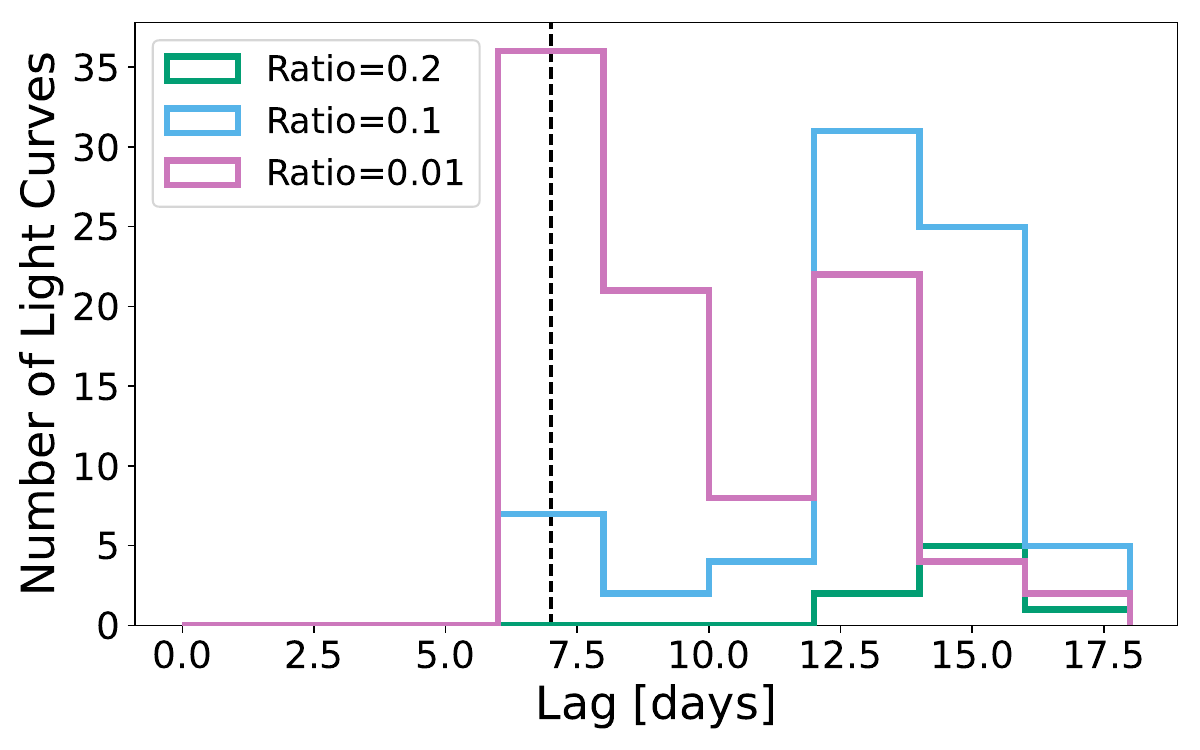}
    \includegraphics[width=0.32\textwidth]{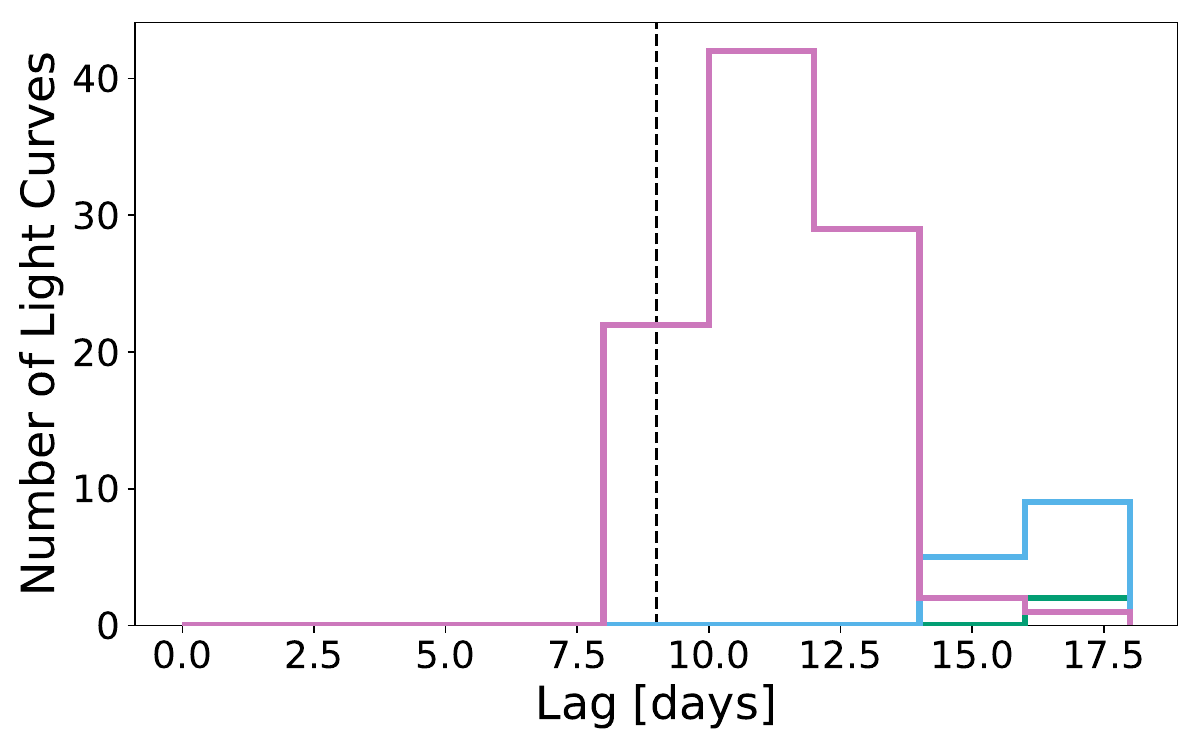}
    \includegraphics[width=0.32\textwidth]{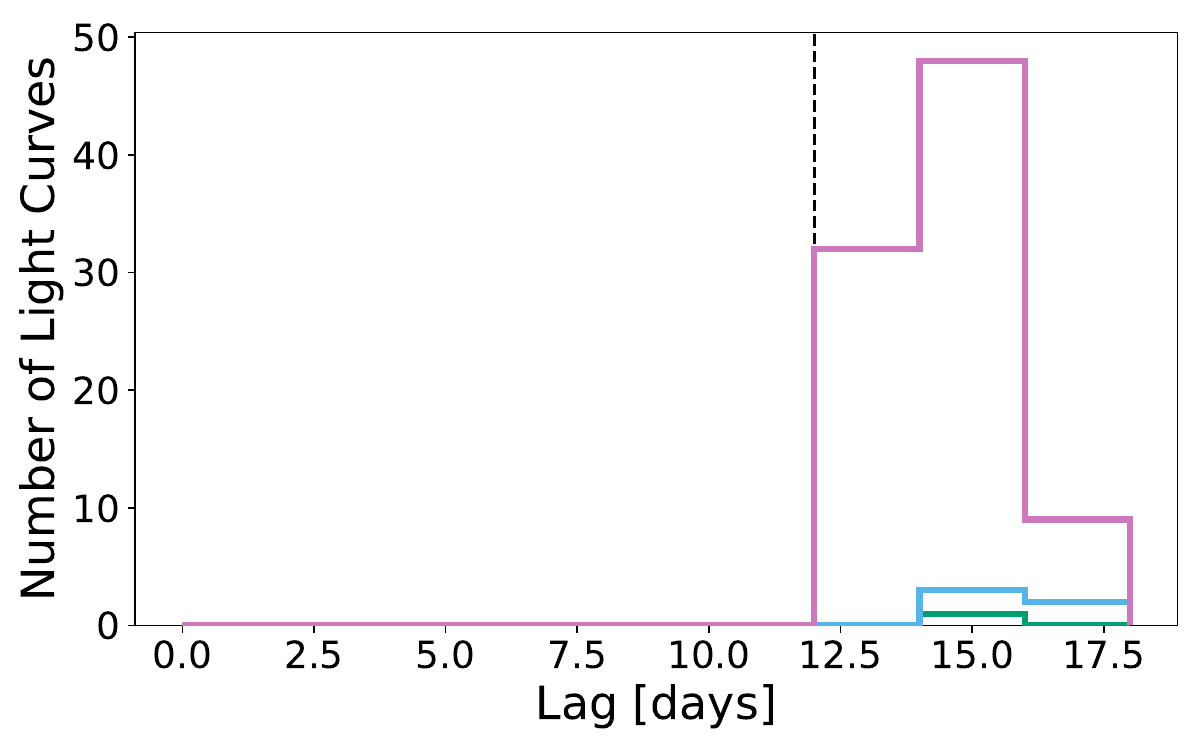}
    \caption{The distribution of the short lags detected by {\sc javelin} for 100 mock long season cadence LSST light curves with different amplitude ratios (shown in the legend). We zoom in on the portion of the distribution between 0 and 20 days to better show the accuracy of the short lag when detected. The input short lags, shown as the vertical dashed line, are 7, 9, and 12 days for the left, middle, and right panels, respectively.}
    \label{fig:jav_short}
\end{figure}

\begin{figure}
    \centering
    \includegraphics[width=0.45\textwidth]{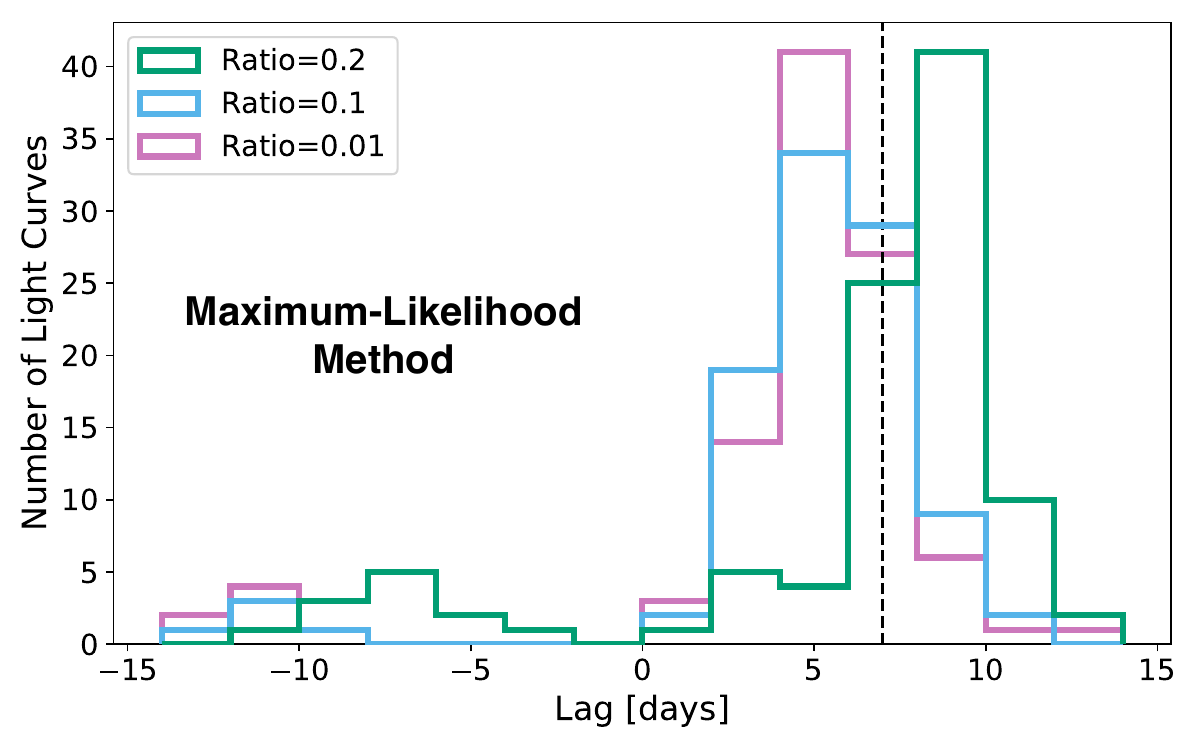}
    \includegraphics[width=0.45\textwidth]{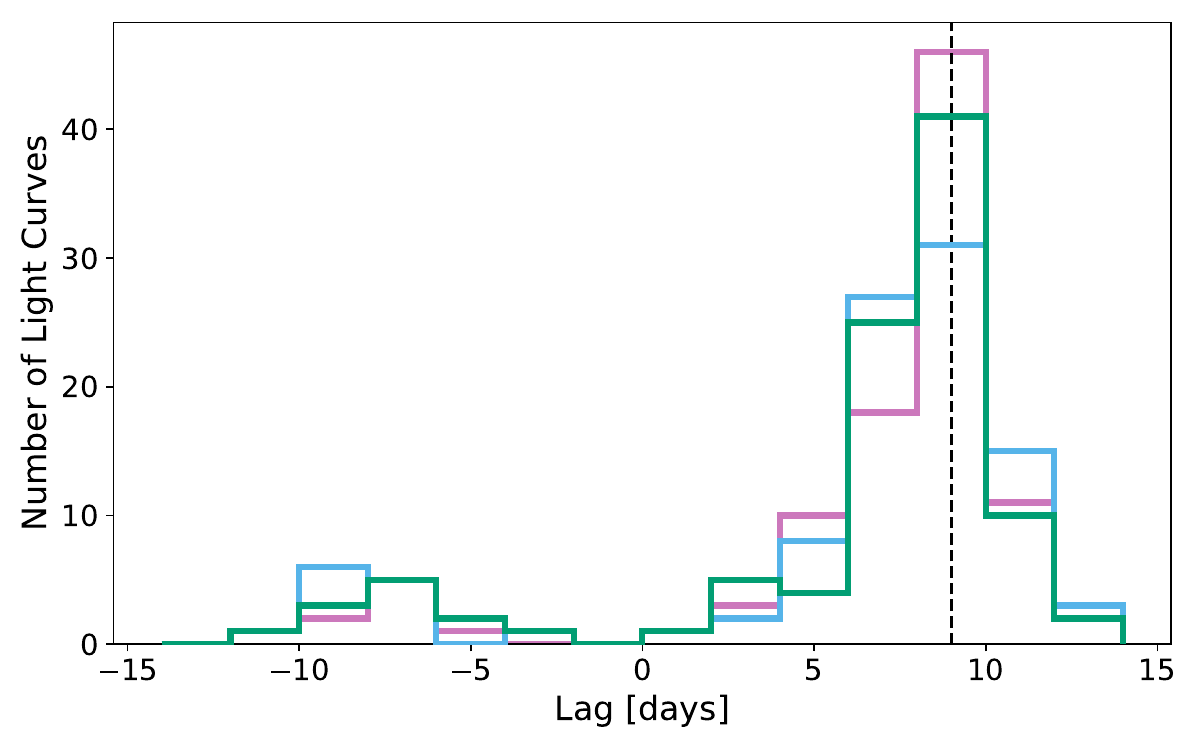}
    \caption{Distribution of the short lags recovered by the maximum-likelihood method for 100 mock long season cadence LSST light curves with different amplitude ratios (shown in the legend). The input short lags, shown as the vertical dashed line, are 7 days in the left panel and 9 days in the right panel.}
    \label{fig:mlm_short}
\end{figure}

We show the short lags detected by {\sc javelin} and the maximum-likelihood method in Figures \ref{fig:jav_short} and \ref{fig:mlm_short}, respectively, for different input lags and amplitude ratios. As discussed in Section \ref{sec:short_long}, {\sc javelin} does a better job detecting the short lag for light curves with an input long lag of $-50$~days and an amplitude ratio of 0.1 than for light curves with other input lags. Otherwise, the accuracy of these distributions for different input lags is similar.

\subsection{Improving the Baseline Cadence}
\label{app:baseline}

\begin{figure}
    \centering
    \includegraphics[width=0.32\textwidth]{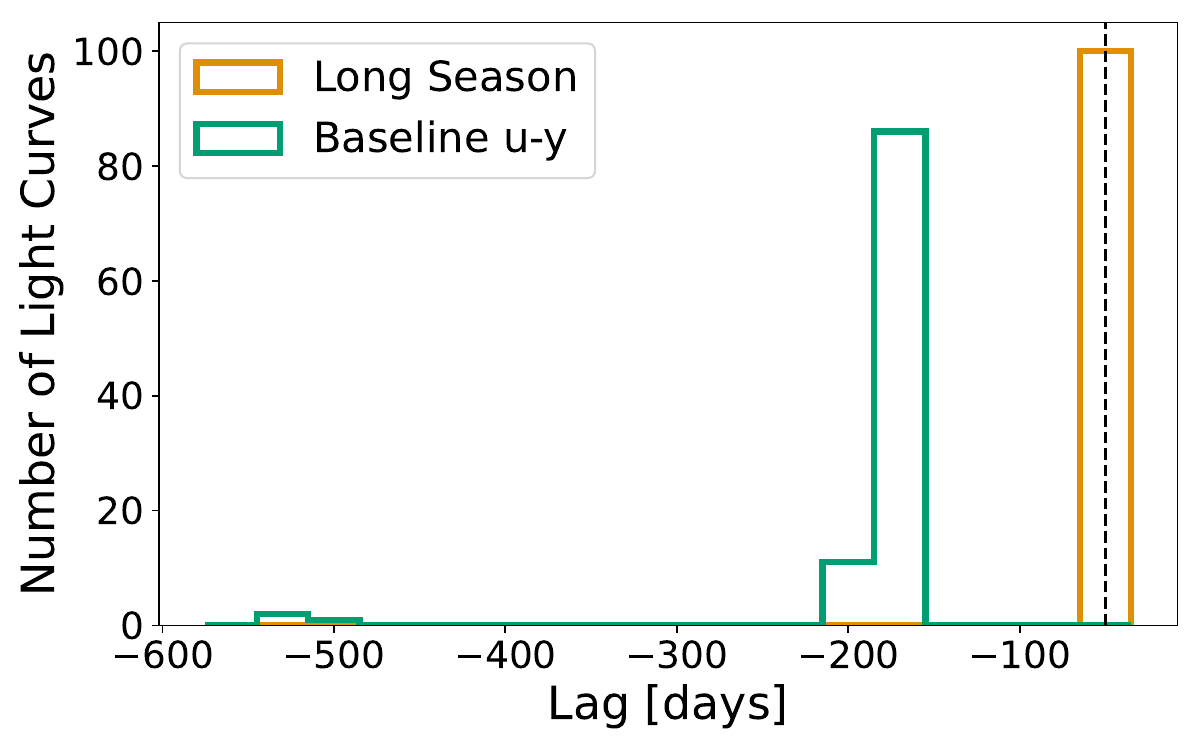}
    \includegraphics[width=0.32\textwidth]{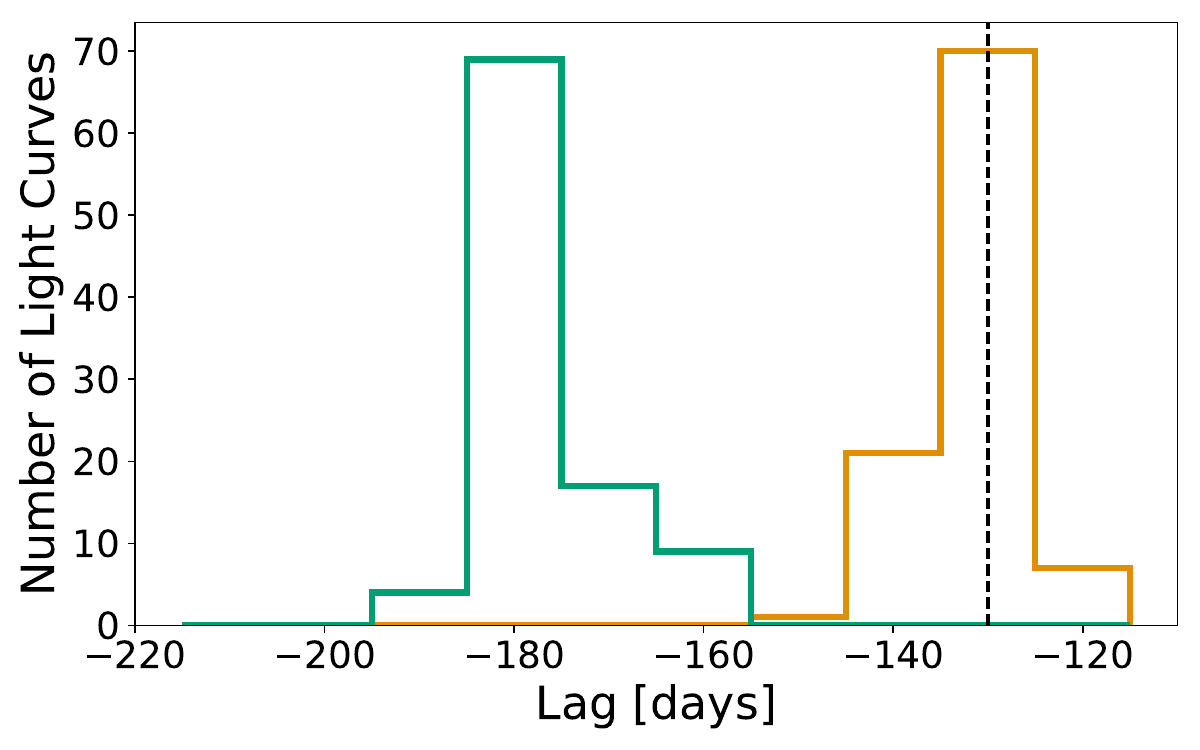}
    \includegraphics[width=0.32\textwidth]{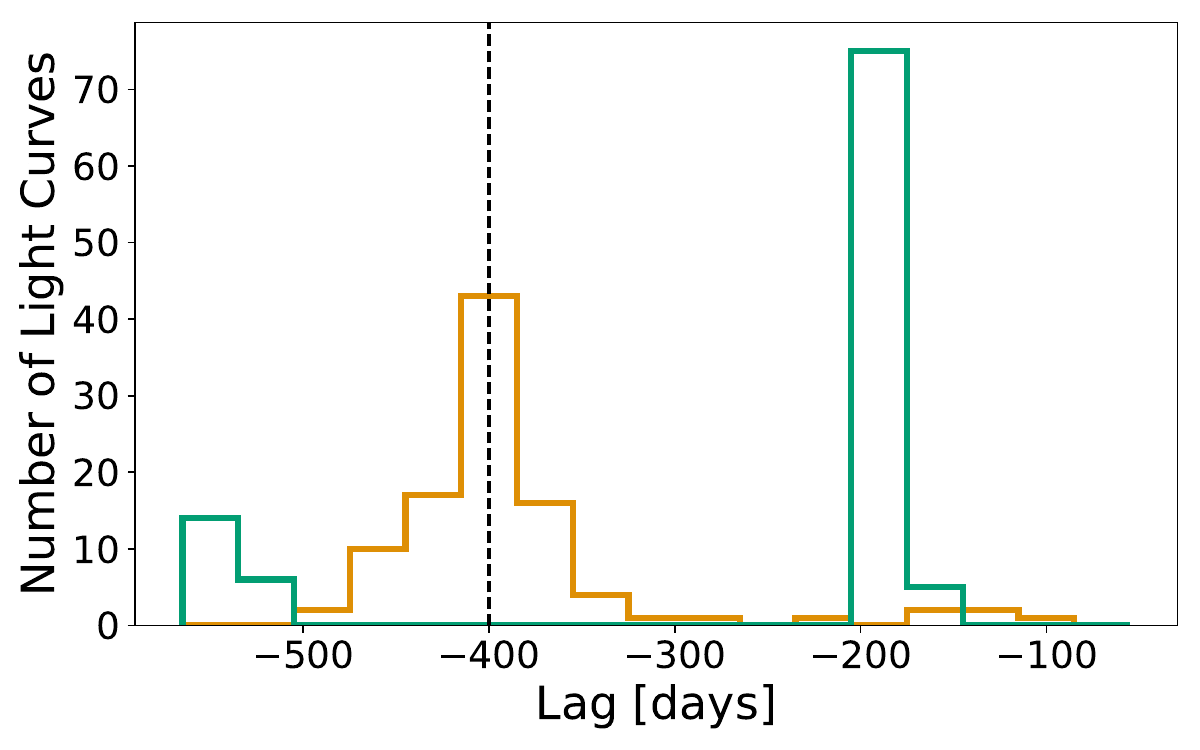}
    \caption{The distributions of the long lags recovered by {\sc javelin} for 100 light curves with input lags of $-50$ and 7 days (left panel), $-130$ and 9 days (middle panel), and $-400$ and 12 days (right panel). The orange and green distributions are for mock light curves with the long season and baseline v2.0 LSST cadences, respectively, for the \emph{u}- and \emph{y}-band. The dashed vertical line shows the input long lag.}
    \label{fig:jav_base}
\end{figure}

\begin{figure}
    \centering
    \includegraphics[width=0.4\textwidth]{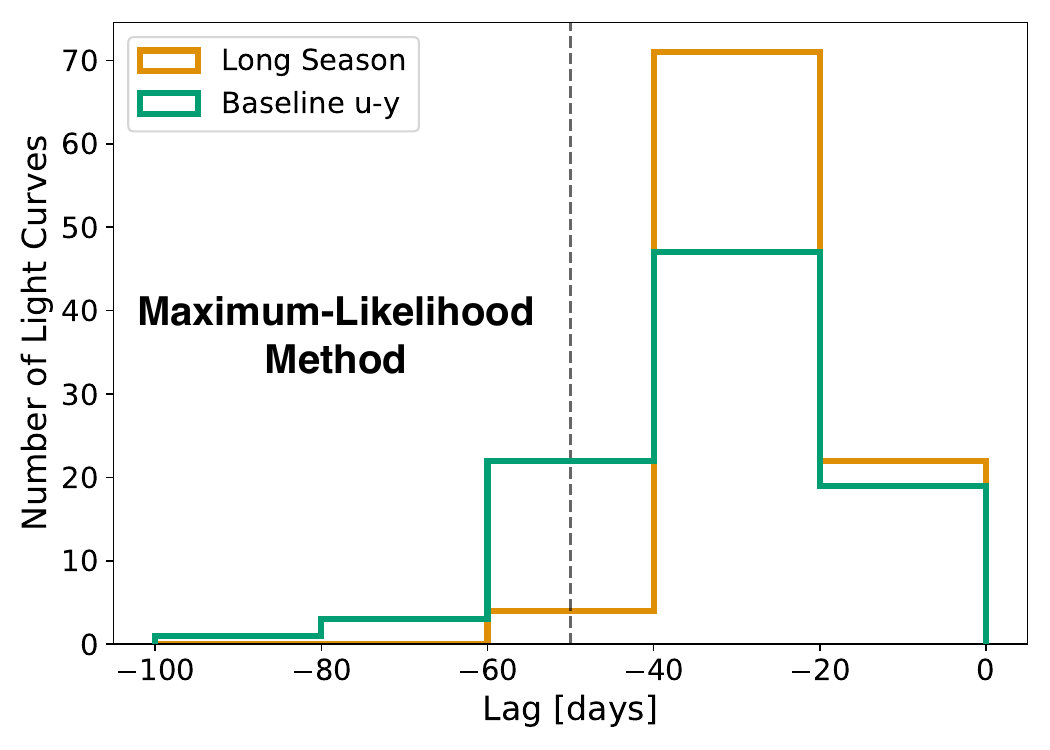}
    \includegraphics[width=0.4\textwidth]{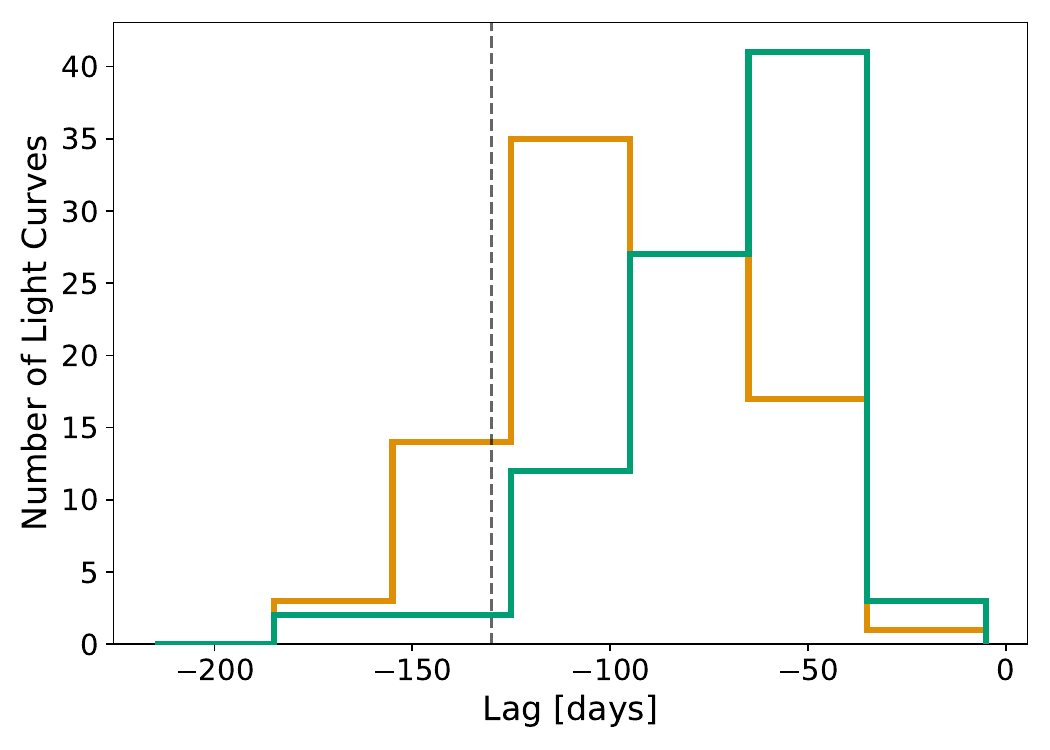}
    \caption{The distributions of the long lags recovered by the maximum-likelihood method for 100 light curves with input lags of $-50$ and 7 days (left panel) and $-130$ and 9 days (right panel). The orange and green distributions are for mock light curves with the long season and baseline v2.0 LSST cadences, respectively, for the \emph{u}- and \emph{y}-band. The dashed vertical line shows the input long lag.}
    \label{fig:mlm_base}
\end{figure}

We compare the accuracy of {\sc javelin} and the maximum-likelihood method for the long season LSST cadence and baseline v2.0 cadence in Figures \ref{fig:jav_base} and \ref{fig:mlm_base}, respectively. {\sc javelin} identifies the seasonal gaps of $-180$~days for light curves with the baseline cadence for all input lags. As mentioned in Section \ref{sec:lco}, the maximum-likelihood method does better at still detecting the input lags for light curves with the baseline cadence than {\sc javelin}, although it does better on light curves with $-50$~day lags than $-130$~day lags, perhaps because $-130$~days is closer to the length of the season gaps.

\begin{figure}
    \centering
    \includegraphics[width=0.32\textwidth]{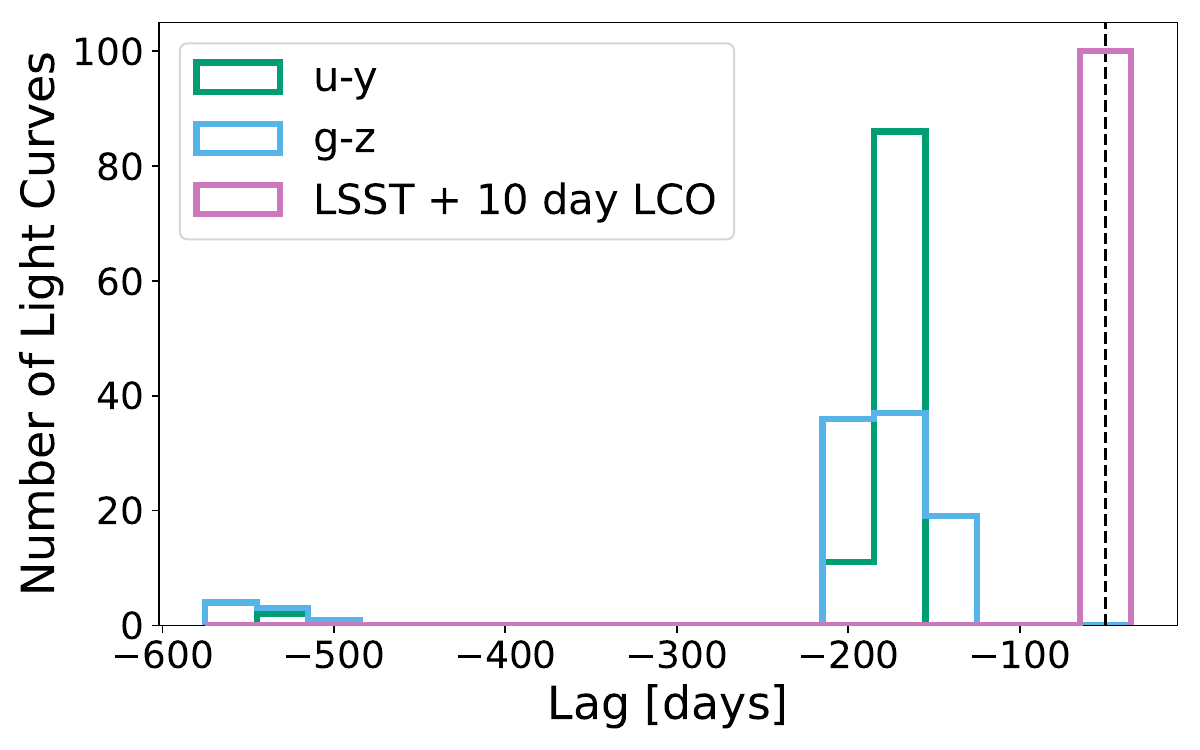}
    \includegraphics[width=0.32\textwidth]{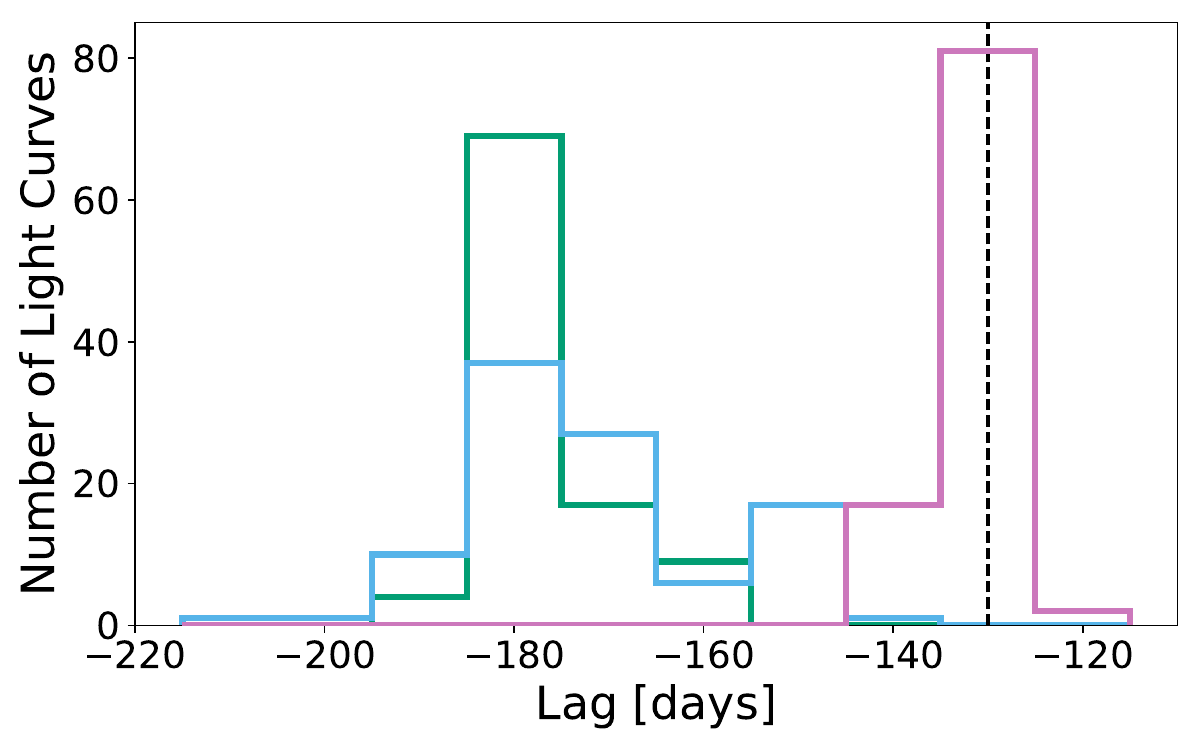}
    \includegraphics[width=0.32\textwidth]{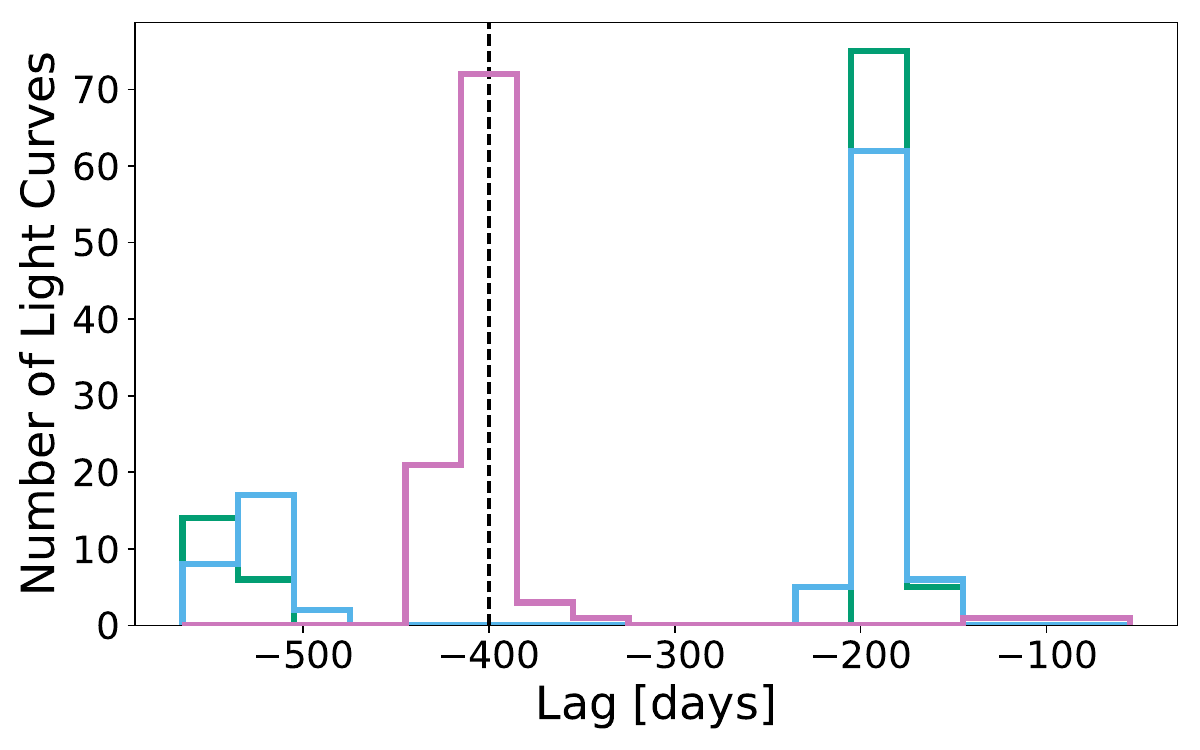}
    \caption{Distribution of the long lags recovered by {\sc javelin} for 100 mock light curves with input lags of $-50$ and 7 days (left panel), $-130$ and 9 days (middle panel), and $-400$ and 12 days (right panel). The dashed vertical line shows the input long lag. The green and light blue distributions are for mock light curves with the baseline v2.0 LSST cadence for the \emph{u}- and \emph{y}-band and \emph{g}- and \emph{z}-band, respectively. In addition to the baseline LSST cadence, the distributions shown in pink have an LCO observation taken roughly every ten days in each LSST observing gap. These LCO observations amount to around 18 observations in each gap, or per year, for a total of around 180 observations.}
    \label{fig:jav_lco}
\end{figure}

\begin{figure}
    \centering
    \includegraphics[width=0.4\textwidth]{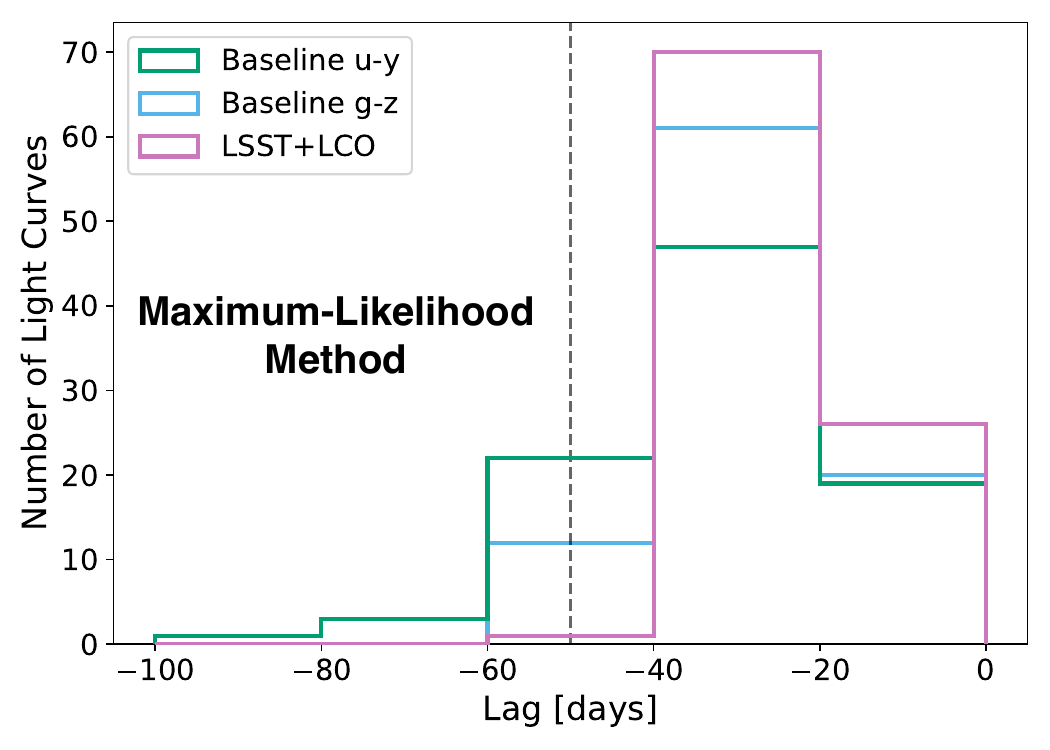}
    \includegraphics[width=0.4\textwidth]{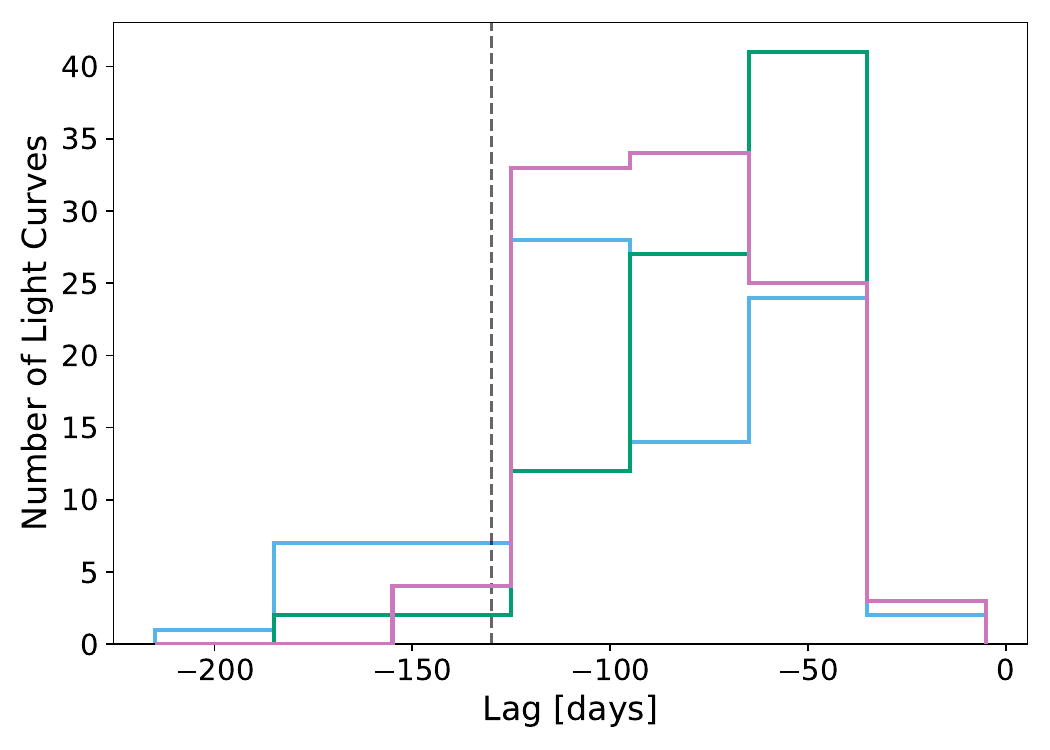}
    \caption{Distribution of the long lag recovered by the maximum-likelihood method for 100 mock light curves with input lags of $-50$ and 7 days (left panel) and  $-130$ and 9 days (right panel). The dashed vertical line shows the input long lag. The green and light blue distributions are for mock light curves with the baseline v2.0 LSST cadence for the \emph{u}- and \emph{y}-band and \emph{g}- and \emph{z}-band, respectively. In addition to an LSST cadence, the distributions shown in pink have an additional LCO observation taken roughly every ten days in each LSST observing gap.}
    \label{fig:mlm_lco}
\end{figure}

\begin{figure}
    \centering
    \includegraphics[width=0.4\textwidth]{final_figs/mlmceven50s.pdf}
    \includegraphics[width=0.4\textwidth]{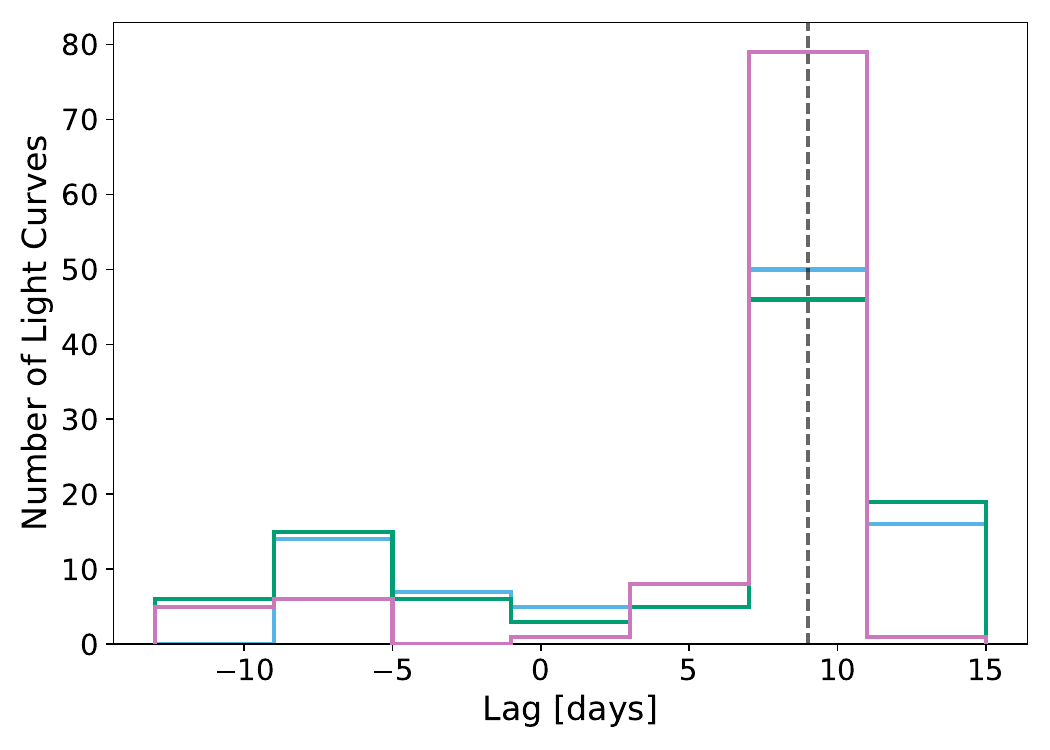}
    \caption{Distribution of the short lag recovered by the maximum-likelihood method for 100 mock light curves with input lags of $-50$ and 7 days (left panel) and $-130$ and 9 days (right panel). The dashed vertical line shows the input short lag. The green and light blue distributions are for mock light curves with the baseline v2.0 LSST cadence for the \emph{u}- and \emph{y}-band and \emph{g}- and \emph{z}-band, respectively. In addition to an LSST cadence, the distributions shown in pink have an additional LCO observation taken roughly every ten days in each LSST observing gap.}
    \label{fig:mlm_lco_short}
\end{figure}

In Figures \ref{fig:jav_lco} and \ref{fig:mlm_lco}, we compare light curves that have both the LSST baseline cadence and a roughly ten day cadence of LCO observations in the LSST season gaps (see Section \ref{sec:lco} for details), to light curves with the LSST baseline cadence for various wavebands. We find that adding LCO observations in the LSST seasonal gaps is very useful for improving the accuracy of {\sc javelin} for all input long lags. The distributions of lags detected by the maximum-likelihood method for light curves with different cadences are consistent with each other for all input long lags. However, Figure \ref{fig:mlm_lco_short} shows that the maximum-likelihood method does a better job detecting short lags with the added LCO cadence for both input lags.

\begin{figure}
    \centering
    \includegraphics[width=0.32\textwidth]{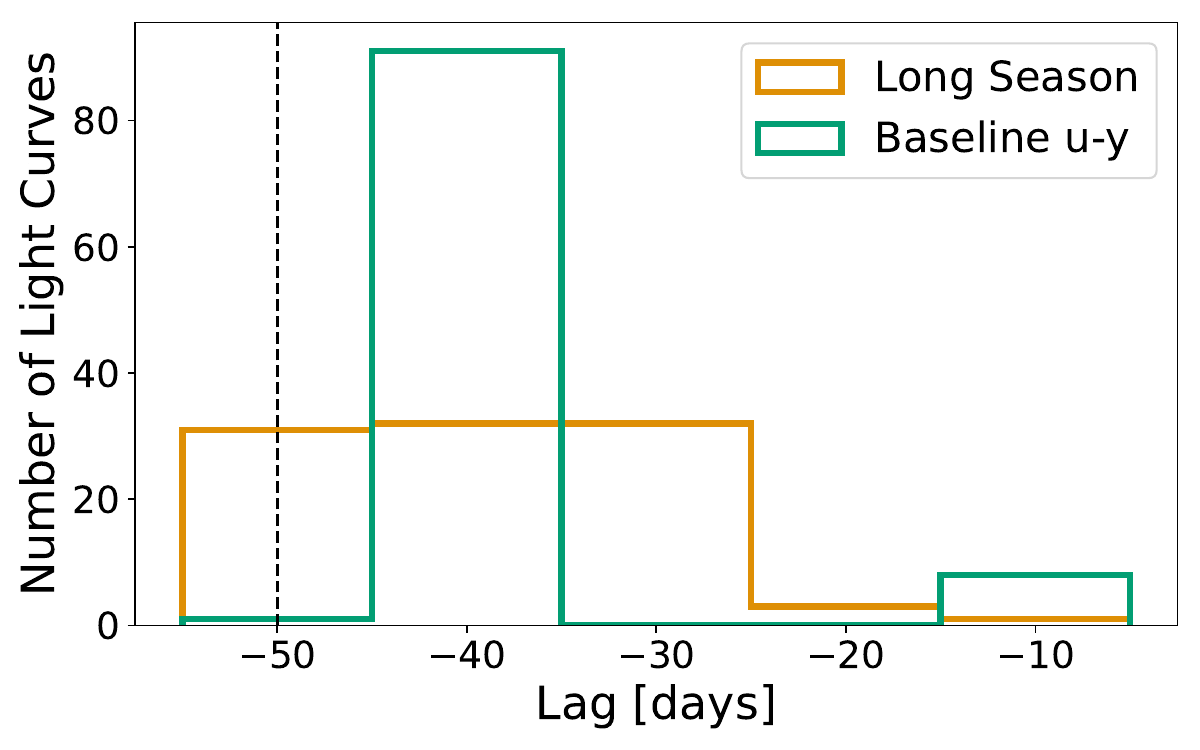}
    \includegraphics[width=0.32\textwidth]{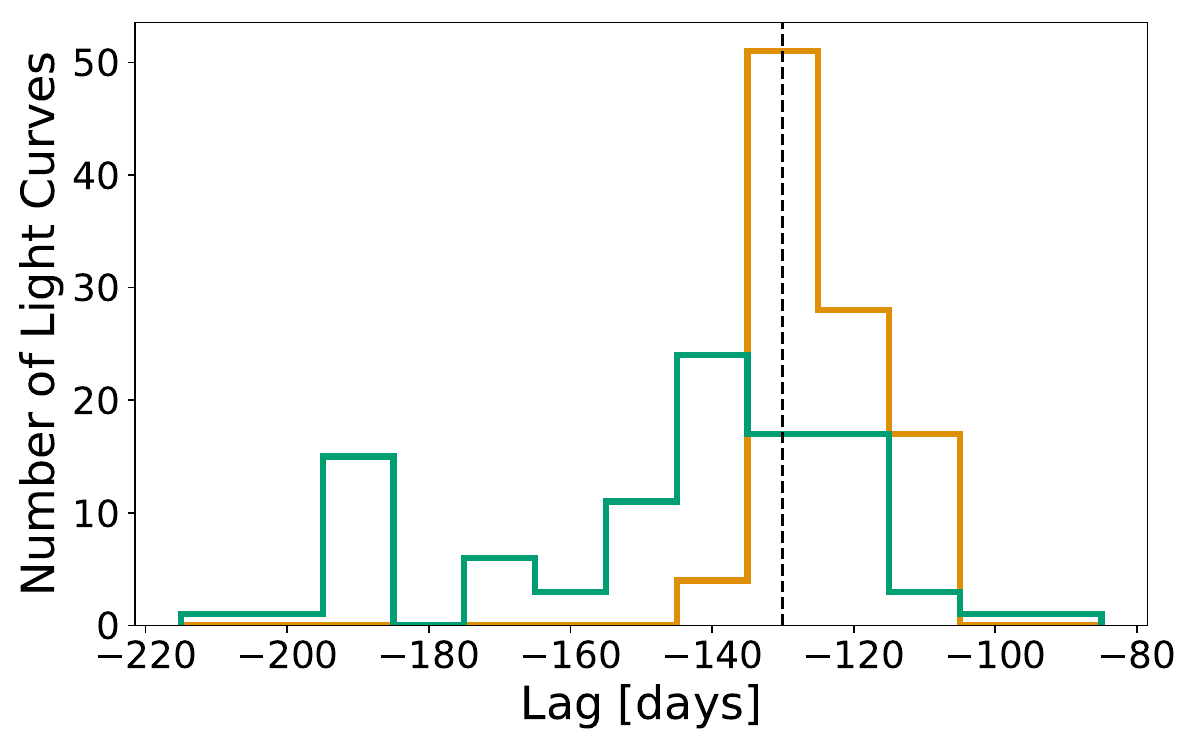}
    \includegraphics[width=0.32\textwidth]{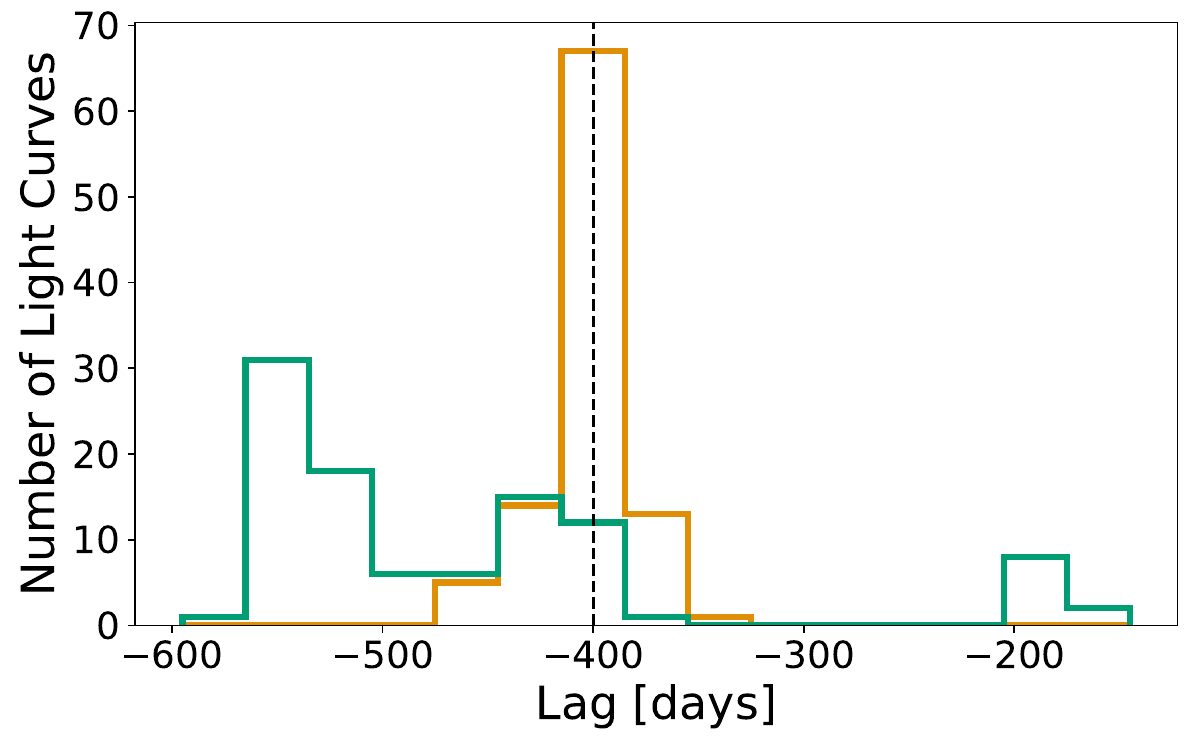}
    \caption{Distribution of the long lag recovered by the Von-Neumann method for 100 mock light curves with input lags of $-50$ and $7$ days (left panel), $-130$ and $9$ days (middle panel), and $-400$ and $12$~days (right panel). The dashed vertical line shows the input long lag. The orange (green) distributions are for mock light curves with the long season (baseline) \emph{u}- and \emph{y}-band cadence.}
    \label{fig:vnm_baseline}
\end{figure}

As mentioned in Section \ref{sec:lco}, the Von-Neumann method performs better on light curves with the baseline cadence than {\sc javelin}. We show the distribution of lags detected by the Von-Neumann method for light curves with the long season (in orange) and baseline (in green) cadences in Figure \ref{fig:vnm_baseline}. The Von-Neumann method detects long lags within 20~days of the input lag for a majority of light curves, except for light curves with an input long lag of $-400$~days, where the distribution peaks around $-550$~days. The median long lags are $-40^{+8}_{-7}$~days and $-40^{+1}_{-2}$~days for an input long lag of $-50$~days, $-130^{+10}_{-0}$~days and $-140^{+10}_{-20}$~days for an input long lag of $-130$~days, and $-400^{+10}_{-10}$~days and $-510^{+80}_{-50}$~days for an input long lag of $-400$~days, for the long season and baseline cadences, respectively.

\section{Null Tests}

\begin{figure*}
    \centering
    \includegraphics[width=0.48\textwidth]{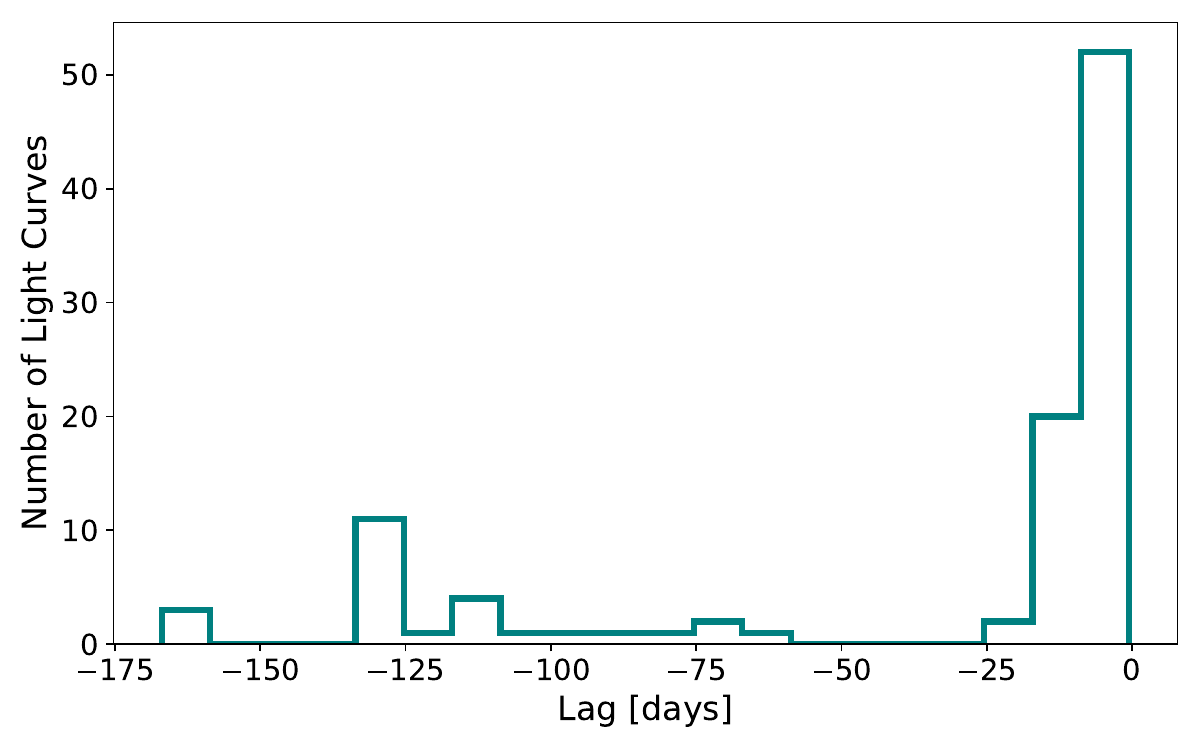}
    \includegraphics[width=0.48\textwidth]{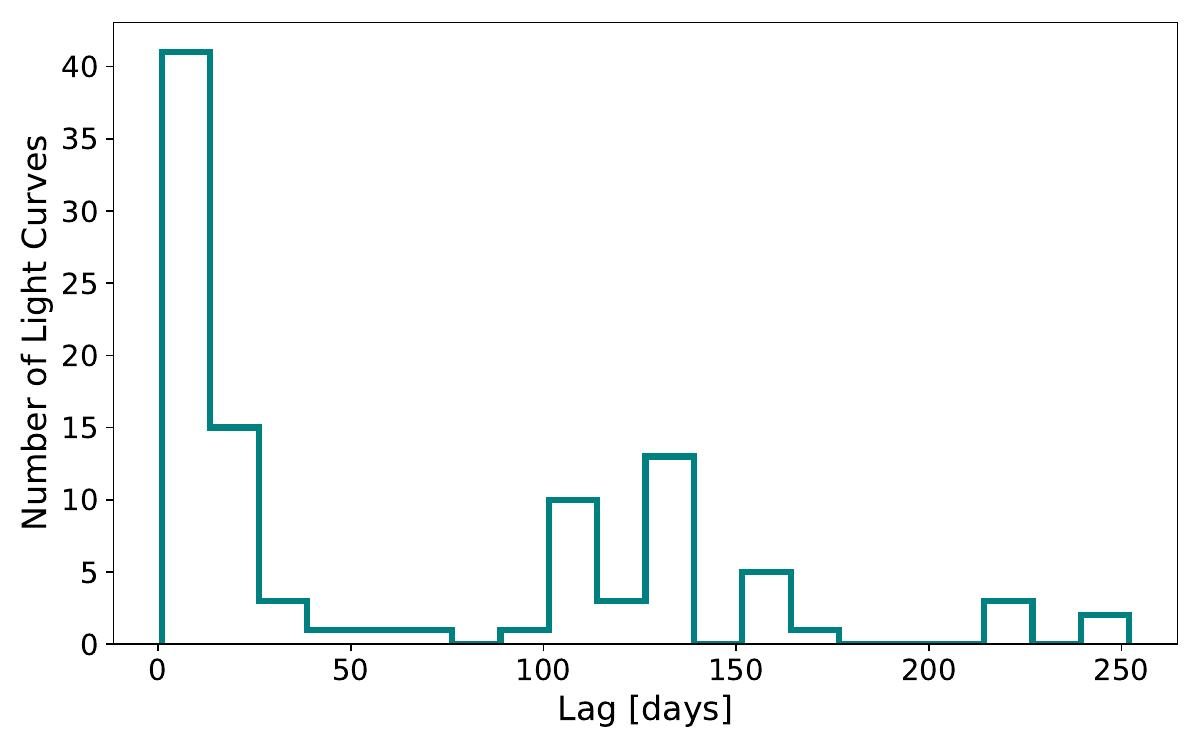} \\
    \includegraphics[width=0.48\textwidth]{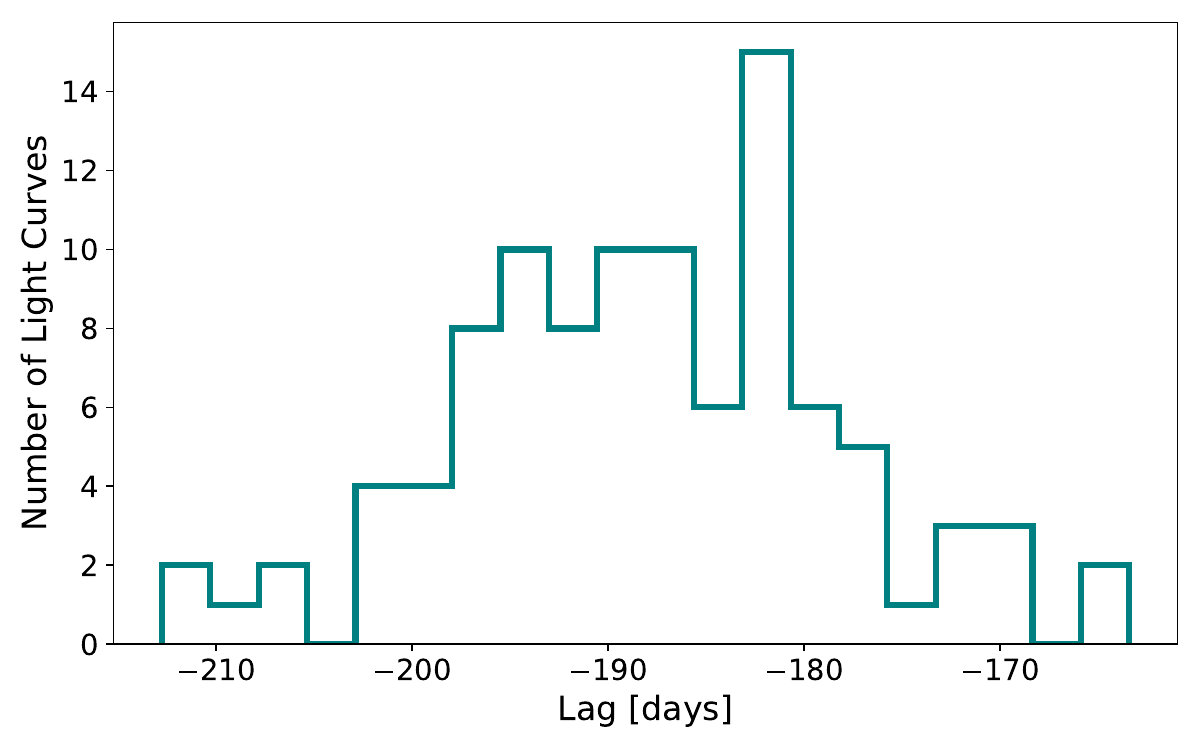}
    \includegraphics[width=0.48\textwidth]{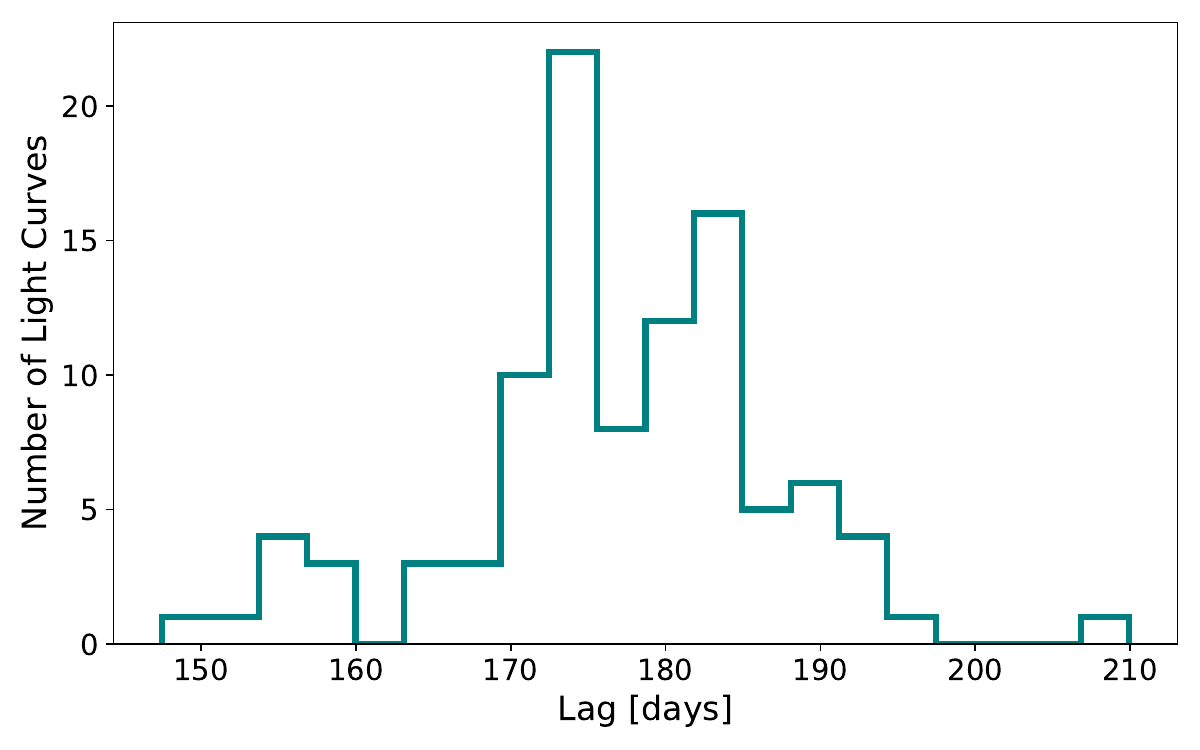} \\
    \includegraphics[width=0.48\textwidth]{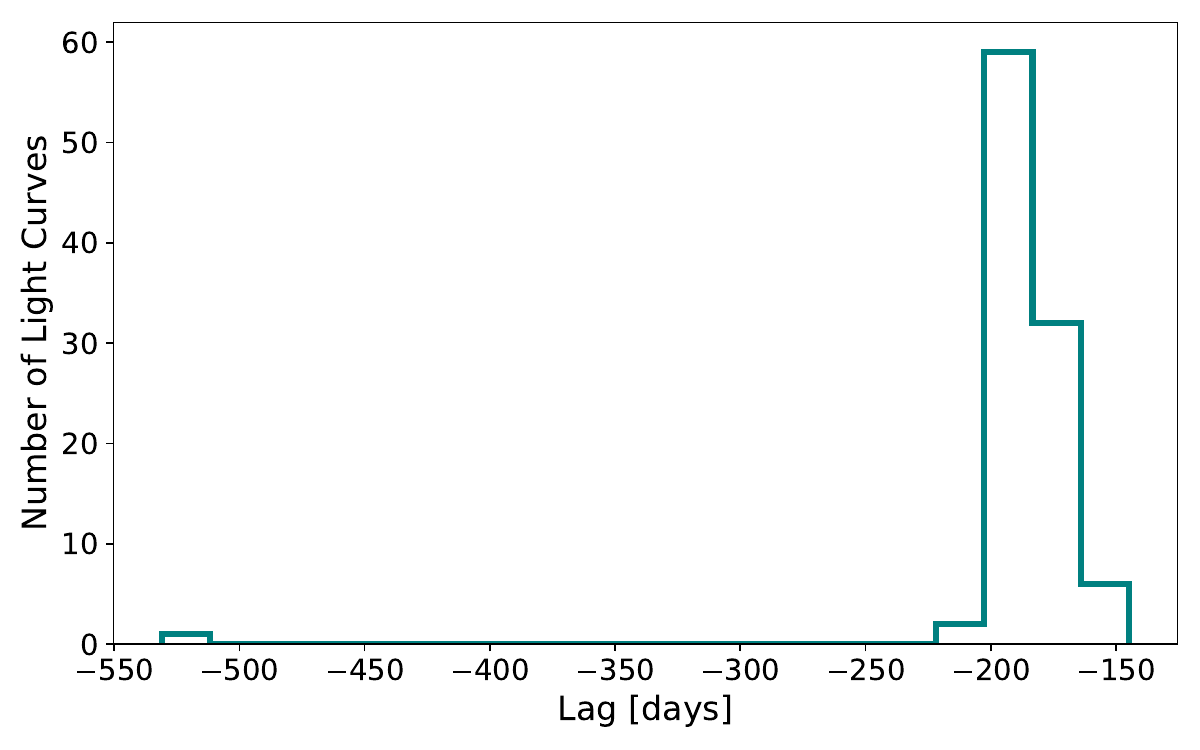}
    \includegraphics[width=0.48\textwidth]{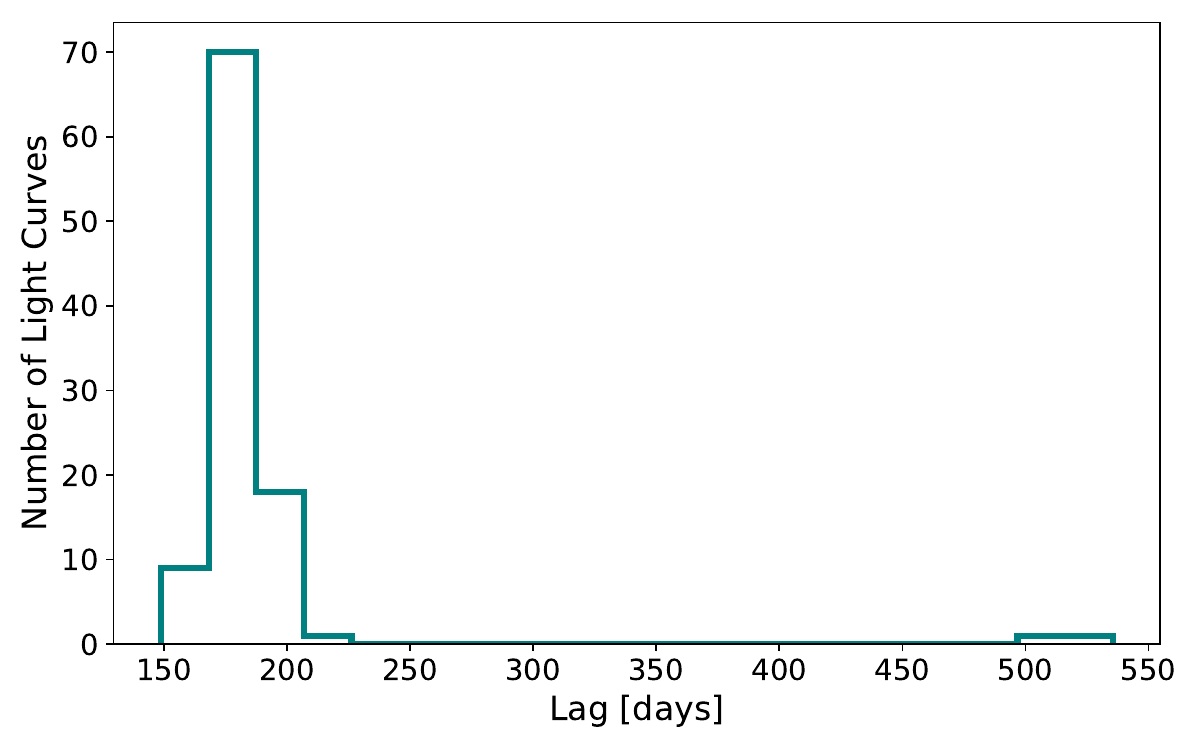} \\
    \includegraphics[width=0.48\textwidth]{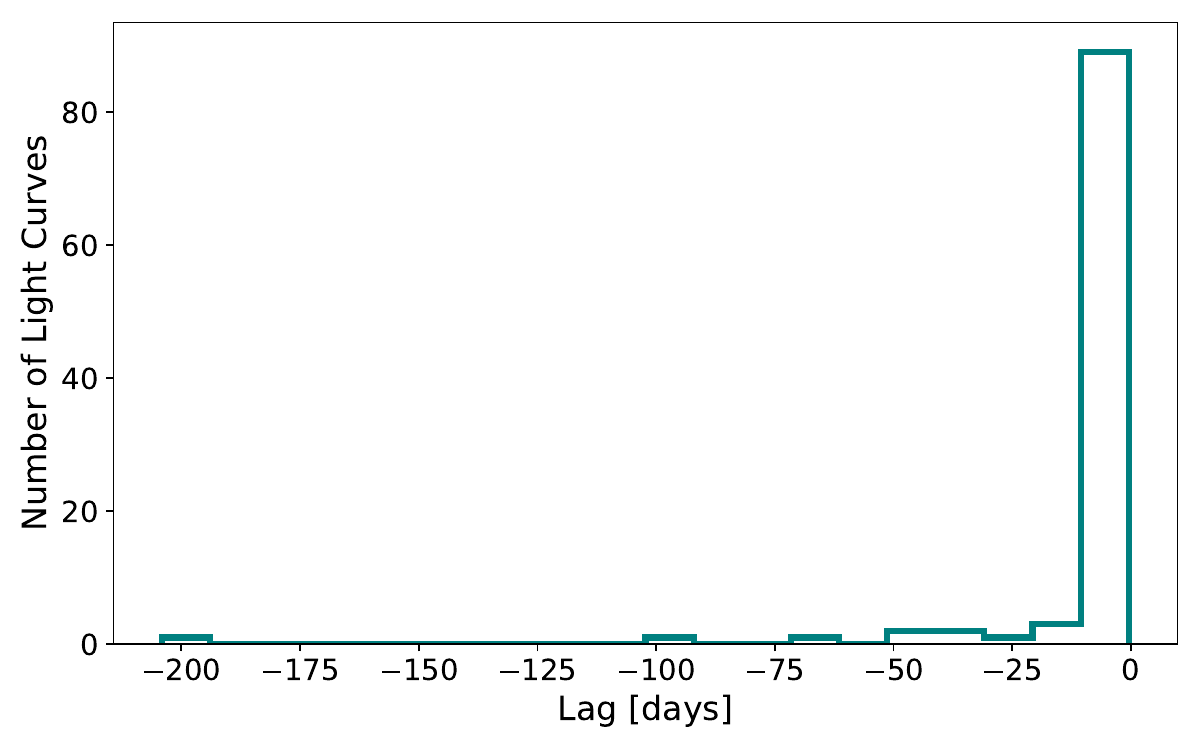}
    \includegraphics[width=0.48\textwidth]{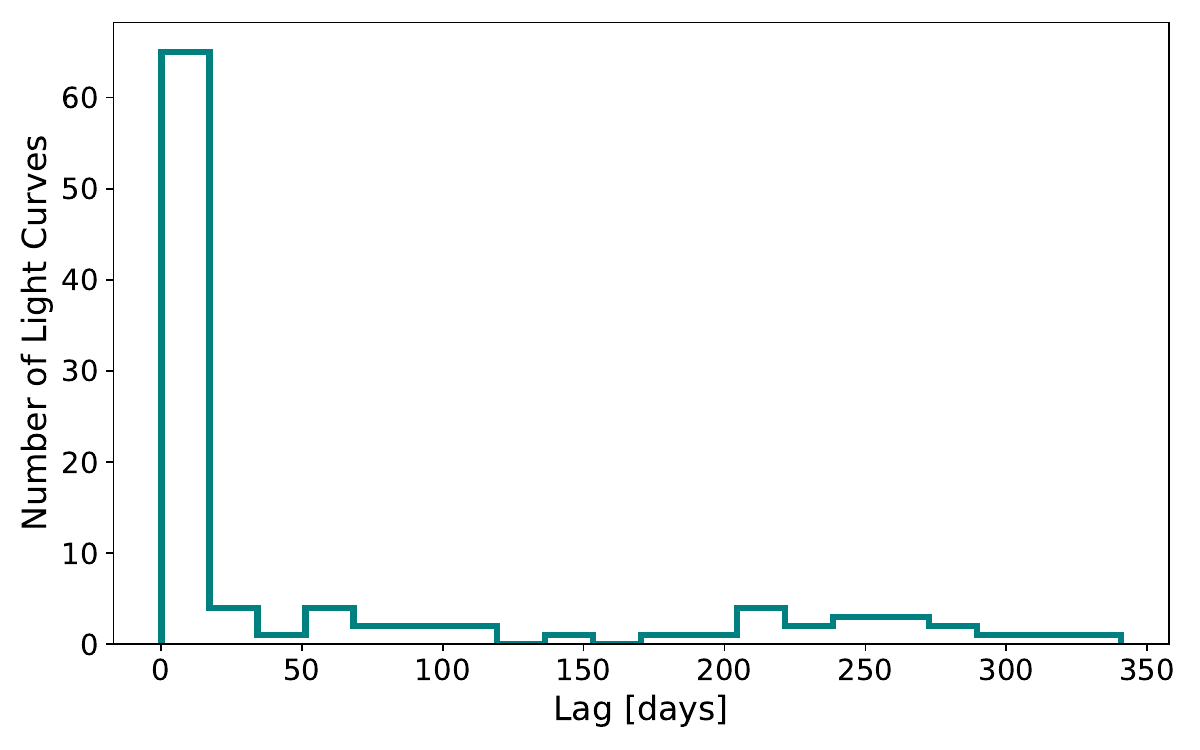}
    \caption{The distribution of the negative (left panels) and positive (right panels) medians of {\sc javelin} distributions for 100 mock light curves with no input lag. From top to bottom, the light curves have been sub-sampled with the long season LSST \emph{u}- and \emph{y}-band cadence, the baseline \emph{u}- and \emph{y}-band cadence, the baseline \emph{g}- and \emph{z}-band cadence, and the baseline \emph{g}- and \emph{z}-band cadence + LCO cadence.}
    \label{fig:app:jav}
\end{figure*}

\begin{figure*}
    \centering
    \includegraphics[width=\textwidth]{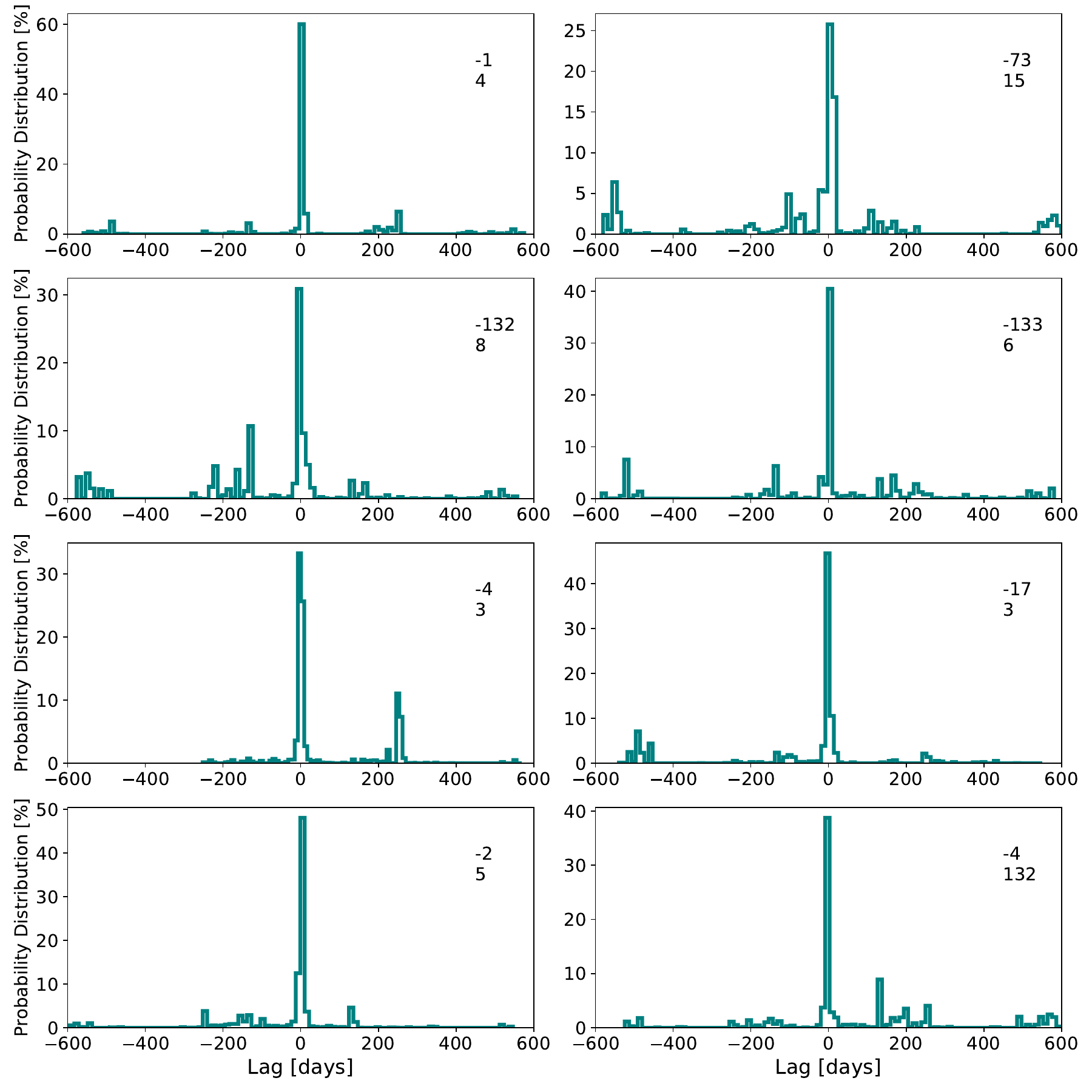}
    \caption{Eight individual {\sc javelin} distributions for light curves with no reprocessing and the long season LSST cadence. Each panel is labeled with the median of the positive and negative values of the distribution.}
    \label{fig:app:javex}
\end{figure*}

To verify that our results are due to the presence of lags we run our two lag detection methods on mock light curves with no input lag for each of the different cadences presented in this paper. To generate these mocks we perform all of the steps outlined in Section \ref{sec:mocks}, but skip Section \ref{sec:reprocess}, i.e. we do not reprocess either light curve with a response function. As a result, we have two identical light curves that are then sub-sampled with the cadence for two different wave bands (\emph{u} and \emph{y} or \emph{g} and \emph{z}, depending on the cadence). Below we discuss the signals recovered by {\sc javelin} (Section \ref{app:jav}) and the maximum-likelihood method (Section \ref{app:mlm}). Any signals recovered could be due to the cadence/gaps in the cadence, or timescales inherent to the light curves themselves, such as the damping timescale.

\subsection{Javelin}
\label{app:jav}

Figure \ref{fig:app:jav} shows the distributions of the medians of the negative (left panel) and positive (right panel) values of the {\sc javelin} distributions for 100 mock DRW light curves with no re-processing. Light curves in the top panel have been mock observed with the long season LSST \emph{u}- and \emph{y}-band cadence used for the light curves in Sections \ref{sec:long}, \ref{sec:short_long}, \ref{sec:light_curves}, and \ref{sec:i21}. In the remaining panels light curves have been mock observed with the baseline \emph{u}- and \emph{y}-band cadence, the baseline \emph{g}- and \emph{z}-band cadence, and the baseline \emph{g}- and \emph{z}-band cadence with the additional LCO cadence, which are all used to sub-sample the light curves in Section \ref{sec:lco}. The distribution of the negative medians for light curves with the long season cadence peaks at zero. The duration of the long lag detected by {\sc javelin} for 70\% of light curves is $<15$~days, suggesting that the lags recovered by {\sc javelin} in the previous sections are due to the input lag signal. On the other hand, over 50\% of positive lags recovered by {\sc javelin} are over 5 days, suggesting that our short lag results could be contaminated.

The individual {\sc javelin} distributions for light curves with the long season cadence that have not been reprocessed with an input lag, several of which are shown in Figure \ref{fig:app:javex}, peak most prominently at zero, but have noise. Most of this noise is either around $\pm 150-200$~days, because that is the length of the season gap, or around the edges of the range over which we run {\sc javelin} ($\pm 600$~days). When we take the median of only the positive or negative parts of the distributions, this noise skews lag values to $>0$~days.

In Figure \ref{fig:app:javex} each panel is labeled with the median of the positive and negative values of the distribution, most of which are very near zero, due to the sharp peak in the distribution around zero. There are several exceptions where the positive or negative median is $\pm 132$~days, due to noise in the distribution. These large median values only amount to 10--15\% of the overall distribution in the top panel of Figure \ref{fig:app:jav}. However, these values are similar to our input lag of $-130$~days and so could contribute somewhat to our ability to detect $-130$~day lags with {\sc javelin}.

The mock light curves made from the simulation in \cite{JiangBlaes2020} presented in Section \ref{sec:light_curves} also have the long season cadence. The top panel of Figure \ref{fig:app:yf_nolag} shows that for mock light curves without an input lag, {\sc javelin} only detects a long lag longer than 10 days and a short lag longer than one day for two light curves each. 

On the other hand, we show in Section \ref{sec:lco} that the baseline cadence leads to contamination from the $\pm 180$~day seasonal gap. For light curves sub-sampled with the baseline cadence that have not been reprocessed, the medians of the positive and negative values of the {\sc javelin} distributions peak around $+180$ and $-180$ days, respectively (see middle panels of Figure \ref{fig:app:jav}). When we add a mock LCO cadence of roughly every ten days to the gaps, however, the results are significantly better (see bottom panel of Figure \ref{fig:app:jav}). 90\% of negative lags detected have a duration of less than 15~days. The same is true for $\sim 60\%$ of the short lags detected. Unsurprisingly, it appears for this cadence the {\sc javelin} results are very robust.

\subsection{Fourier Method}
\label{app:mlm}

\begin{figure}
    \centering
\includegraphics[width=0.48\textwidth]{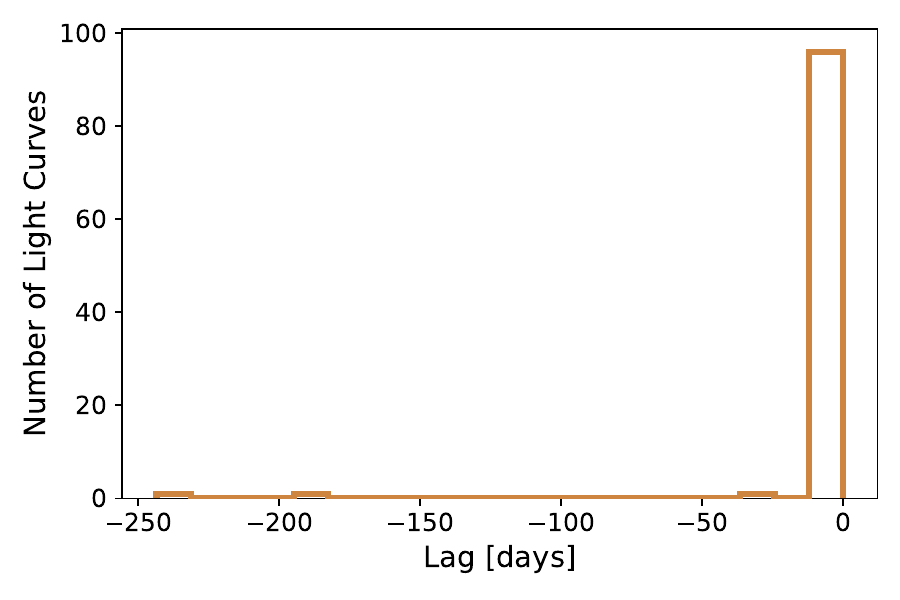}
\includegraphics[width=0.48\textwidth]{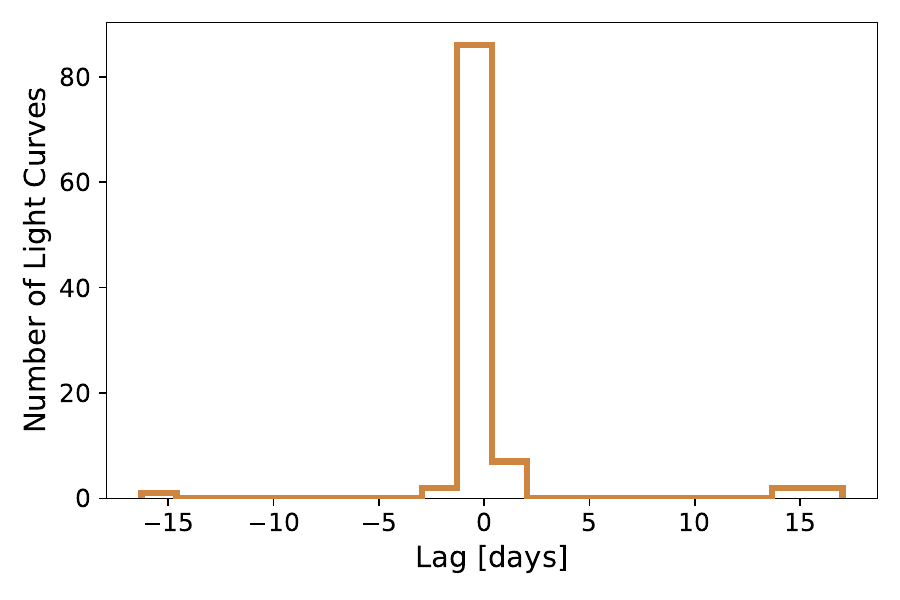} \\
\includegraphics[width=0.48\textwidth]{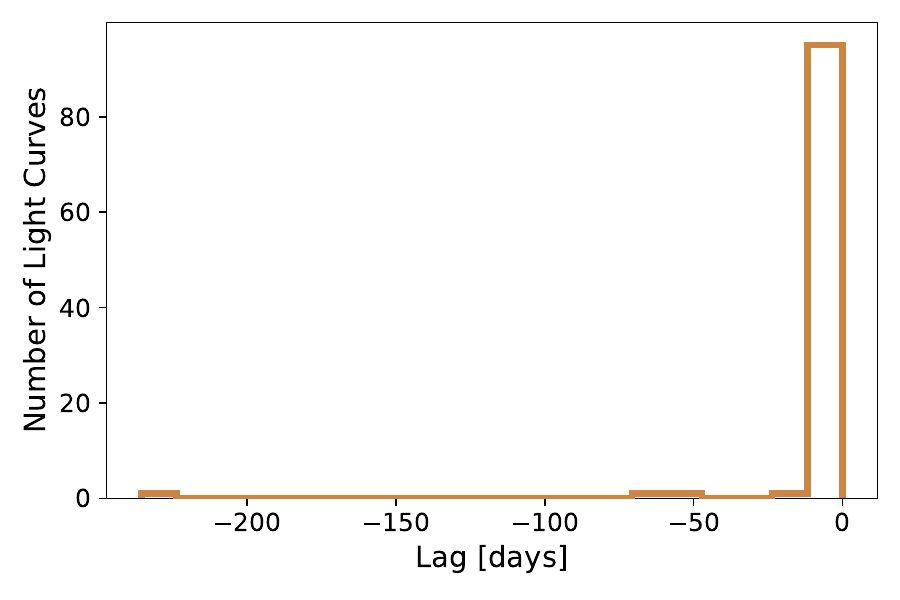}
\includegraphics[width=0.48\textwidth]{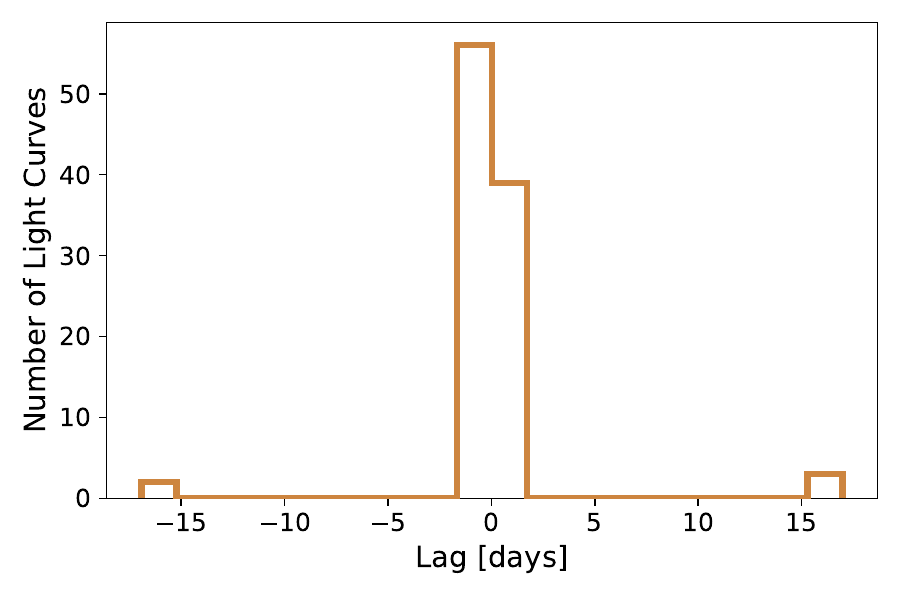}\\
\includegraphics[width=0.48\textwidth]{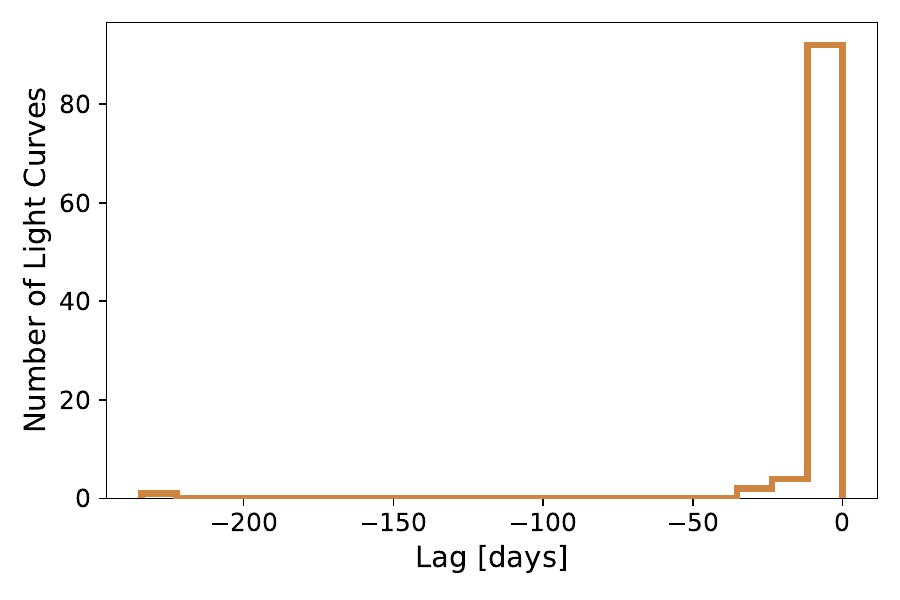}
\includegraphics[width=0.48\textwidth]{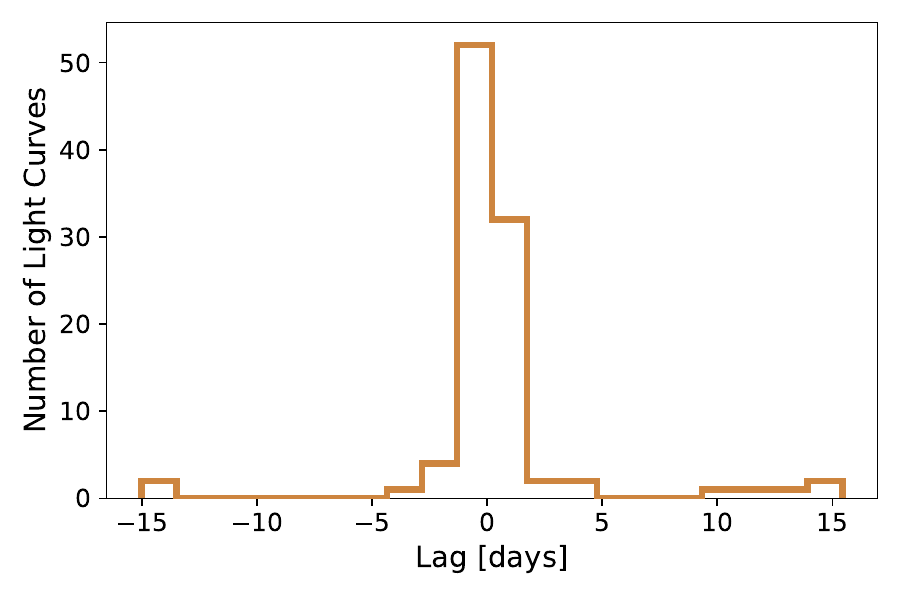} \\
\includegraphics[width=0.48\textwidth]{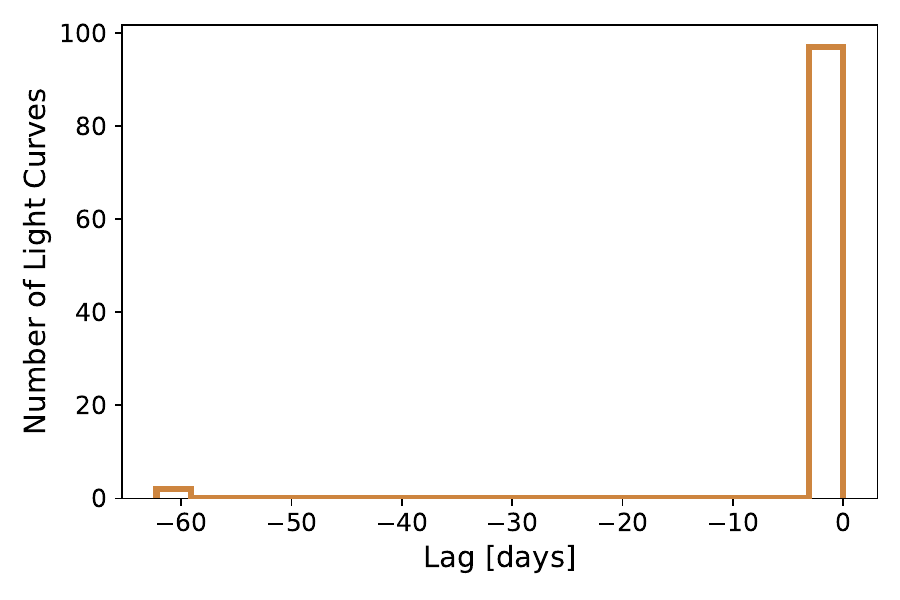}
\includegraphics[width=0.48\textwidth]{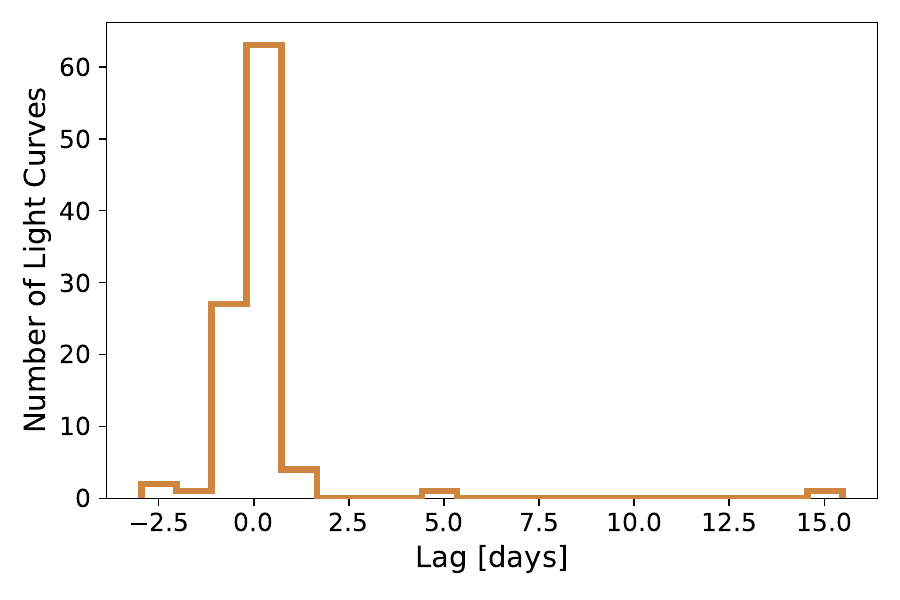} \\
    \caption{The distribution of long lags (left panel) and short lags (right panel) recovered by the maximum-likelihood method for light curves that have not been reprocessed with a lag for all cadences in the paper. From top to bottom, the light curves have been sub-sampled with the long season LSST \emph{u}- and \emph{y}-band cadence, the baseline \emph{u}- and \emph{y}-band cadence, the baseline \emph{g}- and \emph{z}-band cadence, and the baseline \emph{g}- and \emph{z}-band cadence + LCO cadence.}
    \label{fig:app:mlm_nolag}
\end{figure}

\begin{figure}
    \centering
    \includegraphics[width=0.48\textwidth]{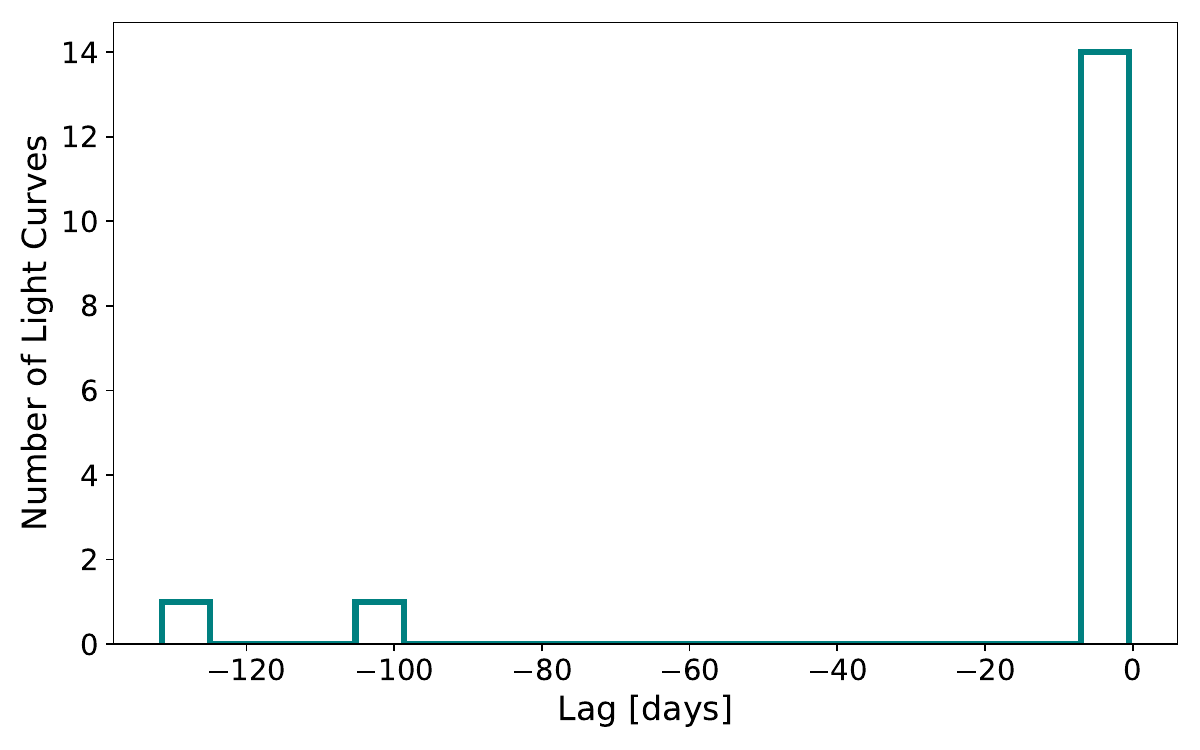}
    \includegraphics[width=0.48\textwidth]{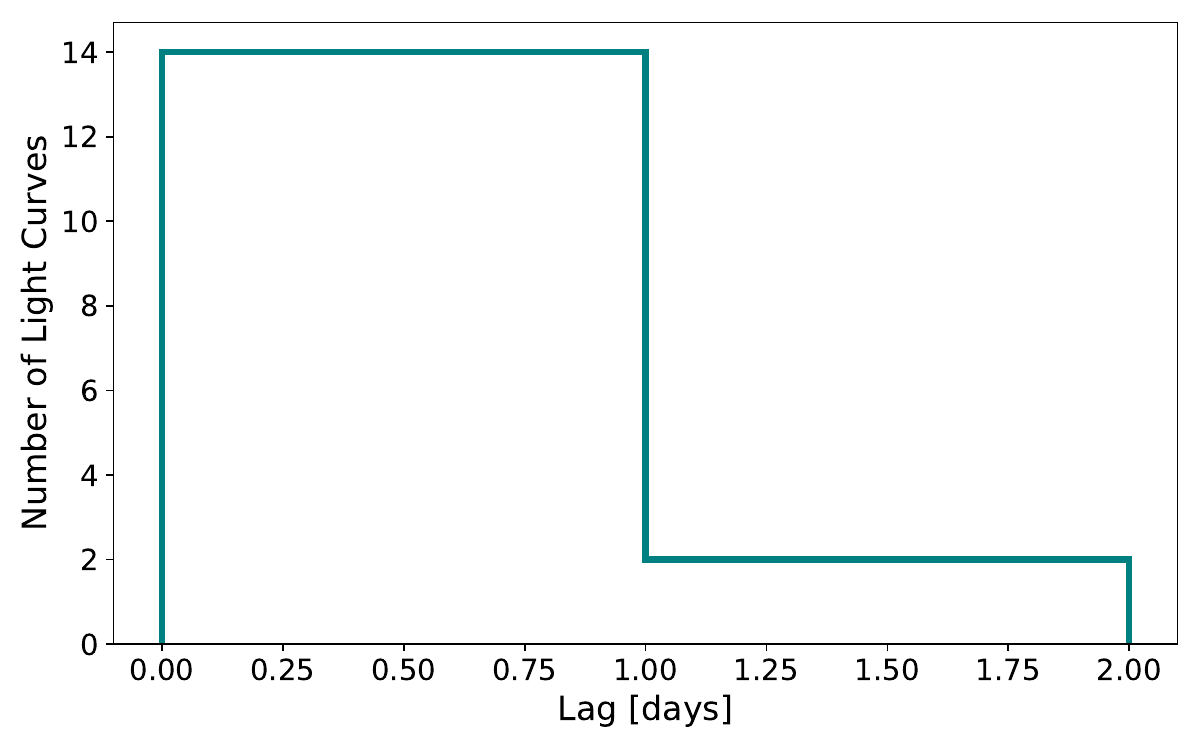} \\
    \includegraphics[width=0.48\textwidth]{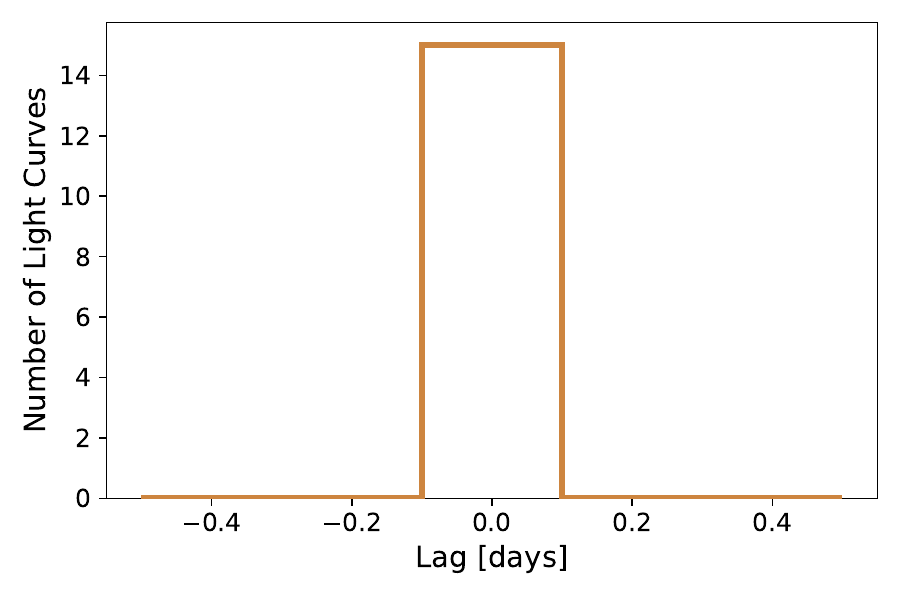}
    \includegraphics[width=0.48\textwidth]{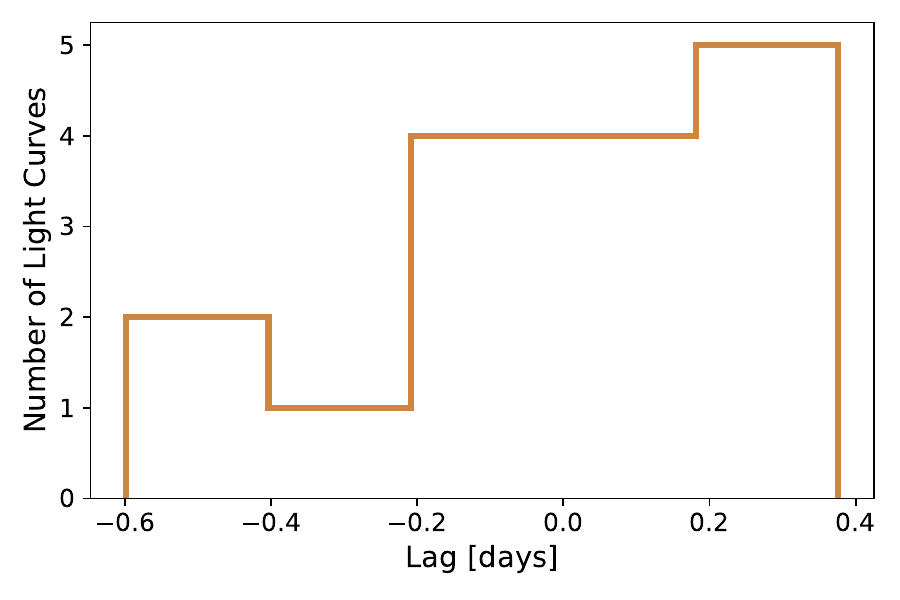} \\
    \caption{The top panel shows the lags recovered by {\sc javelin} and the bottom panel shows the lags recovered by the maximum-likelihood method for 16 mock light curves made from the simulation in \cite{JiangBlaes2020} with no input lag and the long season LSST \emph{u}- and \emph{y}-band cadence.}
    \label{fig:app:yf_nolag}
\end{figure}

Figure \ref{fig:app:mlm_nolag} shows the distribution of the lags recovered by the maximum-likelihood method for light curves sub-sampled with the same cadences used throughout the paper, but with no input lag. Roughly 90\% of long and short lags detected by the maximum-likelihood method are shorter than 10 days for all cadences. For the long season cadence (top panel) the maximum-likelihood method detects only 3 non-zero long lags and only seven short lags longer than 1~day.

For light curves with no reprocessing and the LSST+LCO cadence (bottom panel of Figure \ref{fig:app:mlm_nolag}), the maximum-likelihood method detects lags shorter than 1 day for roughly 90\% of light curves. The maximum-likelihood method performs less well on the two baseline cadences (middle two panels of Figure \ref{fig:app:mlm_nolag}) without the added LCO observations, although 80\% of long (short) lags detected are shorter than 10 (3) days. The lags detected by the maximum-likelihood method for mock light curves made from the light curves in the simulation in \cite{JiangBlaes2020} with no input lag are all less than 1~day (see bottom panel of Figure \ref{fig:app:yf_nolag}). Overall, the lags recovered by the maximum-likelihood method seem unlikely to be contaminated by any of the cadences used in this paper or the light curve properties.

\end{CJK*}
\end{document}